\documentclass[aip,jcp,reprint]{revtex4-2}
\usepackage{graphicx}
\usepackage{appendix}
\usepackage{rotating}
\usepackage{simplewick}
\usepackage{amssymb}
\usepackage{amsmath}
\usepackage{txfonts}
\usepackage{mathrsfs}
\usepackage{bm}
\usepackage{dcolumn}
\usepackage{color}

\begin{document}
\title{Degenerate coupled-cluster theory}

\date{\today}
\author{So \surname{Hirata}}\email[Email:~]{sohirata@illinois.edu}
\affiliation{Department of Chemistry, University of Illinois at Urbana-Champaign, Urbana, Illinois 61801, USA}

\begin{abstract}
A size-extensive, converging, black-box, {\it ab initio} coupled-cluster ($\Delta$CC) ansatz is introduced that computes the energies and wave functions of states from any degenerate or nondegenerate Slater-determinant references with any numbers of $\alpha$- and $\beta$-spin electrons, any patterns of orbital occupancy, any spin multiplicities, and any spatial symmetries. For a nondegenerate reference, it reduces to the single-reference coupled-cluster ansatz. For a degenerate multireference, it is a natural coupled-cluster extension of degenerate M{\o}ller--Plesset perturbation ($\Delta$MP) theory. For ionized and electron-attached references, it is a coupled-cluster Green's function, although the present theory is convergent toward the full-configuration-interaction (FCI) limits, while Feynman--Dyson many-body Green's function (MBGF) theory generally is not. Its single-excitation instance is a projection Hartree--Fock theory as per the Thouless theorem, which may be useful for core ionizations, high-spin states, and possibly electron affinities. A new multireference coupled-cluster theory for a general model space is also developed. This quasidegenerate coupled-cluster (QCC) theory is exactly converging, but not black-box, and intended for strong correlation. Determinant-based, general-order algorithms of $\Delta$CC and QCC theories are implemented and compared with configuration-interaction (CI) and equation-of-motion coupled-cluster (EOM-CC) theories through octuple excitations and with $\Delta$MP and MBGF theories up to the nineteenth order. An algebraic, optimal-scaling algorithm of $\Delta$CC theory is computer-synthesized at the levels of single excitations ($\Delta$CCS) and of single and double excitations ($\Delta$CCSD). The order of performance is:\ QCC $\approx$ $\Delta$CC $>$ EOM-CC $>$ CI at the same order or QCC $\approx$ $\Delta$CC $>$ $\Delta$MP $>$ MBGF at the same cost scaling.
\end{abstract}
\maketitle

\section{Introduction}

Single-reference coupled-cluster (CC) theory\cite{Coester1958,CoesterKummel1960,Cizek1966,Cizek1969,Paldus1972,Kummel1978,bartlettRMP,shavitt,BartlettMP2010,BartlettWIRE2012} 
is widely regarded as the most accurate, generally applicable, size-extensive, black-box, {\it ab initio} method for solving the time-independent Schr\"{o}dinger equations
for atoms, molecules, and solids in their ground electronic states. 

By ``black-box,'' we mean that the method requires no more than
the atomic positions, number of electrons, and spin-multiplicity of a system as input and returns its unambiguous energy and wave function within the Born--Oppenheimer approximation.
(Here, it does not have the negative connotations of an opaque method whose internal workings are unknown or hidden.)
This is essentially the same definition of Pople's Model Chemistry,\cite{Pople} distinguishing its instances from those that are reserved for experts with inscrutable ``chemical intuitions'' 
in selecting multireferences, active orbitals, etc. 

By ``{\it ab initio},'' we insist on the use of the exact electronic Hamiltonian, guaranteeing the systematic convergence of the results toward
the experimental data upon increasing the rank of the method and basis-set size (plus correcting for the relativistic and non-Born--Oppenheimer effects, if necessary). It imbues 
its instances with predictive power.\cite{Bartlett_JPCrev,HirataYagi2008} 
This characteristic is not shared by empirical and semiempirical methods including density-function theory or by various machine-learning models.
In fact, the latter are often parameterized by the results of high-rank CC calculations.
 
Since the development of its core instance in 1982, i.e., coupled-cluster with connected single and double excitations (CCSD),\cite{PurvisCCSD}
following the initial reports of coupled-cluster doubles (CCD),\cite{PopleCCD1978,BartlettCCD1978} CC theory has been extended to include  connected
triples (CCSDT),\cite{NogaCCSDT,ScuseriaCCSDT} quadruples (CCSDTQ),\cite{KucharskiCCSDTQ} and pentuples (CCSDTQP).\cite{musialCCSDTQP} 
A general-order, determinant-based algorithm has been developed,\cite{hirata_cc,kallay_cc,olsen_cc} yielding benchmark data 
through hextuples (CCSDTQPH), septuples (CCSDTQPHS),
and octuples (CCSDTQPHSO).\cite{note_acronym} They are convergent toward the exact basis-set solutions, i.e., the full-configuration-interaction (FCI) limits, at the rate that is
distinctly faster than the same of configuration-interaction (CI) theory or of many-body perturbation theory (MBPT).\cite{BartlettARPC} 

Further developments of CC theory followed, including the capabilities to compute the analytical derivatives of its energy with respect to 
various perturbations,\cite{Adamowicz1984,Scheiner1987,Scuseria1987,Salter1989,Koch1990deriv,Gauss1991_1,Gauss1991_2,Stanton2000,GaussCCSDT2002,Kallay2003,Kallay2004} leading to efficient methods to determine equilibrium structures, vibrational spectra,\cite{PereraRaman1999,Oneill2007,Jagau2013} (hyper)polarizabilities,\cite{Stanton1993polar,Kobayashi1994,Rozyczko1997,Hattig1997,Oneill2007_2}
nuclear magnetic resonance parameters,\cite{Perera1994,GaussNMR1995,GaussNMR1995_2,Perera1996} and so on at unprecedented accuracy.\cite{Bartlett_JPCrev,Helgaker2012}
CCSD under the periodic boundary conditions has been applied to infinite crystals.\cite{HirataCCSDpoly2001,HirataCCSDpoly2004}
One of the weaknesses of CC theory (or of any other {\it ab initio} electron-correlation theories) is the slow convergence of its energies with respect to the one-electron basis-set size.
This too has been overcome by the explicitly correlated ans\"{a}tze\cite{Kutzelnigg:1985,Tenno:2004CPL,Tenno:2004JCP,Klopper:2006,Shiozaki:2009,Kong:2012,tenno_review} fitted to the CCSD,\cite{Shiozaki2008,Kohn2008} CCSDT,\cite{Shiozaki2009} and CCSDTQ methods,\cite{Shiozaki2009}  culminating, for instance,
in the near exact (99.996--100.004\%) determination of the ground-state energy of the water molecule without extrapolation.\cite{Shiozaki2009}

However, the most outstanding question immediately after the CCSD development was, {\it how can an excited state be treated
by CC theory?}\cite{Bartlett_bio} Since excited-state wave functions are often dominated by two or more degenerate Slater determinants,\cite{foresman} they are not expected
to have an exponential structure from a single determinant. On the one hand, CI theory performs diagonalization of the Hamiltonian matrix in the determinant basis, naturally reporting excited-state eigenvectors 
that are automatically orthogonal to the ground-state eigenvector. On the other hand, CC theory is formulated as a set of nonlinear equations to be solved for 
excitation amplitudes. Their numerous roots other than the one corresponding to the ground state do not admit a simple physical interpretation as excited-state roots 
let alone their facile numerical determinations.\cite{Zivkovic1978,KowalskiPRL1998,KowalskiIJQC2000,KowalskiPRA2000,Faulstich2024} 

Nevertheless, the question has been answered decisively in the form of equation-of-motion coupled-cluster (EOM-CC) theory.\cite{Harris1977,Emrich1,Emrich2,Sekino1984,Geertsen1989,Comeau1993,Stanton1993,Kowalski2001EOMCCSDT,kallay_cc,HirataEOM2004,KrylovEOM2008,BartlettWIRE2012} For excitation energies, it is  equivalent to the coupled-cluster linear response\cite{Monkhorst1977,Ghosh1981,Dalgaard1983,Takahashi1986,Koch1990,Koch1990_2,Rico1993}
or symmetry-adapted-cluster configuration-interaction method,\cite{Nakatsuji1979CPL,Nakatsuji1979CPL_2,Nakatsuji1981,Nakatsuji1981_2} although the latter initially invoked drastic approximations
that undermined the equivalence. The theory was formally introduced by Monkhorst\cite{Monkhorst1977} in 1977, while Harris\cite{Harris1977} in the same journal issue 
recognized its similarity to the ``equations-of-motion'' formalism; \cite{Dunning1967,Rowe1968,Shibuya1970,Simons1973,Simons1976} They
also coined the term ``coupled-cluster theory,'' embraced by one of its originators, K\"{u}mmel.\cite{Kummel1978}

EOM-CC theory gains access to excited states via the one-photon excitation processes simulated by the first-order time-dependent perturbation (linear response) theory applied to the initial ground state described by CC theory. 
It leads to a CI-like diagonalization procedure of a CC effective Hamiltonian matrix, whose ground-state eigenvector is decoupled from the rest. 
Hence, the extensive correlation energy in the ground state is captured by an extensive exponential operator, while an intensive excitation energy is described
by an intensive linear operator. Two or more dominant degenerate determinants of an excited-state wave function can be completely accounted for by the linear operator. 

EOM-CC with singles and doubles (EOM-CCSD) can efficiently locate all low-lying excited states of predominantly one-electron excitation character with 
uniform high accuracy, while it suffers from distinctly larger errors for two-electron excitations and lacks roots corresponding to three-electron and higher excitations.\cite{bartlettRMP,shavitt,BartlettMP2010,BartlettWIRE2012,hirata_eomcc} 
Since the one-electron excited states are responsible for intense peaks in one-photon electronic absorption or emission spectra, 
it may be said that EOM-CC theory is ideally suited for such linear spectral simulations (as opposed to nonlinear spectral simulations; see, however, Mosquera\cite{Mosquera2025}). 
By adopting as the linear operator an ionization or electron-attachment operator, EOM-CC theory can probe ionization potentials (IP-EOM-CC)\cite{Stanton_ip,Bartlett_ip,Stanton_ip2,MusialIP2003,Gour2005,kamiya_1} or electron affinities (EA-EOM-CC),\cite{nooijen_eom,MusialEA2003,Gour2005,kamiya_2}
and has often been considered a coupled-cluster Green's function.\cite{Gunnarsson1978,nooijen_gf1,nooijen_gf2,Meissner1993,Kowalski1,Kowalski2,Kowalski3}

However, there is a more obvious and straightforward alternative answer to the above question:\ For a nondegenerate determinant reference --- be it an excited, ionized, or electron-attached type --- the target
state is expressed as an exponential excitation operator acting on the closest reference determinant, i.e., single-reference CC theory itself. There is no reason to believe  
that it would work poorly for reference determinants that do not obey the aufbau principle. Rather, its performance may well be correlated with the presence or absence of 
quasidegeneracy of the reference. 
In fact, these expectations were borne out by recent applications of single-reference 
CCSD for nondegenerate excited states.\cite{Lee2019,Damour2024} Treating every state on an equal footing, it works as well as EOM-CCSD for predominantly 
one-electron excited states, distinctly better than EOM-CCSD for two-electron excited states, and it has roots for three-electron and higher single determinant
references. 

Of course, this non-aufbau single-reference CC theory does not constitute a valid black-box method because 
it cannot be applied to a degenerate multireference. Damour {\it et al.}\cite{Damour2024}\ adopted the two-determinant coupled-cluster singles and doubles (TD-CCSD) method
of Balkov\'{a} and coworkers\cite{Balkova1991JCP,Balkova1992CPL,Balkova1993JCP,Balkova1994JCP,Szalay1994JCP,Lutz2018JCP} as a degenerate counterpart applicable only to
a limited class of doubly degenerate multireferences.  Clearly, this hybrid ansatz also falls short of being a unified, systematic, and general hierarchical theory.

The objective of this study is to introduce just such CC theory --- degenerate coupled-cluster ($\Delta$CC) theory --- that is systematically converging, size-extensive, black-box, {\it ab initio}, and generally applicable to any degenerate and nondegenerate determinant references with any numbers of $\alpha$- and $\beta$-spin electrons, any (aufbau or non-aufbau) patterns of orbital occupancies, 
any spin multiplicities, and any spatial symmetries. It adds to the arsenal of predictive electron-correlation theories based on the CC ansatz, rivaling or even exceeding  
EOM-CC theory in terms of accuracy and general applicability. 

For a nondegenerate reference, this ansatz is identical to the single-reference CC ansatz, which demands that the exponential wave function satisfy the Schr\"{o}dinger 
equation in a certain determinant space. 

For a degenerate multireference, essentially the same projection ansatz is augmented with the provision to determine a linear combination of degenerate 
determinants whose exponential wave functions satisfy the Schr\"{o}dinger equation in prescribed determinant spaces.
It is based on the same logic of derivation as degenerate Rayleigh--Schr\"{o}dinger perturbation theory (RSPT)\cite{Brandow,Sandars1969,Johnson1971,Kuo1971,Lindgren1974,hirschfelder} with the M{\o}ller--Plesset partitioning of the Hamiltonian ($\Delta$MP theory),\cite{Hirata2017,HirataJCP2021} and is the natural 
CC extension of the latter; $\Delta$CC theory is for $\Delta$MP theory 
as CC theory is for MBPT.\cite{BartlettARPC} 

In this article, we present and justify the ansatz of $\Delta$CC theory and derive the working equations of $\Delta$CC with singles ($\Delta$CCS) and with singles and doubles ($\Delta$CCSD). 
We illustrate their diagrammatic representations, placing emphasis on their topological features signifying the size-extensivity of $\Delta$CCSD.
We introduce an analytic (i.e., nondiagrammatic) proof of the size-extensivity of $\Delta$CC theory of all orders, 
which encompasses that of $\Delta$MP theory (surprisingly, the size-extensivity of degenerate RSPT seems to be still an open question today).
On this basis, we discuss the diagrammatic criteria of size-extensivity for a general, degenerate theory, which $\Delta$CCSD seems to satisfy.
As compared to the cases of single-reference theories, which are articulated by the linked-diagram theorems, the diagrammatic criteria are less definitive for degenerate theories. 

We delineate $\Delta$CC theory's similarities to and differences from related 
theories such as various multireference coupled-cluster (MRCC) theories\cite{Jeziorski1981,Meissner1989,Meissner1990,Kucharski1991,PaldusLi2003,LiPaldus2003,LiPaldus2003_2} including two-determinant coupled-cluster (TD-CC) theory,\cite{Balkova1991JCP,Balkova1992CPL,Balkova1993JCP,Balkova1994JCP,Szalay1994JCP,Lutz2018JCP}  EOM-CC theory, and $\Delta$MP theory,
whose ``backward'' or ``folded'' diagrams\cite{Brandow,Sandars1969,Johnson1971,Kuo1971,Lindgren1974,hirschfelder,Kucharski1989} also manifest in $\Delta$CC theory.

As a byproduct of these comparisons, we have conceived a new ansatz of MRCC theory for 
a general model space (GMS) including a degenerate multireference, which is directly related to, but still distinct from the $\Delta$CC ansatz. 
Just like $\Delta$CC theory, it is convergent toward FCI, but unlike $\Delta$CC theory, 
it is 
a non-black-box method. Whether it is size-extensive or not varies depending on the nature of the model space adopted.
While $\Delta$CC theory is not limited to an exact degenerate multireference, the new MRCC theory 
is expected to be better suited to quasidegenerate or strong-correlation problems. We call this theory
quasidegenerate coupled-cluster (QCC) theory.

$\Delta$CC theory can be applied to ionized or electron-attached determinant references of Koopmans and satellite (non-Koopmans) types with any number of electrons removed or added. It offers 
an alternative viewpoint to a coupled-cluster Green's function to the one represented by IP- and EA-EOM-CC theories.\cite{Gunnarsson1978,nooijen_gf1,nooijen_gf2,Meissner1993,Kowalski1,Kowalski2,Kowalski3} While the Feynman--Dyson perturbation expansion of many-body Green's function (MBGF)
is fundamentally nonconvergent at the exact (FCI) limits except for some Koopmans roots,\cite{Hirata_PRA2024} 
$\Delta$MP theory is basically convergent\cite{deltamp,Hirata2017,Hirata_PRA2024} and $\Delta$CC theory is always convergent. The proof of the latter --- the {\it full} $\Delta$CC method
is equivalent to FCI --- is documented in this article.

As per the Thouless theorem,\cite{Thouless1960} a CCS wave function of a single Slater determinant is another single Slater determinant. Therefore, the variational CCS method is identified as
the conventional, variational Hartree--Fock (HF) theory starting from any arbitrary nondegenerate reference. Here, we argue that the projection CCS method, in contrast, 
defines a {\it projection} HF theory for any nondegenerate reference, whose wave function
is still a single deteminant, but with different orbitals from the variational ones. Then, the $\Delta$CCS method can be viewed as its generalization to degenerate multireferences. 
A few pieces of numerical evidence indicate the utility of the $\Delta$CCS method for core ionizations, high-spin states, and possibly electron affinities.  

The $\Delta$CC methods have been implemented into two independent algorithms. 
One is the order-by-order implementations of the matrix-algebraic energy and $T$-amplitude equations at the $\Delta$CCS and $\Delta$CCSD levels, which have the optimal
$O(n^4)$ and $O(n^6)$
dependence, respectively, of the operation cost on the number of spin-orbitals ($n$). These equations have been derived and the corresponding 
codes synthesized by quantum chemists' original  artificial intelligence\cite{Hirata06:TCAReview} called {the tensor contraction engine} or {\sc tce}.\cite{tce,tce2} 
The other is the determinant- or string-based algorithm,\cite{hirata_cc} allowing general-order calculations
through the full $\Delta$CC level at an FCI computational cost at each order. The QCC ansatz has also been implemented into a determinant-based, general-order algorithm.
Their results are compared with those of the 
general-order, determinant-based implementations of the CI,\cite{knowles_fci} EOM-CC,\cite{hirata_eomcc,hirata_ipeomcc} $\Delta$MP,\cite{knowles_mp,deltamp,Hirata2017} and MBGF methods\cite{Hirata2017} 
carried out through the FCI limits or up to the nineteenth order. 

The comparison indicates the following general order of performance for transition energies:\ QCC $\approx$ $\Delta$CC $>$ EOM-CC $>$ CI at the same order or QCC $\approx$ $\Delta$CC $>$ $\Delta$MP $>$ MBGF
at the same cost scaling. In particular, while EOM-CCSD has an average error of {\it ca.}\ 0.1 eV for one-electron transitions, an error of over 1 eV for two-electron transitions, and no roots for three-electron transitions,
$\Delta$CCSD achieves accuracy of 0.1 eV across the board at the same computational cost. 

\section{Theories}

\subsection{The single-reference CC ansatz (review)\label{sec:CC}}

Let us briefly review the ansatz of the single-reference CC theory,\cite{bartlettRMP,shavitt,BartlettMP2010,BartlettWIRE2012}  arguably the most successful
electron-correlation theory developed to date.

It demands that an exponential wave function $e^{\hat{T}_I}|I\rangle$ satisfy the 
Schr\"{o}dinger equation within the space of Slater determinants reachable by $(1+\hat{T}_I)|I\rangle$,
\begin{eqnarray}
\hat{P}_I\hat{H}_I e^{\hat{T}_I} |I\rangle = \hat{P}_I e^{\hat{T}_I} | I \rangle \tilde{E}_{I} , \label{CC_ansatz}
\end{eqnarray}
where $|I\rangle$ is the $I$th determinant, $\hat{T}_I$ is an excitation operator from $|I\rangle$, $\hat{P}_I$ is the projector onto the aforementioned space of determinants, and $\tilde{E}_{I}$ is the CC energy. The number of unknowns ($\tilde{E}_I$ plus all $T$-amplitudes) is equal to the number of equations. 

Equation (\ref{CC_ansatz})
is not expressed in the usual, connected form; it is written in the disconnected, unlinked form. The former is obtained by
first noting\cite{Harris1977} that $e^{-\hat{T}_I^\dagger}(1+\hat{T}_I)|I\rangle$ spans the same determinant space as $(1+\hat{T}_I)|I\rangle$. Therefore, left-multiplying $e^{-\hat{T}_I}$ with 
Eq.\ (\ref{CC_ansatz}) prior to projection does not alter the ansatz, transforming it into
\begin{eqnarray}
\hat{P}_Ie^{-\hat{T}_I}  \hat{H}_I e^{\hat{T}_I} |I\rangle = \hat{P}_I | I \rangle \tilde{E}_{I}, \label{CC_ansatz2}
\end{eqnarray}
whereupon the Baker--Campbell--Hausdorff formula highlights the connectedness (indicated by subscript ``C'') of the CC effective Hamiltonian,
\begin{eqnarray}
e^{-\hat{T}_I}  \hat{H}_I e^{\hat{T}_I} &=& \hat{H}_I + \Big[ \hat{H}_I, \hat{T}_I \Big] + \frac{1}{2!} \Big[ \Big[ \hat{H}_I, \hat{T}_I \Big], \hat{T}_I \Big] + \dots \nonumber\\ \label{BCH}
&=& \Big( \hat{H}_I e^{\hat{T}_I} \Big)_\text{C}.
\end{eqnarray}

The exact ({\it ab initio}) electronic Hamiltonian, $\hat{H}_I$, is independent of the reference $|I\rangle$, but for later convenience, it is brought to
a normal-ordered form (indicated by ``$\{ \dots \}$'') relative to the Fermi vacuum of $|I\rangle$,
\begin{eqnarray}
\hat{H}_I= E_I^{\text{HF}} + \sum_{p,q} (f_I)^p_q \left\{ \hat{p}^\dagger \hat{q} \right\} + \frac{1}{4} \sum_{p,q,r,s} (v)^{pq}_{rs} \left\{ \hat{p}^\dagger \hat{q}^\dagger \hat{s}\hat{r} \right\} ,
\end{eqnarray}
where the HF energy, $E_I^{\text{HF}}$, and Fock matrix elements, $({f}_I)^p_q$, are defined for $|I\rangle$ as
\begin{eqnarray}
E_I^{\text{HF}} &=& E_{\text{nuc.}} + \sum_{i}^{\text{occ.\ in }|I\rangle} (h)^i_i + \frac{1}{2} \sum_{i,j}^{\text{occ.\ in }|I\rangle} (v)^{ij}_{ij}, \label{EHF} \\
(f_I)^p_q &=&  (h)^p_q + \sum_{i}^{\text{occ.\ in }|I\rangle} (v)^{pi}_{qi}. \label{Fock}
\end{eqnarray}
The summation indices, $i$ and $j$, run over spin-orbitals occupied by an electron in $|I\rangle$, and therefore, $E_I^{\text{HF}}$ and $({f}_I)^p_q$ are now dependent on $|I\rangle$.
On the other hand, the nuclear-repulsion energy, $E_{\text{nuc.}}$, core Hamiltonian matrix elements,  $({h})^p_q$, and anti-symmetrized two-electron integrals, $(v)^{pq}_{rs}$, are independent of $|I\rangle$.

Single-reference CC theory is a black-box method, requiring only the nuclear-repulsion energy, reference determinant, and truncation rank of $\hat{T}_I$ to be specified by the user.
Its energy and wave function converge at the exact, basis-set, nonrelativistic (FCI) results as $\hat{T}_I$ becomes complete. It is size-extensive.

\subsection{The $\Delta$CC ansatz\label{sec:DCC}}

$\Delta$CC theory is identical with the foregoing single-reference CC theory when the reference is nondegenerate, 
regardless of the numbers of the $\alpha$- and $\beta$-spin electrons, the occupancies of spin-orbitals in the 
reference Slater determinant, or the definitions of the spin-orbitals.

When the reference is $M$-fold degenerate (as per, e.g., the M{\o}ller--Plesset zeroth-order Hamiltonian), 
the ansatz needs to be generalized, while encompassing 
the foregoing nondegenerate case without any modification. Let us first present the ansatz and then give justifications.

$\Delta$CC theory stipulates that the set of $M$ exponential wave functions in the degenerate subspace, $\{e^{\hat{T}_I}|I\rangle\}$,
together satisfy the Schr\"{o}dinger equation with a matrix form of the energy 
in the respective spaces of determinants reachable by $\{(1+\hat{T}_I)|I\rangle\}$,
\begin{eqnarray}
\hat{P}_I\hat{H}_I e^{\hat{T}_I} |I\rangle = \hat{P}_I \sum_{J=1}^{M} \hat{P}_J e^{\hat{T}_J} | J \rangle E_{JI} \label{DCC_ansatz}
\end{eqnarray}
as well as the biorthogonality and intermediate normalization of the $M$ exponential wave functions,
\begin{eqnarray}
\langle J | \hat{P}_I e^{\hat{T}_I} | I \rangle = \delta_{JI}, \label{DCC_ansatz2}
\end{eqnarray}
where $|I\rangle$ and $|J\rangle$ are degenerate references and $\delta_{JI}$ is Kronecker's delta. Equation (\ref{DCC_ansatz2}) is identified as ``the C condition'' of Li and Paldus's MRCC theory.\cite{LiPaldus2003,LiPaldus2003_2,PaldusLi2003,Paldus2004,LiPaldusIJQC2004}
The projector $\hat{P}_I$ spans determinants reachable by $(1+\hat{T}_I)|I\rangle$ and they include determinants 
that are degenerate with $|I\rangle$ (internal space) as well as those outside the degenerate subspace (external space). 
It is stressed that each ($I$th) degenerate reference has its own $\hat{T}_I$ and $\hat{P}_I$ operators.
(In this article,  ``internal space,'' ``model space,'' ``subspace,'' and ``multireference'' are used interchangeably.)

Defining the $M$-by-$M$ Hamiltonian and overlap matrices, $\bm{H}$ and $\bm{S}$, by
\begin{eqnarray}
H_{JI} &=& \langle J | \hat{P}_I \hat{H} _I e^{\hat{T}_I} | I \rangle, \label{H} \\
S_{JI} &=& \langle J |  \hat{P}_I e^{\hat{T}_I} | I \rangle, \label{S}
\end{eqnarray}
where $|I\rangle$ and $|J\rangle$ are degenerate references, 
the projection of Eq.\ (\ref{DCC_ansatz}) onto the internal space becomes
\begin{eqnarray}
\bm{H} = \bm{S}\bm{E}, \label{HisSE}
\end{eqnarray}
while Eq.\ (\ref{DCC_ansatz2}) is written as
\begin{eqnarray}
\bm{S} = \bm{1}. \label{C-condition1}
\end{eqnarray}
The $T$-amplitudes in the external space are determined by solving Eq.\ (\ref{DCC_ansatz}). 
Those in the internal space are obtained by satisfying Eq.\ (\ref{C-condition1}), whereupon $\bm{E}=\bm{H}$. 
The $I$th-state $\Delta$CC energy $\tilde{E}_I$ is then the $I$th eigenvalue of this non-Hermitian energy matrix $\bm{E}$,
\begin{eqnarray}
\sum_{J=1}^M E_{KJ} C_{JI} =  C_{KI}\tilde{E}_{I} . \label{diagonalization}
\end{eqnarray}
The corresponding $\Delta$CC wave function is given by 
\begin{eqnarray}
|\tilde{I}\rangle = \sum_{J=1}^M \hat{P}_J e^{\hat{T}_J} | J\rangle C_{JI}. \label{wavefunction}
\end{eqnarray}
(While it is customary to define a multireference theory based on the Bloch equation,\cite{Lindgren,shavitt} for an exactly degenerate multireference, we can adopt a more straightforward and transparent
approach using the Schr\"{o}dinger equation instead;\cite{Griffiths,hirschfelder} they are equivalent to each other.) 

We shall call this $\Delta$CC theory, where $\Delta$ standing for both a {\it difference} between two state energies taken to define the transition energy and {\it degeneracy}
of the reference determinants. 

Clearly, the $\Delta$CC ansatz reduces to the single-reference CC ansatz upon letting $M=1$.
$\Delta$CC theory is a black-box method, requiring only the nuclear-repulsion energy, degenerate reference determinants, and truncation rank of the $\hat{T}_I$ operators. 
The degenerate reference determinants are defined with any (single) set of spin-orbitals that are, typically, but not necessarily,
the HF orbitals determined for the neutral ground state, called {\it the orbital reference}, which can be distinct from
any of the degenerate references. 
The degenerate references need not have the same numbers of $\alpha$- and $\beta$-spin electrons
or the same spin-multiplicity as the orbital reference. They need not be the lowest-lying states, either. There is no restrictions about their spatial symmetries (such as
having to transform as a different irreducible representation from the orbital reference). $\Delta$CC theory is convergent at the exact (FCI) results
for all degenerate reference states (whose degeneracy may persist or be partially or fully lifted upon correlation) as the $\hat{T}_I$ operators become complete (see Sec.\ \ref{sec:reverse} for a proof). 
It is size-extensive (see Appendix for a proof).

Let us justify this ansatz. Since any linear combination of degenerate eigenfunctions is also another degenerate eigenfunction,
a sound ansatz should not attach particular significance to a single determinant as a reference when it is degenerate. Instead,
a reference should evolve into a linear combination of all degenerate determinants as more electron correlation is included.
In other words, we do not know {\it a priori} the ``exact'' degenerate references, i.e., which linear combinations of degenerate determinants eventually map onto 
the full $\Delta$CC wave functions (see Sec.\ \ref{sec:reverse}). We determine them via Eq.\ (\ref{diagonalization}), which 
would then approach the exact degenerate references as the $\hat{T}_I$ operators become more complete. This is also consistent with 
$\Delta$MP theory (to which $\Delta$CC theory is intimately related; see Sec.\ \ref{sec:HCPT}), which updates the degenerate references at each perturbation order via diagonalization
of ``the effective Hamiltonian''\cite{Kucharski1989}
akin to the above $\bm{E}$. The diagonalization ensures that the wave functions of Eq.\ (\ref{wavefunction}) be biorthogonal to one another in the internal space.

Each degenerate determinant is entitled to its own set of $T$-amplitudes. 
This is a natural generalization of the single-reference CC ansatz to degenerate references, where the only essential difference is the need for 
determining the linear combinations of degenerate references in the latter case. In terms of the physics of electron correlation also, it does not make much sense for two  
reference determinants, even if they are degenerate, to share the same set of excitation amplitudes. Shared amplitudes would also go counter to the $\Delta$MP ansatz. 
These justify Eqs.\ (\ref{DCC_ansatz}) and (\ref{wavefunction}) in which
$|J\rangle$ is acted on by its own, independent $e^{\hat{T}_J}$ operator.

Equation (\ref{C-condition1}) ensures that there be no conflict within the internal space, i.e., that $e^{\hat{T}_J}|J\rangle$ does not overlap with any other degenerate 
reference determinant $|I\rangle$, when $I \neq J$.
If $|I\rangle$ {\it were} spanned by $e^{\hat{T}_J}|J\rangle$, the exact $\Delta$CC wave function cannot be written uniquely in the form of Eq.\ (\ref{wavefunction}), and
its convergence toward FCI with increasing the rank of $\hat{T}$ is no longer assured. 

Until the convergence of iterative algorithms to satisfy Eqs.\ (\ref{DCC_ansatz})
and (\ref{DCC_ansatz2}), $\bm{S} \neq \bm{1}$ and the energy matrix that enters Eq.\ (\ref{DCC_ansatz}) needs to be evaluated by $\bm{E} = \bm{S}^{-1}\bm{H}$ (not by $\bm{E}=\bm{H}$). 

Perhaps the thorniest issue of the $\Delta$CC ansatz is whether or not to include the second projector $\hat{P}_J$ in  Eq.\ (\ref{DCC_ansatz}). 
It has been concluded that limiting the scope of the action of $e^{\hat{T}_J}$ by $\hat{P}_J$ in Eq.\ (\ref{DCC_ansatz}) is consistent with the same in Eq.\ (\ref{DCC_ansatz2})
and is deemed more sensible. This choice furthermore allows the use of the same, uniform formula and computer code to evaluate $\hat{P}_J e^{\hat{T}_J} | J\rangle$
at a given order of $\hat{T}_J$, explicitly maintaining the black-box nature of the method. 
[In Sec.\ \ref{sec:newSUMRCC}, we shall consider yet another, non-black-box ansatz that does away with 
both the second projector $\hat{P}_J$ in Eq.\ (\ref{DCC_ansatz}) and the projector $\hat{P}_I$ in Eq.\ (\ref{DCC_ansatz2}).]
 


\subsection{The $\Delta$CCS method (algebraic)}\label{sec:DCCS}

As the first concrete example of $\Delta$CC theory, let us consider the degenerate coupled-cluster singles ($\Delta$CCS) method.
Since degenerate references are often singly excited from one another and $\Delta$CC theory must take into account coupling among these degenerate references as well as
electron correlation in each reference, the lowest meaningful instance should be $\Delta$CCS rather than the degenerate coupled-cluster doubles method.\cite{Harris1977} 

In $\Delta$CCS, $\hat{P}_I$ is the projector onto the $I$th degenerate reference determinant, $|I\rangle$, plus the set of all singly excited determinants from $|I\rangle$, i.e., $\{ |I^{p_2}_{h_1} \rangle\}$.
The left-hand side of Eq.\ (\ref{DCC_ansatz}) projected onto the singles is the sum of the connected (first) and unlinked (second) terms,
\begin{eqnarray}
\langle I^{p_2}_{h_1} | \hat{H}_I e^{\hat{T}_I} | I\rangle &=&  (\Lambda_I) ^{ p_{2} } _{ h_{1} } + (t_I)^{p_2}_{h_1} H_{II}, \label{DCCS1}
\end{eqnarray}
where $(\Lambda_I)^{ p_{2} } _{ h_{1} }$ defines the connected $T_1$-amplitude equation of single-reference CCS for the $|I\rangle$ reference. It reads
\begin{eqnarray}
(\Lambda_I) ^{ p_{2} } _{ h_{1} } &=&  (f_I) ^{ p_{2} } _{ h_{1} } -  (f_I) ^{ h_{3} } _{ h_{1} }(t_I)^{ p_{2} } _{ h_{3} } +  (f_I) ^{ p_{2} } _{ p_{3} }(t_I)^{ p_{3} } _{ h_{1} } \nonumber\\
&& - (t_I)^{ p_{3} } _{ h_{4} }(v)^{ h_{4} p_{2} } _{ h_{1} p_{3} }- (t_I)^{ p_{3} } _{ h_{1} }(t_I)^{ p_{2} } _{ h_{4} } (f_I) ^{ h_{4} } _{ p_{3} } \nonumber\\
&& + (t_I)^{ p_{2} } _{ h_{3} }(t_I)^{ p_{4} } _{ h_{5} }(v)^{ h_{5} h_{3} } _{ h_{1} p_{4} }+ (t_I)^{ p_{3} } _{ h_{1} }(t_I)^{ p_{4} } _{ h_{5} }(v)^{ h_{5} p_{2} } _{ p_{4} p_{3} } \nonumber\\
&&- (t_I)^{ p_{2} } _{ h_{3} }(t_I)^{ p_{4} } _{ h_{1} }(t_I)^{ p_{5} } _{ h_{6} }(v)^{ h_{6} h_{3} } _{ p_{5} p_{4} }.  \label{DCCS1_2}
\end{eqnarray}
(This and some of the following equations were derived and typeset by {\sc tce}.\cite{tce,tce,tce2})
Henceforth, Einstein's convention of implied summations over repeated indices is employed, and $p$ and $h$ as in $(t_I)^p_h$ stand for a virtual and occupied spin-orbital, respectively, in $|I\rangle$. The last factor in Eq.\ (\ref{DCCS1}) is the diagonal element of the Hamiltonian matrix $\bm{H}$ of Eq.\ (\ref{H}) and is given by
\begin{eqnarray}
H_{II} &=& \langle I | \hat{H}_I e^{\hat{T}_I} | I\rangle \nonumber\\
&=&
  E_I^{\text{HF}} + (f_I) ^{ h_{1} } _{ p_{2} } (t_I) ^{ p_{2} } _{ h_{1} }  + \frac{1}{2} (t_I) ^{ p_{1} } _{ h_{2} } (t_I) ^{ p_{3} } _{ h_{4} } (v) ^{ h_{4} h_{2} } _{ p_{3} p_{1} },  \label{DCCS2} 
\end{eqnarray}
which is connected. It should be cautioned that $E_I^{\text{HF}}$ and $f_I$ are dependent on $I$. 

When $|I^{p_2}_{h_1}\rangle$ is in the internal space, i.e., it is degenerate with $|I\rangle$, Eq.\ (\ref{DCCS1}) defines the off-diagonal elements of 
$\bm{H}$; for $|J\rangle$ that differs from $|I\rangle$ by more than one spin-orbital, $H_{JI}=0$. 

When $|I^{p_2}_{h_1}\rangle$ is in the external space, Eq.\ (\ref{DCCS1}) becomes the left-hand side of the $T_1$-amplitude equation of $\Delta$CCS  [Eq.\ (\ref{DCC_ansatz})], 
which is unlinked through the last term. 

The left-hand side of Eq.\ (\ref{DCC_ansatz2}) is expanded as 
\begin{eqnarray}
\langle I^{p_2}_{h_1} | e^{\hat{T}_I} | I\rangle &=&  (t_I) ^{ p_{2} } _{ h_{1} }, \label{DCCS3}\\
\langle I | e^{\hat{T}_I} | I\rangle &=&  1.\label{DCCS4}
\end{eqnarray}
When $|I^{p_2}_{h_1}\rangle$ is in the internal space, Eqs.\ (\ref{DCCS3}) and (\ref{DCCS4}) together define the overlap matrix $\bm{S}$ of Eq.\ (\ref{S}). When $|J\rangle$ is more than one-spin-orbital different from $|I\rangle$, $S_{JI}=0$. These equations also define factors appearing in the right-hand side of the $T_1$-amplitude equation [Eq.\ (\ref{DCC_ansatz})].
The C condition is written as
\begin{eqnarray}
 (t_I) ^{ p_{2} } _{ h_{1} } &=& 0. \label{DCCS_C}
\end{eqnarray}

The right-hand side of the $T_1$-amplitude equation in the external space [Eq.\ (\ref{DCC_ansatz})] then becomes
\begin{eqnarray}
&&\sum_{J=1}^M \langle I^{p_2}_{h_1} | \hat{P}_J e^{\hat{T}_J} | J\rangle E_{JI} = \langle I^{p_2}_{h_1} |  e^{\hat{T}_I} | I \rangle E_{II} 
\nonumber\\&& 
+ \sum_{J\neq I} \sum_{h_3}^{\text{occ. in }|J\rangle}\sum_{p_4}^{\text{vir. in }|J\rangle} \langle I^{p_2}_{h_1} | J^{p_4}_{h_3} \rangle \langle J^{p_4}_{h_3} |  e^{\hat{T}_J} | J\rangle E_{JI}, 
\label{DCCS5}
\end{eqnarray}
where the parity $\langle I^{p_2}_{h_1} | J^{p_4}_{h_3} \rangle$ takes the value of $0$ or $\pm1$. 
All the other Dirac brackets, $\langle \dots \rangle$, have already been defined in terms of the $T$-amplitudes and molecular integrals. 



\subsection{The $\Delta$CCSD method (algebraic)} \label{sec:DCCSD}

Next, let us consider the degenerate coupled-cluster singles and doubles ($\Delta$CCSD) method. Its projector $\hat{P}_I$ spans the zero, singly, and doubly excited determinants from $|I\rangle$, leading to an energy equation and 
$T_1$- and $T_2$-amplitude equations. 

The projection onto the doubles is the sum of the connected (first), disconnected (second), and unlinked (third) terms,
\begin{eqnarray}
\langle I^{p_3p_4}_{h_1h_2} | \hat{H}_I e^{\hat{T}_I} | I\rangle&=& (\Lambda_I)^{ p_{3} p_{4} } _{ h_{1} h_{2} }
 + \hat{A}\left[(t_I)^{p_3}_{h_1}  (\Lambda_I)^{p_4}_{h_2} \right] \nonumber\\
&& + \left( (t_I) ^{ p_{3} p_{4} } _{ h_{1} h_{2} } + \hat{A}\left[ (t_I) ^{ p_{3} } _{ h_{1} } (t_I) ^{ p_{4} } _{ h_{2} }\right]  \right) H_{II}, \label{DCCSD1}
\end{eqnarray}
where $(\Lambda_I)^{ p_{3} p_{4} } _{ h_{1} h_{2} }$ denotes the connected $T_2$-amplitude equation of single-reference CCSD for the $|I\rangle$ reference. It is given by
\begin{widetext}
\begin{eqnarray}
(\Lambda_I)^{p_3p_4}_{h_1h_2} &=& (v)^{ p_{3} p_{4} } _{ h_{1} h_{2} }
- \hat{A}\left[(t_I)^{ p_{3} } _{ h_{5} }(v)^{ h_{5} p_{4} } _{ h_{1} h_{2} }\right]
+ \hat{A}\left[(t_I)^{ p_{5} } _{ h_{2} }(v)^{ p_{3} p_{4} } _{ h_{1} p_{5} }\right]
- \hat{A}\left[(f_I)^{ h_{5} } _{ h_{1} }(t_I)^{ p_{3} p_{4} } _{ h_{5} h_{2} }\right]
- \hat{A}\left[(f_I)^{ p_{4} } _{ p_{5} }(t_I)^{ p_{5} p_{3} } _{ h_{1} h_{2} }\right]
+ \frac{1}{2}(t_I)^{ p_{3} p_{4} } _{ h_{5} h_{6} }(v)^{ h_{5} h_{6} } _{ h_{1} h_{2} }
\nonumber\\&&
+ \hat{A}\left[(t_I)^{ p_{5} p_{3} } _{ h_{6} h_{2} }(v)^{ h_{6} p_{4} } _{ h_{1} p_{5} }\right]
+ \frac{1}{2}(t_I)^{ p_{5} p_{6} } _{ h_{1} h_{2} }(v)^{ p_{3} p_{4} } _{ p_{5} p_{6} }
+ (t_I)^{ p_{3} } _{ h_{5} }(t_I)^{ p_{4} } _{ h_{6} }(v)^{ h_{5} h_{6} } _{ h_{1} h_{2} }
- \hat{A}\left[(t_I)^{ p_{5} } _{ h_{2} }(t_I)^{ p_{3} } _{ h_{6} }(v)^{ h_{6} p_{4} } _{ h_{1} p_{5} }\right]
+ (t_I)^{ p_{5} } _{ h_{1} }(t_I)^{ p_{6} } _{ h_{2} }(v)^{ p_{3} p_{4} } _{ p_{5} p_{6} }
\nonumber\\&&
- \hat{A}\left[(f_I)^{ h_{5} } _{ p_{6} }(t_I)^{ p_{3} p_{4} } _{ h_{5} h_{2} }(t_I)^{ p_{6} } _{ h_{1} }\right]
+ \hat{A}\left[(f_I)^{ h_{5} } _{ p_{6} }(t_I)^{ p_{6} p_{3} } _{ h_{1} h_{2} }(t_I)^{ p_{4} } _{ h_{5} }\right]
+ \frac{1}{2}\hat{A}\left[(t_I)^{ p_{3} p_{4} } _{ h_{5} h_{6} }(t_I)^{ p_{7} } _{ h_{2} }(v)^{ h_{5} h_{6} } _{ h_{1} p_{7} }\right]
- \hat{A}\left[(t_I)^{ p_{5} p_{3} } _{ h_{6} h_{2} }(t_I)^{ p_{4} } _{ h_{7} }(v)^{ h_{6} h_{7} } _{ h_{1} p_{5} }\right]
\nonumber\\&&
- \hat{A}\left[(t_I)^{ p_{3} p_{4} } _{ h_{5} h_{2} }(t_I)^{ p_{6} } _{ h_{7} }(v)^{ h_{5} h_{7} } _{ h_{1} p_{6} }\right]
- \hat{A}\left[(t_I)^{ p_{5} p_{3} } _{ h_{6} h_{2} }(t_I)^{ p_{7} } _{ h_{1} }(v)^{ h_{6} p_{4} } _{ p_{5} p_{7} }\right]
- \frac{1}{2}\hat{A}\left[(t_I)^{ p_{5} p_{6} } _{ h_{1} h_{2} }(t_I)^{ p_{3} } _{ h_{7} }(v)^{ h_{7} p_{4} } _{ p_{5} p_{6} }\right]
+ \hat{A}\left[(t_I)^{ p_{5} p_{3} } _{ h_{1} h_{2} }(t_I)^{ p_{6} } _{ h_{7} }(v)^{ h_{7} p_{4} } _{ p_{5} p_{6} }\right]
\nonumber\\&&
+ \frac{1}{2}\hat{A}\left[(t_I)^{ p_{5} p_{4} } _{ h_{1} h_{2} }(t_I)^{ p_{6} p_{3} } _{ h_{7} h_{8} }(v)^{ h_{7} h_{8} } _{ p_{5} p_{6} }\right]
+ \frac{1}{4}(t_I)^{ p_{5} p_{6} } _{ h_{1} h_{2} }(t_I)^{ p_{3} p_{4} } _{ h_{7} h_{8} }(v)^{ h_{7} h_{8} } _{ p_{5} p_{6} }
- \frac{1}{2}\hat{A}\left[(t_I)^{ p_{3} p_{4} } _{ h_{5} h_{1} }(t_I)^{ p_{6} p_{7} } _{ h_{8} h_{2} }(v)^{ h_{5} h_{8} } _{ p_{6} p_{7} }\right]
- \hat{A}\left[(t_I)^{ p_{5} p_{4} } _{ h_{6} h_{1} }(t_I)^{ p_{7} p_{3} } _{ h_{8} h_{2} }(v)^{ h_{6} h_{8} } _{ p_{5} p_{7} }\right]
\nonumber\\&&
+ \hat{A}\left[(t_I)^{ p_{5} } _{ h_{2} }(t_I)^{ p_{3} } _{ h_{6} }(t_I)^{ p_{4} } _{ h_{7} }(v)^{ h_{6} h_{7} } _{ h_{1} p_{5} }\right]
- \hat{A}\left[(t_I)^{ p_{5} } _{ h_{1} }(t_I)^{ p_{6} } _{ h_{2} }(t_I)^{ p_{3} } _{ h_{7} }(v)^{ h_{7} p_{4} } _{ p_{5} p_{6} }\right]
+ \frac{1}{2}(t_I)^{ p_{5} } _{ h_{1} }(t_I)^{ p_{6} } _{ h_{2} }(t_I)^{ p_{3} p_{4} } _{ h_{7} h_{8} }(v)^{ h_{7} h_{8} } _{ p_{5} p_{6} }
\nonumber\\&&
+ \hat{A}\left[(t_I)^{ p_{5} } _{ h_{1} }(t_I)^{ p_{4} } _{ h_{6} }(t_I)^{ p_{7} p_{3} } _{ h_{8} h_{2} }(v)^{ h_{6} h_{8} } _{ p_{5} p_{7} }\right]
+ \hat{A}\left[(t_I)^{ p_{5} } _{ h_{1} }(t_I)^{ p_{6} } _{ h_{7} }(t_I)^{ p_{3} p_{4} } _{ h_{8} h_{2} }(v)^{ h_{7} h_{8} } _{ p_{5} p_{6} }\right]
+ \frac{1}{2}(t_I)^{ p_{3} } _{ h_{5} }(t_I)^{ p_{4} } _{ h_{6} }(t_I)^{ p_{7} p_{8} } _{ h_{1} h_{2} }(v)^{ h_{5} h_{6} } _{ p_{7} p_{8} }
\nonumber\\&&
- \hat{A}\left[(t_I)^{ p_{4} } _{ h_{5} }(t_I)^{ p_{6} } _{ h_{7} }(t_I)^{ p_{8} p_{3} } _{ h_{1} h_{2} }(v)^{ h_{5} h_{7} } _{ p_{6} p_{8} }\right]
+ (t_I)^{ p_{5} } _{ h_{1} }(t_I)^{ p_{6} } _{ h_{2} }(t_I)^{ p_{3} } _{ h_{7} }(t_I)^{ p_{4} } _{ h_{8} }(v)^{ h_{7} h_{8} } _{ p_{5} p_{6} } , 
\label{DCCSD1_2}
\end{eqnarray}
where $\hat{A}[\dots]$ is the anti-symmetrizer. The projection onto the singles leads to
\begin{eqnarray}
\langle I^{p_2}_{h_1} | \hat{H}_I e^{\hat{T}_I} | I\rangle &=& (\Lambda_I)^{ p_{2} } _{ h_{1} }
+ (t_I)^{p_2}_{h_1} H_{II}, \label{DCCSD2}
\end{eqnarray}
where $(\Lambda_I)^{p_2}_{h_1}$ [also appearing in Eq.\ (\ref{DCCSD1}) and distinct from Eq.\ (\ref{DCCS1_2})] is the connected $T_1$-amplitude equation of single-reference CCSD, which reads
\begin{eqnarray}
(\Lambda_I)^{p_2}_{h_1} &=& (f_I)^{ p_{2} } _{ h_{1} }
- (f_I)^{ h_{3} } _{ h_{1} }(t_I)^{ p_{2} } _{ h_{3} }
+ (f_I)^{ p_{2} } _{ p_{3} }(t_I)^{ p_{3} } _{ h_{1} }
- (t_I)^{ p_{3} } _{ h_{4} }(v)^{ h_{4} p_{2} } _{ h_{1} p_{3} }
+ (f_I)^{ h_{3} } _{ p_{4} }(t_I)^{ p_{4} p_{2} } _{ h_{3} h_{1} }
+ \frac{1}{2}(t_I)^{ p_{3} p_{2} } _{ h_{4} h_{5} }(v)^{ h_{4} h_{5} } _{ h_{1} p_{3} }
+ \frac{1}{2}(t_I)^{ p_{3} p_{4} } _{ h_{5} h_{1} }(v)^{ h_{5} p_{2} } _{ p_{3} p_{4} }\nonumber\\
&&- (t_I)^{ p_{3} } _{ h_{1} }(t_I)^{ p_{2} } _{ h_{4} }(f_I)^{ h_{4} } _{ p_{3} }
- (t_I)^{ p_{2} } _{ h_{3} }(t_I)^{ p_{4} } _{ h_{5} }(v)^{ h_{3} h_{5} } _{ h_{1} p_{4} }
- (t_I)^{ p_{3} } _{ h_{1} }(t_I)^{ p_{4} } _{ h_{5} }(v)^{ h_{5} p_{2} } _{ p_{3} p_{4} }
- \frac{1}{2}(t_I)^{ p_{3} p_{2} } _{ h_{4} h_{5} }(t_I)^{ p_{6} } _{ h_{1} }(v)^{ h_{4} h_{5} } _{ p_{3} p_{6} }
- \frac{1}{2}(t_I)^{ p_{3} p_{4} } _{ h_{5} h_{1} }(t_I)^{ p_{2} } _{ h_{6} }(v)^{ h_{5} h_{6} } _{ p_{3} p_{4} }\nonumber\\
&&+ (t_I)^{ p_{3} p_{2} } _{ h_{4} h_{1} }(t_I)^{ p_{5} } _{ h_{6} }(v)^{ h_{4} h_{6} } _{ p_{3} p_{5} }
- (t_I)^{ p_{3} } _{ h_{1} }(t_I)^{ p_{2} } _{ h_{4} }(t_I)^{ p_{5} } _{ h_{6} }(v)^{ h_{4} h_{6} } _{ p_{3} p_{5} }. \label{DCCSD2_2}
\end{eqnarray}
\end{widetext}
The $H_{II}$ factor in these equations represents the energy of single-reference CCSD, i.e.,
\begin{eqnarray}
H_{II} &=& \langle I | \hat{H}_I e^{\hat{T}_I} | I\rangle \nonumber\\
&=& E^{\text{HF}}_I +  (f_I) ^{ h_{1} } _{ p_{2} } (t_I) ^{ p_{2} } _{ h_{1} } 
+ \frac{1}{4} (t_I) ^{ p_{1} p_{2} } _{ h_{3} h_{4} } (v) ^{ h_{3} h_{4} } _{ p_{1} p_{2} } \nonumber\\
&& + \frac{1}{2} (t_I) ^{ p_{1} } _{ h_{2} } (t_I) ^{ p_{3} } _{ h_{4} } (v) ^{ h_{2} h_{4} } _{ p_{1} p_{3} } ,\label{DCCSD3} 
\end{eqnarray}
which is connected. 

When $|I^{p_3p_4}_{h_1h_2}\rangle$ and $|I^{p_2}_{h_1}\rangle$ are in the internal space, Eqs.\ (\ref{DCCSD1}) and (\ref{DCCSD2}) define the off-diagonal elements of $\bm{H}$; for $|J\rangle$ that is more than two-spin-orbital different from $|I\rangle$, 
$H_{JI}=0$. Otherwise
they become the left-hand sides of the $T_2$- and $T_1$-amplitude equations, respectively. 

The overlap matrix elements entering the C condition and the right-hand side of the $T_2$- and $T_1$-amplitude equations  are given by
\begin{eqnarray}
\langle I^{p_3p_4}_{h_1h_2} | e^{\hat{T}_I} | I\rangle &=&  (t_I) ^{ p_{3} p_{4} } _{ h_{1} h_{2} } + \hat{A}\left[ (t_I) ^{ p_{3} } _{ h_{1} } (t_I) ^{ p_{4} } _{ h_{2} }\right]   \label{DCCSD4} \\
&=& (t_I) ^{ p_{3} p_{4} } _{ h_{1} h_{2} } + (t_I) ^{ p_{3} } _{ h_{1} } (t_I) ^{ p_{4} } _{ h_{2} } -  (t_I) ^{ p_{3} } _{ h_{2} } (t_I) ^{ p_{4} } _{ h_{1} } , \label{antisymdef}\\
\langle I^{p_2}_{h_1} | e^{\hat{T}_I} | I\rangle &=&  (t_I) ^{ p_{2} } _{ h_{1} }, \label{DCCSD5}\\
\langle I | e^{\hat{T}_I} | I\rangle &=&  1,\label{DCCSD6}
\end{eqnarray}
where the action of $\hat{A}$ is spelled out in Eq.\ (\ref{antisymdef}). 
When $|I^{p_3p_4}_{h_1h_2} \rangle$ and $|I^{p_2}_{h_1}\rangle$ are in the internal space, these define the elements of $\bm{S}$; for $|J\rangle$ that is more than two-spin-orbital different from $|I\rangle$, $S_{JI}=0$. The C condition then becomes
\begin{eqnarray}
(t_I) ^{ p_{3} p_{4} } _{ h_{1} h_{2} } + \hat{A}\left[ (t_I) ^{ p_{3} } _{ h_{1} } (t_I) ^{ p_{4} } _{ h_{2} }\right] &=& 0 , \label{DCCSD_C1}\\
(t_I) ^{ p_{2} } _{ h_{1} } &=& 0. \label{DCCSD_C2}
\end{eqnarray}

The right-hand sides of the $T_2$- and $T_1$-amplitude equations read,
\begin{eqnarray}
&& \sum_{J=1}^M \langle I^{p_3p_4}_{h_1h_2} | \hat{P}_J e^{\hat{T}_J} | J\rangle E_{JI} = 
 \langle I^{p_3p_4}_{h_1h_2} |  e^{\hat{T}_I} | I \rangle E_{II} 
\nonumber\\&& 
+ \sum_{J\neq I} \sum_{h_5}^{\text{occ. in }|J\rangle}\sum_{p_6}^{\text{vir. in }|J\rangle} \langle I^{p_3p_4}_{h_1h_2} | J^{p_6}_{h_5} \rangle \langle J^{p_6}_{h_5} |  e^{\hat{T}_J} | J\rangle E_{JI} 
\nonumber\\&& 
+ \sum_{J\neq I} \sum_{h_5<h_6}^{\text{occ. in }|J\rangle}\sum_{p_7<p_8}^{\text{vir. in }|J\rangle} \langle I^{p_3p_4}_{h_1h_2} | J^{p_7p_8}_{h_5h_6} \rangle \langle J^{p_7p_8}_{h_5h_6} |  e^{\hat{T}_J} | J\rangle E_{JI} 
, \label{DCCSD7} 
\end{eqnarray}
and
\begin{eqnarray}
&&\sum_{J=1}^M \langle I^{p_2}_{h_1} | \hat{P}_J e^{\hat{T}_J} | J\rangle E_{JI} = 
  \langle I^{p_2}_{h_1} |  e^{\hat{T}_I} | I \rangle E_{II} 
\nonumber\\&& 
+ \sum_{J\neq I} \sum_{h_3}^{\text{occ. in }|J\rangle}\sum_{p_4}^{\text{vir. in }|J\rangle} \langle I^{p_2}_{h_1} | J^{p_4}_{h_3} \rangle \langle J^{p_4}_{h_3} |  e^{\hat{T}_J} | J\rangle E_{JI} 
\nonumber\\&& 
+ \sum_{J\neq I} \sum_{h_3<h_4}^{\text{occ. in }|J\rangle}\sum_{p_5<p_6}^{\text{vir. in }|J\rangle} \langle I^{p_2}_{h_1} | J^{p_5p_6}_{h_3h_4} \rangle \langle J^{p_5p_6}_{h_3h_4} |  e^{\hat{T}_J} | J\rangle E_{JI} 
, \label{DCCSD8}
\end{eqnarray}
respectively, where the parities --- $\langle I^{p_3p_4}_{h_1h_2} | J^{p_6}_{h_5} \rangle$, 
$\langle I^{p_3p_4}_{h_1h_2} | J^{p_7p_8}_{h_5h_6} \rangle$, $\langle I^{p_2}_{h_1} | J^{p_4}_{h_3} \rangle$, and
$\langle I^{p_2}_{h_1} | J^{p_5p_6}_{h_3h_4} \rangle$ --- take the value of $0$ or $\pm1$,
and all the other Dirac brackets have already been expanded above.

\subsection{The $\Delta$CCSD method (diagrammatic)\label{sec:size}}

Here, we introduce
a diagrammatic representation of the algebraic $\Delta$CCSD equations given above, outlining the diagrammatic conditions of a general size-extensive degenerate theory, which $\Delta$CCSD theory
seems to obey. In Appendix, we give an analytic (nondiagrammatic) proof of the size-extensivity of $\Delta$CC theory of all orders, 
and also justify these diagrammatic conditions.

A size-extensive degenerate theory is characterized by the following conditions:\ (1) All energy equations
(including off-diagonal energies) must be connected; (2) All amplitude (wave-function) equations must be linked; (3) The definition of linkedness is broadened:\cite{Sandars1969,HoseKaldor1979,Kucharski1989,shavitt}\ A linked diagram is a disconnected diagram at least one of whose disconnected parts is closed; an apparently open 
disconnected part is considered closed if the open part corresponds to an excitation within the internal space. 
See Appendix for their justifications. 

\begin{figure}
\includegraphics[scale=0.6]{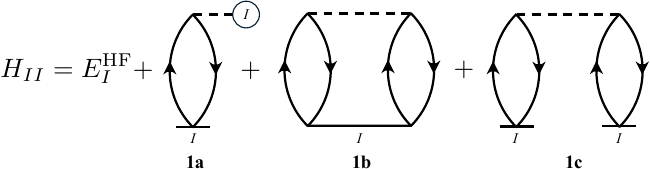}
\caption{The diagonal element $H_{II}$ of $\Delta$CCSD [Eq.\ (\ref{DCCSD3})], which is connected.
A dashed-line vertex with a circled $I$ represents $(f_I)^{\text{out}}_{\text{in}}$, while a dashed-line vertex with four lines denotes $(v)^{\text{left out, right out}}_{\text{left in, right in}}$. A solid-line vertex with an $I$ designates $(t_I)^{\text{out}}_{\text{in}}$ or $(t_I)^{\text{left out, right out}}_{\text{left in, right in}}$.
}
\label{fig:HII}
\end{figure}

The diagrams representing the diagonal element $H_{II}$ of $\Delta$CCSD are given in Fig.\ \ref{fig:HII}, which are connected.
Upon satisfaction of the C condition, $\bm{S}=\bm{1}$ and $\bm{E} = \bm{S}^{-1}\bm{H} = \bm{H}$, and therefore Fig.\ \ref{fig:HII} 
depicts the diagrams for $E_{II}$ also. 

\begin{figure}
\includegraphics[scale=0.6]{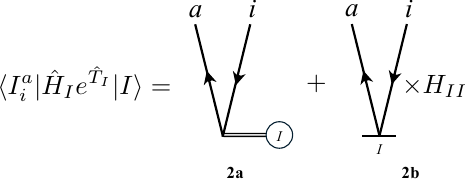}
\caption{The left-hand side of the $T_1$-amplitude equation of $\Delta$CCSD or $\langle I^{a}_{i} |\hat{H}_Ie^{\hat{T}_I}|I\rangle$ of Eq.\ (\ref{DCCSD2})
in the external single-excitation space ($\epsilon_i -\epsilon_a \neq 0$). 
A double-line vertex with a circled $I$ ({\bf 2a}) is the $T_1$-amplitude equation of single-reference CCSD or $(\Lambda_I)^a_i$ of Eq.\ (\ref{DCCSD2_2}), which is connected, 
whereas diagram {\bf 2b} is unlinked. This unlinked diagram {\bf 2b} will be canceled by diagram {\bf 11a} in the right-hand side.
}
\label{fig:HJI_singles_ext}
\end{figure}

\begin{figure}
\includegraphics[scale=0.6]{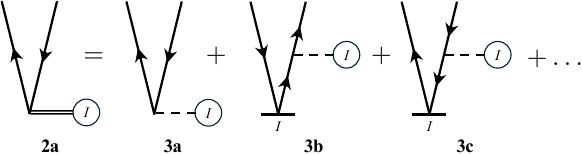}
\caption{The first few diagrams defining diagram {\bf 2a} or $(\Lambda_I)^a_i$ of Eq.\ (\ref{DCCSD2_2}), which are connected.
See, e.g., Fig.\ 10.2 of Shavitt and Bartlett\cite{shavitt} for a full list of diagrams.
}
\label{fig:Lambda1}
\end{figure}

Figure \ref{fig:HJI_singles_ext} is a diagrammatic representation of the left-hand side of the $T_1$-amplitude equation of $\Delta$CCSD, i.e., $\langle I^{a}_{i} |\hat{H}_Ie^{\hat{T}_I}|I\rangle$ of Eq.\ (\ref{DCCSD2}) in the external single-excitation space. Here, 
we revert to the convention that $i$, $j$, $k$, etc.\ refer to occupied spin-orbitals and $a$, $b$, $c$, etc.\ to virtual (unoccupied) ones in $|I\rangle$ (these line indices
will be suppressed unless necessary for explanations).
Diagram {\bf 2a} is the $T_1$-amplitude equation of single-reference CCSD or $(\Lambda_I)^a_i$ of Eq.\ (\ref{DCCSD2_2}), which is connected.
The first few diagrams defining diagram {\bf 2a} [Eq.\ (\ref{DCCSD2_2})] are drawn in Fig.\ \ref{fig:Lambda1}. 
Diagram {\bf 2b} is unlinked, but will be canceled by the identical unlinked diagram ({\bf 11a}) in the right-hand side of the $T_1$-amplitude equation (see below). 
Upon this cancellation, the left-hand side of the $T_1$-amplitude equation of $\Delta$CCSD is connected and thus linked.

\begin{figure}
\includegraphics[scale=0.6]{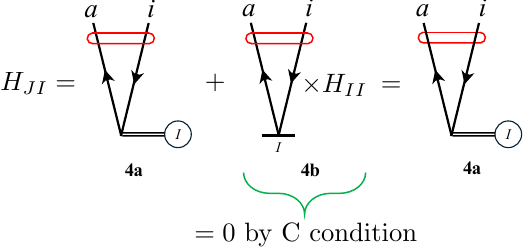}
\caption{The off-diagonal element $H_{JI}$ with $|J\rangle = |I^a_i\rangle$ or $\langle I^{a}_{i} |\hat{H}_Ie^{\hat{T}_I}|I\rangle$
 of Eq.\ (\ref{DCCSD2}) in the internal single-excitation space. 
 A red oblong denotes the fictitious resolvent line that demands $\epsilon_i -\epsilon_a = 0$. A double-line vertex with a circled $I$ ({\bf 4a}) now designates the 
one-electron part of the $\Delta$CCSD effective Hamiltonian, which is isomorphic with the connected $T_1$-amplitude equation of single-reference CCSD.}
\label{fig:HJI_singles_int}
\end{figure}

\begin{figure}
\includegraphics[scale=0.6]{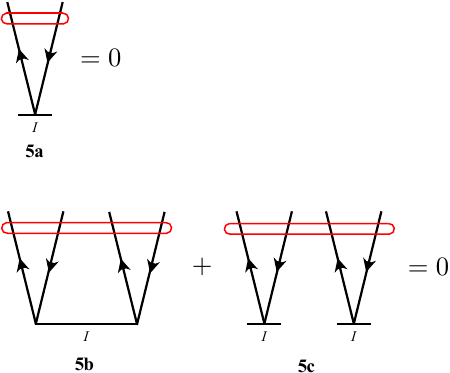}
\caption{The C condition of $\Delta$CCSD [Eqs.\ (\ref{DCCSD_C1}) and (\ref{DCCSD_C2})], i.e., $S_{JI}=0$ with $|J\rangle = |I^a_i\rangle$ and $\epsilon_i -\epsilon_a = 0$ ({\bf 5a})
or with $|J\rangle = |I^{ab}_{ij}\rangle$ and $\epsilon_i +\epsilon_j -\epsilon_a -\epsilon_b = 0$ ({\bf 5b} and {\bf 5c}).}
\label{fig:SJI_singles}
\end{figure}

\begin{figure}
\includegraphics[scale=0.6]{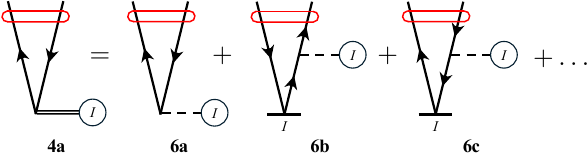}
\caption{The one-electron part of the effective Hamiltonian (diagram {\bf 4a}) of $\Delta$CCSD, $H_{JI}$ with $|J\rangle = |I^a_i\rangle$ and $\epsilon_i -\epsilon_a = 0$, which is isomorphic with the connected $T_1$-amplitude equation of the single-reference CCSD (Fig.\ \ref{fig:Lambda1}). See, e.g., Fig.\ 10.2 of Shavitt and Bartlett\cite{shavitt} for a full list.}
\label{fig:Lambda1_2}
\end{figure}

Figure \ref{fig:HJI_singles_int} draws the same matrix element, $\langle I^{a}_{i} |\hat{H}_Ie^{\hat{T}_I}|I\rangle$ of Eq.\ (\ref{DCCSD2}), in the internal single-excitation space. 
It defines the off-diagonal element of the $\Delta$CCSD Hamiltonian matrix $H_{JI}$ with $|J\rangle = |I^{a}_{i} \rangle$ and $\epsilon_i - \epsilon_a = 0$. 
This figure is the same as Fig.\ \ref{fig:HJI_singles_ext} except for two important differences:\ First, the dangling lines are intersected by a fictitious resolvent line (red oblong),
whose fictitious denominator is zero. It is ``fictitious'' because it does not perform any division but indicates that the ranges of orbital indices are restricted
to the ones that satisfy $\epsilon_i - \epsilon_a = 0$. Second, the unlinked diagram ({\bf 4b}) vanishes 
by virtue of the C condition [Eq.\ (\ref{DCCSD_C2})], which is diagrammatically depicted in Fig.\ \ref{fig:SJI_singles}. 
As a result, its double-line vertex with a circled $I$ ({\bf 4a}) now denotes the one-electron part of the $\Delta$CCSD effective Hamiltonian. The latter is 
further expanded in Fig.\ \ref{fig:Lambda1_2}. It is isomorphic with the $T_1$-amplitude equation of single-reference CCSD (Fig.\ \ref{fig:Lambda1}) and is connected.

\begin{figure}
\includegraphics[scale=0.6]{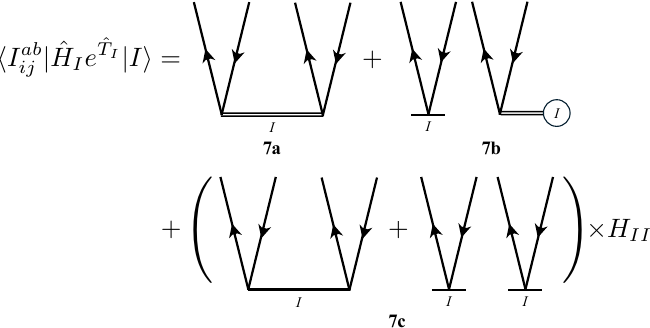}
\caption{The left-hand side of the $T_2$-amplitude equation of $\Delta$CCSD or $\langle I^{ab}_{ij} |\hat{H}_Ie^{\hat{T}_I}|I\rangle$ of Eq.\ (\ref{DCCSD1}) in the external double-excitation space ($\epsilon_i+\epsilon_j -\epsilon_a -\epsilon_b \neq 0$). 
A double-line vertex with an $I$ ({\bf 7a}) is the $T_2$-amplitude equation of single-reference CCSD or $(\Lambda_I)^{ab}_{ij}$ of Eq.\ (\ref{DCCSD1_2}), which is connected. Diagram {\bf 7b} is disconnected and partially unlinked. Product {\bf 7c} is wholly unlinked, but cancels the unlinked product {\bf 15a}. 
}
\label{fig:HJI_doubles_ext}
\end{figure}

\begin{figure}
\includegraphics[scale=0.6]{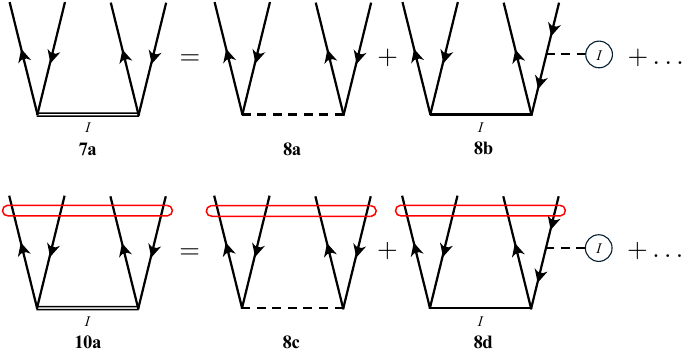}
\caption{Top row:\ The first few diagrams defining diagram {\bf 7a} or $(\Lambda_I)^{ab}_{ij}$ of Eq.\ (\ref{DCCSD1_2}), which are connected. 
Bottom row:\ The two-electron part of the effective Hamiltonian (diagram {\bf 10a}) of $\Delta$CCSD, $H_{JI}$ with $|J\rangle = |I^{ab}_{ij}\rangle$ and $\epsilon_i+\epsilon_j -\epsilon_a -\epsilon_b= 0$, which is isomorphic with the diagrams in the top row and connected. See Figs.\ 9.2 and 10.3 of Shavitt and Bartlett\cite{shavitt} for a full list.}
\label{fig:Lambda2}
\end{figure}

Figure \ref{fig:HJI_doubles_ext} illustrates the diagrams for the left-hand side of the $T_2$-amplitude equation of $\Delta$CCSD, i.e., $\langle I^{ab}_{ij} |\hat{H}_Ie^{\hat{T}_I}|I\rangle$ of Eq.\ (\ref{DCCSD1}) in the external double-excitation space ($\epsilon_i+\epsilon_j -\epsilon_a -\epsilon_b \neq 0$). 

Diagram {\bf 7a}  stands for $(\Lambda_I)^{ab}_{ij}$ of Eq.\ (\ref{DCCSD1_2}), i.e., the connected diagrams of the $T_2$-amplitude equation of single-reference CCSD, whose first few diagrams 
are drawn in the top row of Fig.\ \ref{fig:Lambda2}. 

\begin{figure}
\includegraphics[scale=0.6]{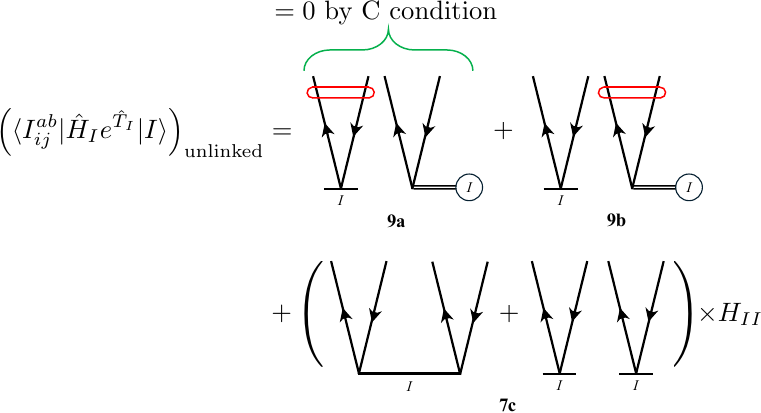}
\caption{The unlinked contributions to the left-hand side of the $T_2$-amplitude equation of $\Delta$CCSD (Fig.\ \ref{fig:HJI_doubles_ext}).
Diagram {\bf 9a} vanishes by the C condition, diagram {\bf 9b} will be {\it substantively} canceled by diagram {\bf 16a}, and product {\bf 7c} will be canceled by product {\bf 15a}.}
\label{fig:HJI_doubles_unlinked}
\end{figure}

Diagram {\bf 7b} is disconnected and partially unlinked. It is partially unlinked because it {\it contains} unlinked contributions whose disconnected part is 
intersected by a fictitious resolvent line at the top and is thus closed.  
These unlinked contributions (including {\bf 7c}) are isolated in Fig.\ \ref{fig:HJI_doubles_unlinked}.
The unlinked diagram {\bf 9a} is zero by virtue of the C condition (Fig.\ \ref{fig:SJI_singles}).
The unlinked diagram {\bf 9b} will be {\it substantively} canceled by the corresponding unlinked diagram {\bf 16a} in the 
right-hand side of the $T_2$-amplitude equation. See below and Appendix \ref{app:DCC} for a full explanation. 

Product {\bf 7c} is wholly unlinked, but will be canceled by the unlinked product {\bf 15a} in the right-hand side of the $T_2$-amplitude equation (see below).
After these disappearance and cancellations, the left-hand side of the $T_2$-amplitude equation 
is linked (but disconnected). 

\begin{figure}
\includegraphics[scale=0.6]{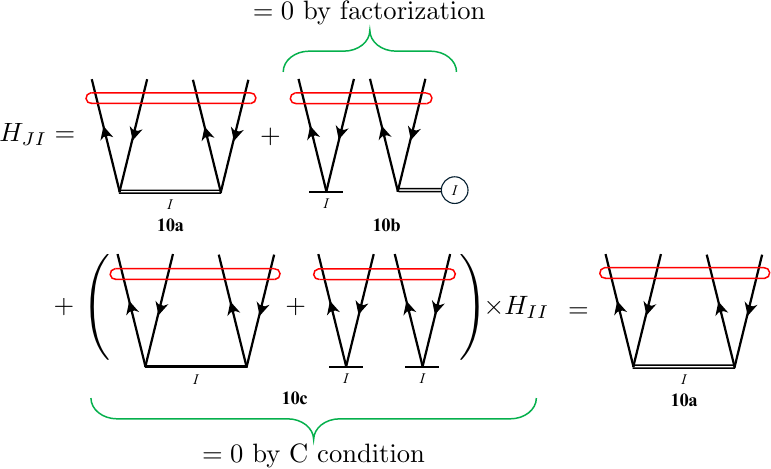}
\caption{The off-diagonal element $H_{JI}$ with $|J\rangle = |I^{ab}_{ij}\rangle$ or $\langle I^{ab}_{ij} |\hat{H}_Ie^{\hat{T}_I}|I\rangle$ of Eq.\ (\ref{DCCSD1}) in the internal double-excitation space.  
A red oblong denotes the fictitious resolvent line that demands $\epsilon_i+\epsilon_j -\epsilon_a -\epsilon_b= 0$. 
A double-line vertex with an $I$ ({\bf 10a}) now designates the two-electron part of the $\Delta$CCSD effective Hamiltonian, which is 
isomorphic with the connected $T_2$-amplitude equation of single-reference CCSD. 
}
\label{fig:HJI_doubles_int}
\end{figure}

Figure \ref{fig:HJI_doubles_int} is the diagrams of $\langle I^{ab}_{ij} | \hat{H}_Ie^{\hat{T}_I}|I\rangle$ of Eq.\ (\ref{DCCSD1}) in the internal double-excitation space, defining the off-diagonal element $H_{JI}$ with $|J\rangle = |I^{ab}_{ij}\rangle$ with $\epsilon_i+\epsilon_j -\epsilon_a -\epsilon_b = 0$. The disconnected diagram {\bf 10b} either 
vanishes by the factorization theorem\cite{Frantz} and symmetry responsible for the degeneracy\cite{Jeziorski1981} or does not violate size-extensivity because of multiple 
fictitious resolvent lines in it. A full explanation is given in Appendices  \ref{app:DMP} and  \ref{app:DCC} . The unlinked product {\bf 10c} is zero by virtue of the C condition (Fig.\ \ref{fig:SJI_singles}). As a result of these disappearances, $H_{JI}$ is equal to 
the connected diagram {\bf 10a}, which is isomorphic 
with the diagrams of the connected $T_2$-amplitude equation of single-reference CCSD. Its first few diagrams are drawn in the bottom row of Fig.\ \ref{fig:Lambda2}.
Together with Figs.\ \ref{fig:HII} and \ref{fig:HJI_singles_int}, all the diagonal and off-diagonal $\Delta$CCSD energies are connected. 

Let us now examine the right-hand sides of the $T$-amplitude equations of $\Delta$CCSD. 

\begin{figure}
\includegraphics[scale=0.6]{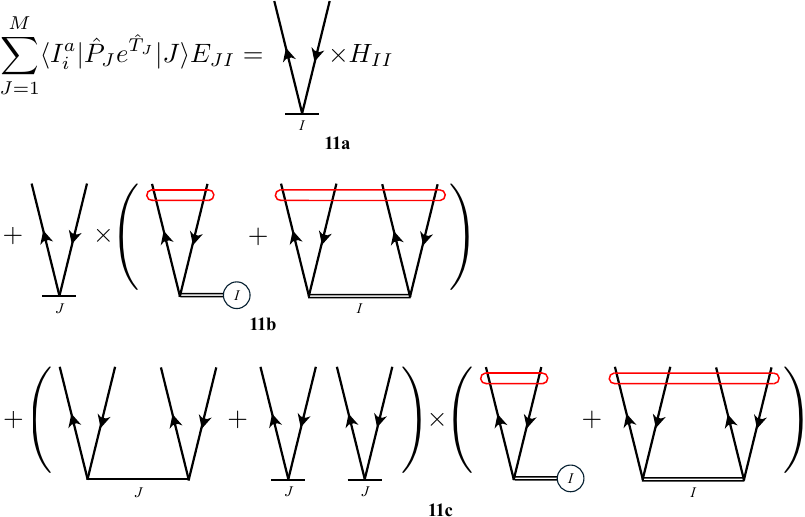}
\caption{The right-hand side of the $T_1$-amplitude equation of $\Delta$CCSD [Eq.\ (\ref{DCCSD8})]. 
Diagram {\bf 11a} is unlinked, but cancels the unlinked diagram {\bf 2b}.
All subsequent products have the wrong topology because 
dangling lines are yet to be contracted (see the following figures and discussions in the main text
on how the correct topology can be restored).}
\label{fig:RHS_singles}
\end{figure}

The right-hand side of the $T_1$-amplitude equation is diagrammatically depicted in Fig.\ \ref{fig:RHS_singles}.
It corresponds to Eq.\ (\ref{DCCSD8}) line-by-line. Recall that $\bm{E}=\bm{H}$ upon 
satisfaction of $\bm{S}=\bm{1}$ and that $E_{JI}$ ($J\neq I$) is the sum of the connected diagrams {\bf 4a} and {\bf 10a}.
The first term ({\bf 11a}) is unlinked and cancels the unlinked diagram {\bf 2b} in the left-hand side.
Subsequent products ({\bf 11b} and {\bf 11c}) do not have the correct topology; they have too many dangling lines as compared with 
diagram {\bf 11a}, which is upward open with just two dangling lines (we avoid the term ``external lines'' to prevent confusion with external space). 

We shall illustrate how these dangling lines are contracted with one another so that these products are transformed into diagrams with the correct topology. 

\begin{figure}
\includegraphics[scale=0.6]{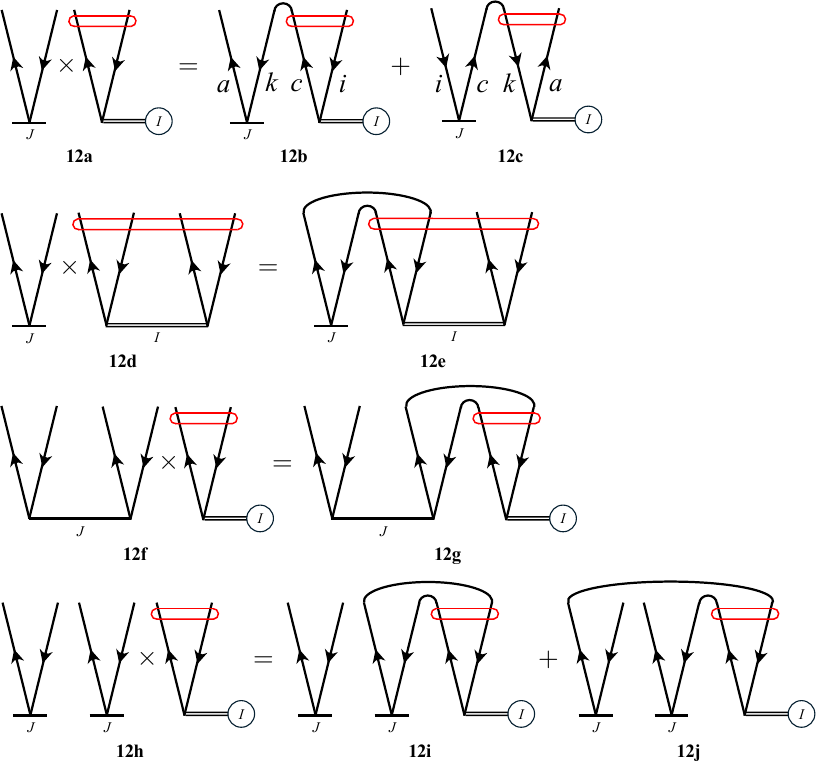}
\caption{A transformation of products {\bf 11b} and {\bf 11c} of Fig.\ \ref{fig:RHS_singles} into sums of folded diagrams with the correct topology by contracting excess dangling lines with warped lines.
}
\label{fig:RHS_singles_folded1}
\end{figure}

Consider the first product of {\bf 11b}, i.e., diagram {\bf 12a} of Fig.\ \ref{fig:RHS_singles_folded1}. 
It represents $\langle I^{p_2}_{h_1} | J^{p_4}_{h_3} \rangle \langle J^{p_4}_{h_3} |e^{\hat{T}_J} | J\rangle \langle J|\hat{H}_I e^{\hat{T}_I}|I\rangle$ of Eq.\ (\ref{DCCSD8}), where
$|J\rangle$ and $|I\rangle$ are degenerate. Since $\langle I^{p_2}_{h_1} | J^{p_4}_{h_3} \rangle = \pm1$, it is either that (i) $|J\rangle = \pm |I^c_i\rangle$ with $\epsilon_i-\epsilon_c = 0$ (but $c\neq i$) 
and $|I^a_i \rangle = \pm |J^a_k\rangle$ with $c=k$ but $k\neq i$ or that (ii) $|J\rangle=\pm |I^a_k\rangle$ with $\epsilon_k-\epsilon_a = 0$ (but $a\neq k$) and 
$|I^a_i\rangle = \pm|J^c_i\rangle$ with $c=k$ but $c \neq a$. In case (i), the virtual orbital (particle line)  labeled $c$ of the $E_{JI}$ (double-line) vertex 
is identified as the occupied orbital (hole line) labeled $k$ 
of the $(t_J)^a_k$ vertex, and hence
they should be one warped line. The contraction of these two lines results in diagram {\bf 12b}, an example of a backward diagram of Sandars\cite{Sandars1969,Lindgren1974,Kucharski1989} or a folded diagram of Brandow\cite{Brandow,Johnson1971,Kuo1971,Lindgren1974,Kucharski1989} in 
multireference perturbation theory (hereafter a folded diagram). Case (ii) corresponds to folded diagram {\bf 12c},
where the occupied orbital $k$ of the $E_{JI}$ (double-line) vertex turns out to be the virtual orbital $c$ of the $(t_J)^c_i$ vertex. So each warped line changes its hole/particle attribute as it goes from one vertex/vacuum to 
another vertex/vacuum. 

At this stage, the two diagrams {\bf 12b} and {\bf 12c} have the correct topology as they are upward open with two dangling lines. They are connected.

\begin{figure}
\includegraphics[scale=0.6]{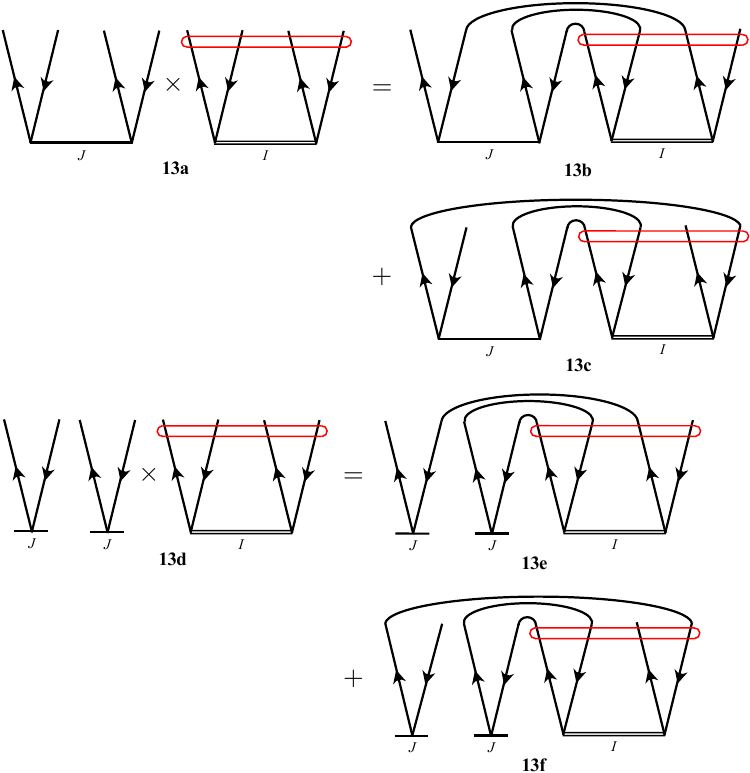}
\caption{Continued from Fig.\ \ref{fig:RHS_singles_folded1}.}
\label{fig:RHS_singles_folded2}
\end{figure}

\begin{figure}
\includegraphics[scale=0.6]{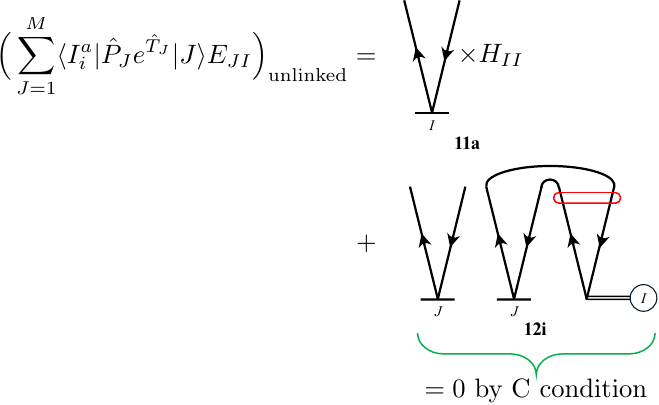}
\caption{The unlinked contributions to the right-hand side of the $T_1$-amplitude equation of $\Delta$CCSD. Diagram {\bf 11a} cancels diagram {\bf 2b} in the left-hand side.}
\label{fig:RHS_singles_unlinked}
\end{figure}

The remaining products of {\bf 11b} and {\bf 11c} are transformed into folded diagrams with the correct topology
in Figs.\ \ref{fig:RHS_singles_folded1} and  \ref{fig:RHS_singles_folded2}. They are all connected except for diagram {\bf 12i}, which is not only disconnected but also unlinked. 
It is, however, easy to see that this diagram is zero by the C condition (Fig.\ \ref{fig:SJI_singles}). All unlinked diagrams
in the right-hand side are isolated in Fig.\ \ref{fig:RHS_singles_unlinked}. They are either canceled by the unlinked diagram {\bf 2b} in the left-hand side
or zero by the C condition. The whole $T_1$-amplitude equation of $\Delta$CCSD is, therefore, connected and thus linked. 

\begin{figure}
\includegraphics[scale=0.6]{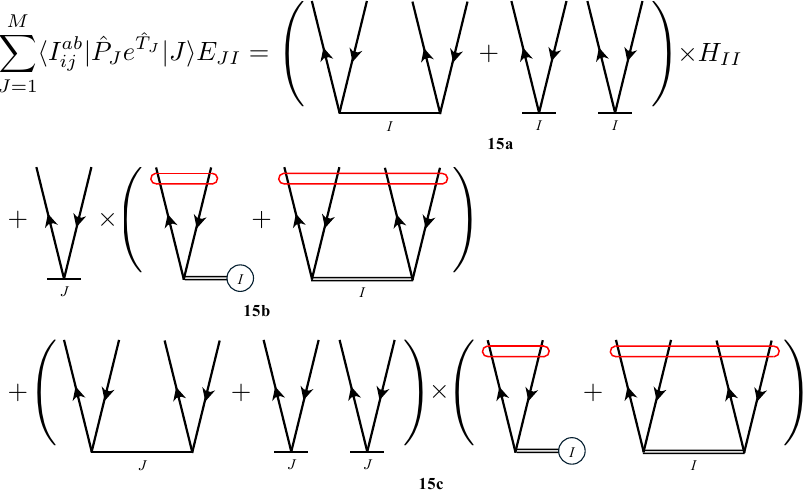}
\caption{The right-hand side of the $T_2$-amplitude equation of $\Delta$CCSD [Eq.\ (\ref{DCCSD7})]. 
Product {\bf 15a} is unlinked, but cancels the unlinked product {\bf 7c}.
All subsequent diagrams have wrong topology, but will be transformed into folded diagrams with the correct topology.
}
\label{fig:RHS_doubles}
\end{figure}

\begin{figure}
\includegraphics[scale=0.6]{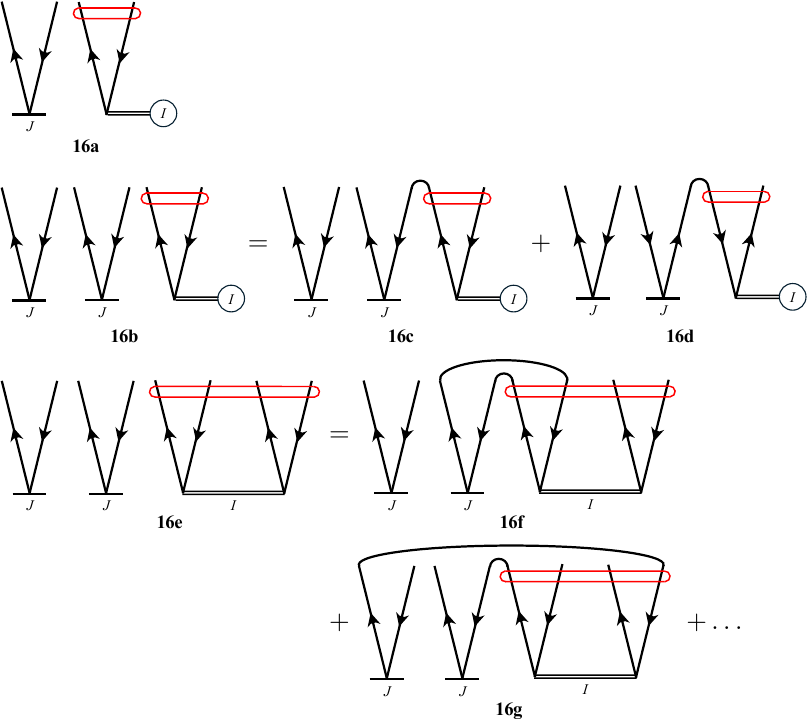}
\caption{A transformation of products {\bf 15b} and {\bf 15c} into folded diagrams with the correct topology, i.e., which are upward open with four dangling lines.
Only those products that lead to disconnected folded diagrams are shown.}
\label{fig:RHS_doubles_folded}
\end{figure}

The diagrammatic representation of the right-hand side of the $T_2$-amplitude equation is shown in Fig.\ \ref{fig:RHS_doubles}. 
It is the literal diagrammatic translation of Eq.\ (\ref{DCCSD7}). The first product {\bf 15a} is unlinked, but it cancels exactly the unlinked product {\bf 7c} in the left-hand side. 
The remaining products tend to have the wrong topology, but they are systematically folded to become topologically correct. 
Some of these folded diagrams are disconnected, which are shown  
in Fig.\ \ref{fig:RHS_doubles_folded}. They are either wholly unlinked (diagram {\bf 16a}) or partially unlinked (diagrams {\bf 16c}, {\bf 16d}, and {\bf 16f}). 

\begin{figure}
\includegraphics[scale=0.6]{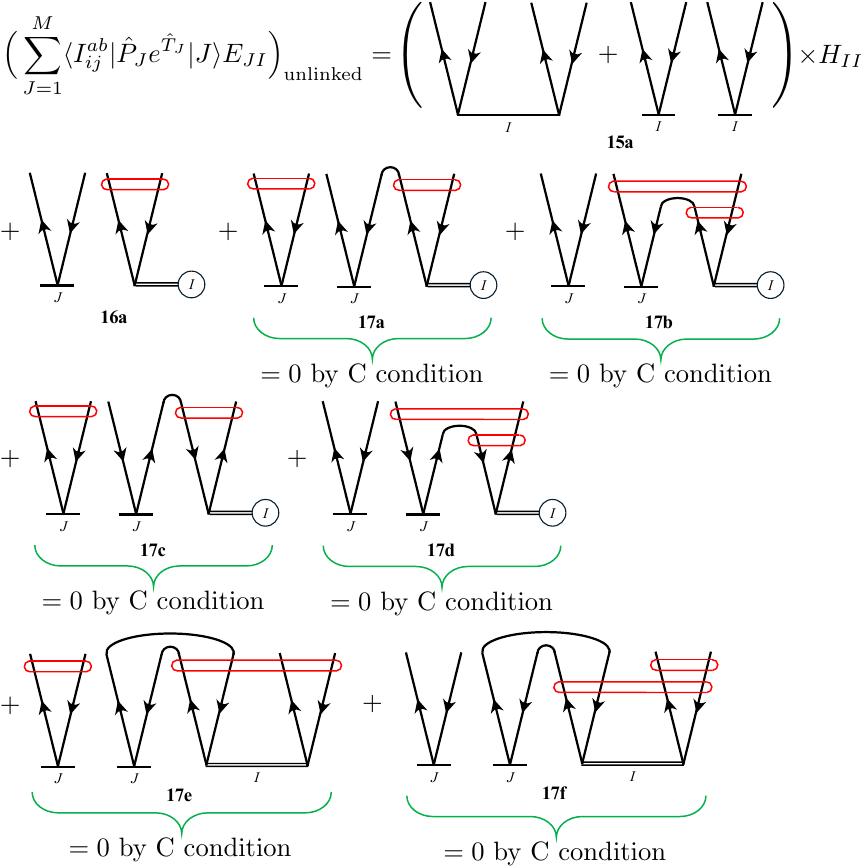}
\caption{The unlinked contributions to the right-hand side of the $T_2$-amplitude equation of $\Delta$CCSD.
Product {\bf 15a} cancels product {\bf 7c}, diagram {\bf 16a} {\it substantively} cancels diagram {\bf 9b}, and the rest are zero by the C condition.}
\label{fig:RHS_doubles_unlinked}
\end{figure}

Figure \ref{fig:RHS_doubles_unlinked} isolates the unlinked contributions to the right-hand side of the $T_2$-amplitude equation. 
The disconnected diagram {\bf 16c} contains the unlinked contributions {\bf 17a} and {\bf 17b}, whose disconnected parts are intersected by 
a fictitious resolvent line at the top and are thus closed. Owing to the very fictitious resolvent lines, however, these unlinked contributions 
are zero by the C condition (Fig.\ \ref{fig:SJI_singles}). Likewise, 
diagram {\bf 16d} contains the unlinked diagrams {\bf 17c} and {\bf 17d}, which are also zero by the C condition. 
Diagram {\bf 16f} has unlinked diagrams {\bf 17e} and {\bf 17f} as its part, but
they too vanish. 

Comparing Figs.\ \ref{fig:HJI_doubles_unlinked} and \ref{fig:RHS_doubles_unlinked}, which are 
the unlinked diagrams in the left- and right-hand sides of the $T_2$-amplitude equations, respectively,
we see that the surviving unlinked products {\bf 7c} and {\bf 15a} annihilate each other, while
another pair of surviving unlinked diagrams {\bf 9b} and {\bf 16a} {\it substantively} cancel out. By ``substantively,'' we mean that the cancellation is not exact because 
a $(t_I)$-amplitude appears in {\bf 9b}, whereas a $(t_J)$-amplitude enters {\bf 16a}, but that the remnant of the cancellation is connected, not 
altering the fact that the whole $T_2$-amplitude equation is linked. 
See Appendix \ref{app:DCC} for a full explanation. 

In summary, all diagonal and off-diagonal elements of $\bm{E}= \bm{H}$ are connected (and linked) 
and the $T_1$- and $T_2$-amplitude equations are linked in $\Delta$CCSD theory 
in the broadened definition of linkedness. The connectedness and linkedness are expected (but not proven directly) to be true 
at all higher orders of $\Delta$CC theory. 
See an alternative, analytic proof of the size-extensivity of $\Delta$CC theory of all orders in Appendix \ref{app:analytic}.

\subsection{Relationship to $\Delta$MP theory\label{sec:HCPT}}

Single-reference RSPT (Refs.\ \onlinecite{shavitt,BartlettARPC}) stipulates its $n$th-order correction to the $I$th-state energy to be 
\begin{eqnarray}
E^{(n)}_I = \langle I^{(0)} | \hat{V}_I | I^{(n-1)} \rangle,
\end{eqnarray}
where the Hamiltonian is partitioned 
into the zeroth-order Hamiltonian and perturbation operator as $\hat{H}_I = \hat{H}_0 + \hat{V}_I$, and 
$|I^{(n)}\rangle$ is the $n$th-order correction to the $I$th-state wave function and $|I^{(0)}\rangle = |I\rangle$ (a single Slater determinant). The $|I^{(n)}\rangle$  is obtained recursively as 
\begin{eqnarray}
|I^{(n)}\rangle = \hat{R}_I \Big( \hat{V}_I |I^{(n-1)} \rangle - \sum_{i=1}^{n-1} |I^{(n-i)}\rangle E^{(i)}_I \Big),
\end{eqnarray}
where the resolvent, $\hat{R}_I = (E_I^{(0)} - \hat{H}_0)^{-1}$, is understood to project out all determinants that lead to a division by zero
(the summation above is also understood to vanish when $n=1$).

Let us confine ourselves to the M{\o}ller--Plesset partitioning of the Hamiltonian without losing generality:
\begin{eqnarray}
\hat{H}_0 &=& E_{I}^{(0)} + \sum_p \epsilon_p \left\{ \hat{p}^\dagger \hat{p} \right\}, \\
\hat{V}_I &=& E_{II}^{(1)} + \sum_{p, q} \Big( (f_I)^p_q -\epsilon_p \delta_{pq} \Big) \left\{ \hat{p}^\dagger \hat{q} \right\} \nonumber\\
&&+ \frac{1}{4}\sum_{p,q,r,s} (v)^{pq}_{rs} \left\{ \hat{p}^\dagger \hat{q}^\dagger \hat{s}\hat{r} \right\}
\end{eqnarray}
with
\begin{eqnarray}
E_{I}^{(0)} &=& E_{\text{nuc.}} +  \sum_i^{\text{occ.\ in }|I\rangle} \epsilon_i, \label{MP0} \\
E_{II}^{(1)} &=& E_I^{\text{HF}} - E_{I}^{(0)}, \label{MP1}
\end{eqnarray}
where $E^{\text{HF}}_I$ and $(f_I)^p_q$ are given by Eqs.\ (\ref{EHF}) and (\ref{Fock}). 
This single-reference $n$th-order M{\o}ller--Plesset perturbation (MP$n$) theory\cite{moller} is convergent at exactness (FCI) as $n \to \infty$ unless divergent. 
It is equivalent to the $n$th-order MBPT or MBPT($n$) for $n \geq 2$,\cite{shavitt,BartlettARPC} although the latter is formulated with linked diagrams only 
from the outset and explicitly size-extensive.\cite{shavitt,GellmannLow1951,Brueckner1955,Goldstone1957,Hugenholtz1957,Frantz,Manne}

This ansatz was generalized by Hirschfelder and Certain\cite{hirschfelder} to an exactly degenerate multireference. %
Other groups developed essentially the same theory under different names.\cite{Brandow,Sandars1969,Johnson1971,Kuo1971,Lindgren1974,klein,HoseKaldor1979,Kucharski1989}
We call this in conjunction with the  M{\o}ller--Plesset partitioning $\Delta$MP$n$ theory,\cite{deltamp,Hirata2017} where $n$ is the perturbation order 
(which will be dropped when referring to the whole series) and $\Delta$ carries the dual meaning of {\it difference} and {\it degeneracy}.

The perturbative corrections to the degenerate state energies constitute an $M$-by-$M$ non-Hermitian\cite{Lindgren1974} matrix of the form,
\begin{eqnarray}
E^{(n)}_{JI}= \langle J^{(0)} | \hat{V} | I^{(n-1)} \rangle.\label{HCPTrecursion1}
\end{eqnarray}
The perturbation corrections to the wave functions are given recursively by 
\begin{eqnarray}
|I^{(n)}\rangle = \hat{R}_I \Big( \hat{V} |I^{(n-1)} \rangle - \sum_{i=1}^{n-1} \sum_{J=1}^M |J^{(n-i)}\rangle  E^{(i)}_{JI} \Big), \label{HCPTrecursion2}
\end{eqnarray}
where the resolvent, $\hat{R}_I= (E_I^{(0)} - \hat{H}_0)^{-1}$ projects out all degenerate reference determinants that would cause a division by zero.
In other words, $\hat{R}_I$ ensures that $|I^{(n)}\rangle$ have no overlap with any of the degenerate references, thus playing an 
analogous role as the C condition of $\Delta$CC theory (see below).
The $n$th-order perturbation correction to the $I$th-state energy is an eigenvalue of the matrix $\bm{E}^{(n)}$:
\begin{eqnarray}
\sum_{J=1}^M E_{KJ}^{(n)}{C}_{JI} =  C_{KI}\tilde{E}_{I}^{(n)}, \label{HCPT_diagonalization}
\end{eqnarray}
which also mirrors the $\Delta$CC energy of Eq.\ (\ref{diagonalization}).

$\Delta$MP$n$ theory reduces to MP$n$ theory when $M = 1$.
It is also convergent at exactness for all reference states as $n \to \infty$ unless divergent.\cite{hirschfelder} 
The initial degeneracy may persist or be partially or fully lifted with increasing $n$.\cite{hirschfelder} 
It can be applied to any degenerate or nondegenerate determinant references with any numbers of $\alpha$- and $\beta$-spin electrons, 
any spin multiplicities, and any spatial symmetries, without any modification to its  working equations. 

It is evident that $\Delta$CC theory is a faithful coupled-cluster extension of $\Delta$MP$n$ theory outlined above;\ $\Delta$CC theory is to 
$\Delta$MP$n$ theory as  CC theory is to  MP$n$ theory. 
 $\Delta$CC theory is an infinite partial summation of the $\Delta$MP$n$ diagrams; the two theories
realize two different ways of summing over the same set of diagrams exhaustively, eventually converging at the same exact solutions of the Schr\"{o}dinger equation.


Surprisingly, degenerate RSPT (encompassing $\Delta$MP$n$ theory) is only {\it believed}, but not proven, to be size-extensive. A proof 
of its linked-diagram theorem was given by Hose and Kaldor\cite{HoseKaldor1979} on the basis of the factorization theorem 
of Frantz and Mills.\cite{Frantz} Shortly thereafter, the proof was challenged by Jeziorski and Monkhorst,\cite{Jeziorski1981} who showed that an unlinked diagram 
emerges at the third order. The latter authors nevertheless speculated that such unlinked diagrams systematically vanish by symmetry if the degeneracy 
is caused by the symmetry and not by accident. On this basis, they concluded that degenerate RSPT is ``probably'' size-extensive.\cite{Jeziorski1981} 
(Note that the linked-diagram theorems for quasidegenerate RSPT by Brandow,\cite{Brandow} Sandars,\cite{Sandars1969}  and Lindgren\cite{Lindgren1974} are 
for a complete model space or CMS, and not for an exactly degenerate multireference.) In Appendix \ref{app:analytic}, we provide an analytic 
proof of the size-extensivity of degenerate RSPT including $\Delta$MP theory. The size-extensivity of $\Delta$CC theory, which is an infinite partial summation of 
the $\Delta$MP diagrams, follows immediately. 

Let us further examine the relationship between $\Delta$CC theory and low-order instances of 
$\Delta$MP theory. The first-order correction to the $I$th reference wave function, $|I^{(1)}\rangle$, consists of 
one- and two-electron excitations from $|I^{(0)}\rangle = |I\rangle$,
\begin{eqnarray}
|I^{(1)}\rangle &=&
 \sum_{i}^{\text{occ. in }|I\rangle}\sum_{a}^{\text{vir. in }|I\rangle}|I^{a}_{i}\rangle \frac{({f}_I)^a_i}{\epsilon_i - \epsilon_a} \nonumber\\
&& + \sum_{i<j}^{\text{occ. in }|I\rangle}\sum_{a<b}^{\text{vir. in }|I\rangle}|I^{ab}_{ij}\rangle \frac{(v)^{ab}_{ij}}{\epsilon_i + \epsilon_j - \epsilon_a - \epsilon_b }, 
\end{eqnarray}
where $\epsilon_p$ is the $p$th spin-orbital energy of the orbital reference.
Consequently, the first iterative cycle (indicated by superscript ``[1]'') of the $\Delta$CCSD calculation with zero initial $T$-amplitudes 
yields the $\Delta$MP1 wave functions for all degenerate references, i.e.,
\begin{eqnarray}
(t_I^{[1]})^a_i &=&  \frac{({f}_I)^a_i}{\epsilon_i - \epsilon_a}, \label{T1} \\
(t_I^{[1]})^{ab}_{ij} &=& \frac{(v)^{ab}_{ij}}{\epsilon_i + \epsilon_j - \epsilon_a - \epsilon_b },\label{T2}
\end{eqnarray}
as well as the $\Delta$MP2 energies (if the $\hat{T}_1^2/2!$ contribution to the energy, which is a third-order correction, is excluded; note that $(t_I^{[1]})^a_i$ is generally
nonzero when the $I$th determinant is not the orbital reference). Therefore, $\Delta$MP2 energy is recovered in the first iterative cycle of $\Delta$CCSD for each state 
(with zero initial $T$-amplitudes and null $\hat{T}_1^2/2!$).
This familiar relationship between CCSD and MP2 is maintained 
between $\Delta$CCSD and $\Delta$MP2. 

Next, let us examine the same relationship diagrammatically. 
We adopt Hose and Kaldor's definition\cite{HoseKaldor1979} of the Fermi vacuum:\ When 
an energy or amplitude (wave-function) equation for the $I$th degenerate reference is being considered, the Fermi vacuum is $|I\rangle$ and 
the vertexes  are $E_I^\text{HF}$ and $(f_I)^p_q$ of Eqs.\ (\ref{EHF}) and (\ref{Fock}) as well as $(v)^{pq}_{rs}$.
It may be reminded that a size-extensive degenerate theory is characterized by connected energy diagrams and linked amplitude (wave-function) diagrams in the broadened
definition of linkedness.\cite{Sandars1969,HoseKaldor1979,Kucharski1989,shavitt} 

\begin{figure}
\includegraphics[scale=0.6]{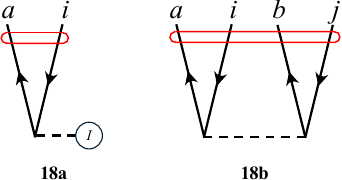}
\caption{The off-diagonal first-order correction to energy, $E_{JI}^{(1)}$, is represented by diagram {\bf 18a} when $|J\rangle = |I^a_i \rangle$ or by {\bf 18b} when $|J\rangle = |I^{ab}_{ij} \rangle$.
A dashed-line vertex with a circled $I$  denotes $(f_I)^{\text{out}}_{\text{in}} - \epsilon_{\text{out}}\delta_{\text{out,in}}$, while a dashed-line vertex with four lines 
represents $(v)^{\text{left out, right out}}_{\text{left in, right in}}$.
A red oblong is a fictitious resolvent line with zero fictitious denominator, i.e., $\epsilon_i - \epsilon_a= 0$ in {\bf 18a} or $\epsilon_i + \epsilon_j - \epsilon_a - \epsilon_b = 0$
in  {\bf 18b}.}
\label{fig:18}
\end{figure}

The off-diagonal first-order corrections to energy, $E_{JI}^{(1)} = \langle J | \hat{V}_I | I \rangle$, are diagrammatically depicted in Fig.\ \ref{fig:18}.
Despite its appearance to the contrary, both diagrams are closed as they are intersected by a fictitious resolvent line at the top. They are connected.
Diagrams {\bf 18a} and {\bf 18b} are identified as diagrams {\bf 6a} and {\bf 8c} of $\Delta$CCSD 
and are the first-order approximations to the one- and two-electron parts of 
$E_{JI} = H_{JI}$ of Figs.\ \ref{fig:Lambda1_2} and \ref{fig:Lambda2}.

\begin{figure}
\includegraphics[scale=0.6]{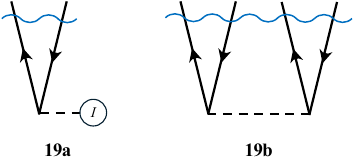}
\caption{The first-order correction to the $I$th reference wave function, $|I^{(1)}\rangle$. Its projection onto the external single-excitation space is represented by diagram {\bf 19a},
while its projection onto the external double-excitation space by {\bf 19b}.
A blue wiggly line denotes a resolvent line.}
\label{fig:WF1}
\end{figure}

The first-order corrections to the $I$th reference wave function, $|I^{(1)}\rangle = \hat{R}_I \hat{V}_I |I^{(0)}\rangle$, are represented by diagrams in Fig.\ \ref{fig:WF1}. 
Diagrams {\bf 19a} and {\bf 19b} are the first-order approximations to the $T_1$- and $T_2$-amplitudes, respectively,
and are connected (Figs.\ \ref{fig:Lambda1} and \ref{fig:Lambda2}). The connectedness proves the size-extensivity of the $\Delta$MP1 wave function. 
It does not immediately imply the connectedness of the diagonal and off-diagonal $\Delta$MP2 energies, however (see Appendix \ref{app:DMP}).

\begin{figure}
\includegraphics[scale=0.6]{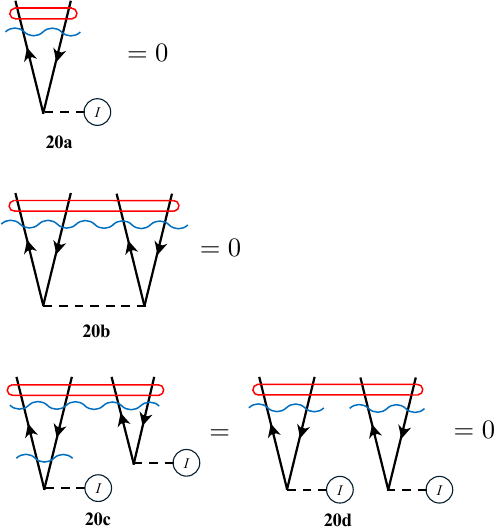}
\caption{The resolvent operators in $\Delta$MP theory play an analogous role as the C condition of $\Delta$CC theory.
The transformation from  {\bf 20c} to {\bf 20d} used the factorization theorem; see Appendix \ref{app:DMP}.}
\label{fig:CconditionMBPT}
\end{figure}

The resolvent lines (blue wiggly lines) project out any excitation amplitudes within the internal space. Therefore, stacking a fictitious resolvent line 
onto an actual resolvent line with the same span annihilates the diagram, as indicated in Fig.\ \ref{fig:CconditionMBPT}. These diagrammatic equalities 
support the notion that a resolvent operator in $\Delta$MP theory plays an analogous role as  the C condition of $\Delta$CC theory (Fig.\ \ref{fig:SJI_singles}).

\begin{figure}
\includegraphics[scale=0.6]{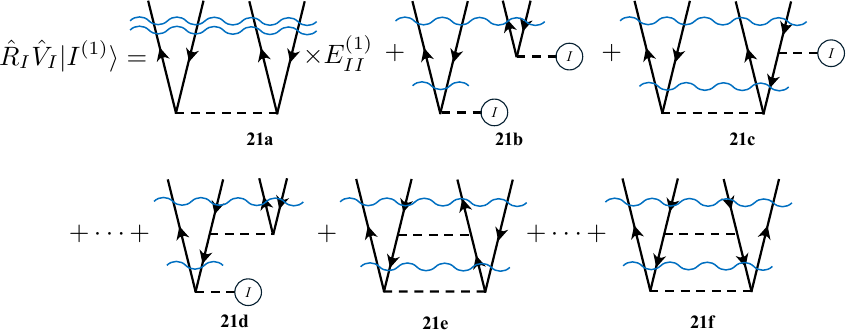}
\caption{The diagrammatic representation of the principal part of the second-order correction to the $I$th reference wave function, i.e., the first term in the right-hand side of Eq.\ (\ref{WF2}), in the external double-excitation space. See, e.g., Fig.\ 8.10 of Shavitt and Bartlett\cite{shavitt} for a full list.}
\label{fig:WF2_part1}
\end{figure}

\begin{figure}
\includegraphics[scale=0.6]{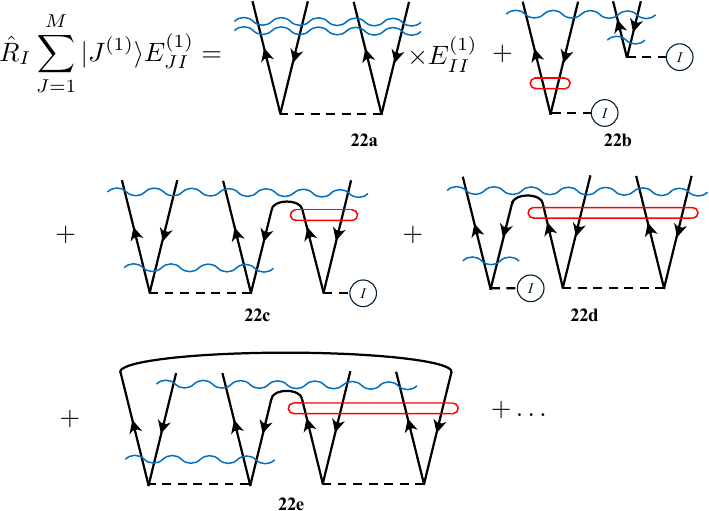}
\caption{The diagrammatic representation of the renormalization part of the second-order correction to the $I$th reference wave function, i.e., the second term in the right-hand side of Eq.\ (\ref{WF2}), 
in the external double-excitation space. See, e.g., Fig.\ 8.11 of Shavitt and Bartlett\cite{shavitt} for a full list.}
\label{fig:WF2_part2}
\end{figure}

The second-order correction to the $I$th reference wave function, $|I^{(2)}\rangle$, is written algebraically as
\begin{eqnarray}
|I^{(2)}\rangle &=& \hat{R}_I \hat{V}_I |I^{(1)}\rangle - \hat{R}_I \sum_{J=1}^{M} |J^{(1)}\rangle E_{JI}^{(1)}. \label{WF2}
\end{eqnarray}
Figure \ref{fig:WF2_part1} draws the diagrammatic representation of the first term (the principal part) of the right-hand side projected onto the external double-excitation space. 
The principal part corresponds to the left-hand side of the $T$-amplitude equations of $\Delta$CC theory.
Therefore, they correspond to the second-order approximation to the left-hand side of the $T_2$-amplitude equation of $\Delta$CCSD (Fig.\ \ref{fig:HJI_doubles_ext}). 
The unlinked diagram {\bf 21a} of $|I^{(2)}\rangle$ is summed over by the unlinked product {\bf 7c}. The disconnected diagram {\bf 21b} 
is included in the disconnected diagram {\bf 7b}. The rest of the diagrams of $|I^{(2)}\rangle$ are connected and accounted for by diagram {\bf 7a}.

The second term (the renormalization part) of the right-hand side of Eq.\ (\ref{WF2}) in the external double-excitation space
is diagrammatically depicted in Fig.\ \ref{fig:WF2_part2}. Its first diagram ({\bf 22a}) is unlinked and cancels the unlinked diagram {\bf 21a} in the principal part.
(An analogous cancellation of unlinked diagrams takes place in the external single-excitation space.) 
It should be recalled that in the broadened definition of linkedness,\cite{Sandars1969,HoseKaldor1979,Kucharski1989,shavitt}  
the disconnected diagram {\bf 21b} contains an unlinked contribution. This too will be canceled by the unlinked diagram {\bf 22b} in the right-hand side, leaving 
a fully linked equation for $|I^{(2)}\rangle$ and ensuring the size-extensivity of the $\Delta$MP2 wave function. 
A proof of the connectedness of the diagonal and off-diagonal energies of $\Delta$MP3 requires a considerably more effort, which will be expounded on in Appendix \ref{app:DMP}. 

The renormalization part corresponds to the right-hand side of the $T$-amplitude equations of $\Delta$CC theory. 
Therefore, the diagrams in Fig.\ \ref{fig:WF2_part2} are summed over by the diagrams in the right-hand side of the $T_2$-amplitude equation 
of $\Delta$CCSD shown in Fig.\ \ref{fig:RHS_doubles}. For instance, diagram {\bf 22a} of $|I^{(2)}\rangle$ 
is accounted for by product {\bf 15a},
while diagram {\bf 22b} is summed over by diagram {\bf 16a}. Diagrams {\bf 22c}--{\bf 22e} are folded diagrams\cite{Brandow,Sandars1969,Johnson1971,Kuo1971,Lindgren1974} 
and are included in folded diagrams originating from products {\bf 15b} and {\bf 15c}.

\begin{figure}
\includegraphics[scale=0.6]{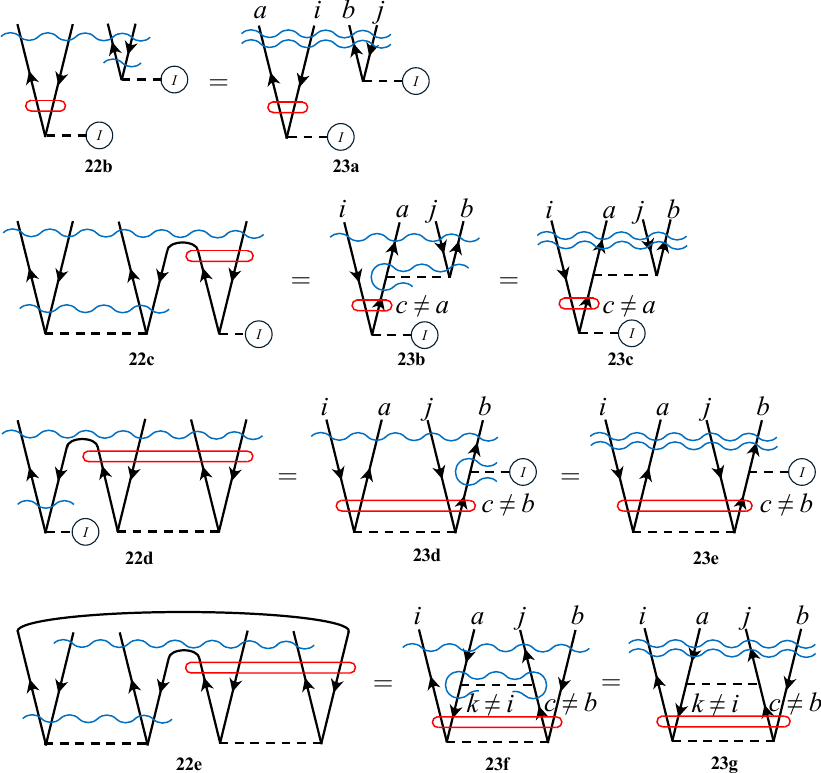}
\caption{A transformation of folded diagrams with warped lines in Fig.\ \ref{fig:WF2_part2} into folded-resolvent diagrams of Kucharski and Bartlett\cite{Kucharski1989} and
then into renormalization diagrams of finite-temperature MBPT.\cite{HirataJCP2021}}
\label{fig:WF2_part2_2}
\end{figure}

These folded diagrams\cite{Brandow,Sandars1969,Johnson1971,Kuo1971,Lindgren1974} can be 
transformed further into ``folded-resolvent'' diagrams of Kucharski and Bartlett\cite{Kucharski1989} with the warped lines straightened 
and the resolvent lines warped. To keep with the $|I\rangle$ vacuum, we lift
the $|J^{(1)}\rangle$ subdiagram and reverse the direction of the warped line anchored to it, as shown in Fig.\ \ref{fig:WF2_part2_2}.
We then arrive at diagrams that are isomorphic with usual single-reference MBPT diagrams, whose resolvent lines are now folded.\cite{Kucharski1989} 
The line directions unambiguously indicate the hole/particle distinction relative to the $|I\rangle$ vacuum.

Furthermore, in each diagram, the folded resolvent line can be shifted up, piling on the top resolvent line by virtue of the zero orbital-energy differences 
represented by the fictitious resolvent lines. 
For example, the folded resolvent line $(\epsilon_c + \epsilon_j - \epsilon_a - \epsilon_b)^{-1}$ of diagram {\bf 23b} can be straightened, shifted up, and stacked onto
the top resolvent line $(\epsilon_i + \epsilon_j - \epsilon_a - \epsilon_b)^{-1}$ because the fictitious resolvent line demands $\epsilon_i - \epsilon_c =0$.
It then becomes diagram {\bf 23c}. (The index restriction ``$c\neq a$'' comes from the fact that $a$ is virtual, but $c$ is occupied in $|J\rangle$.) 
We thus arrive at the diagrams that are isomorphic with the renormalization diagrams of finite-temperature MBPT (so called 
because they too originate from the renormalization term of the Rayleigh--Schr\"{o}dinger recursion),\cite{HirataJCP2021}
in which resolvent lines corresponding to a zero denominator are shifted up. 

\subsection{Relationship to MRCC theories\label{sec:MRCC}}

Table \ref{tab:qcc} summarizes the comparison of various MRCC theories, including $\Delta$CC theory for a degenerate multireference and QCC theory
for a GMS. 

\begin{table*}
\caption{Comparison of MRCC theories.}
\label{tab:qcc}
\begin{ruledtabular}
\begin{tabular}{lcccccc}
Theory  & Jeziorski--Monkhorst & Kucharski--Bartlett & Li--Paldus & Meissner {\it et al.}& $\Delta$CC & QCC\\
& MRCC\cite{Jeziorski1981} & MRCC\cite{Kucharski1991} & MRCC\cite{PaldusLi2003,LiPaldus2003,LiPaldus2003_2} & MRCC\cite{Meissner1989,Meissner1990} 
& (this work) & (this work)\\ 
\hline
Model space\footnotemark[1] & Complete & Complete & General & Special  & Degenerate & General \\
Left-multiply $e^{-\hat{T}}$\footnotemark[2] & Yes & No & Yes & Yes & No & No \\
The C condition\footnotemark[3] & No & No & Yes & No  & Yes & Yes \\
Size-extensive\footnotemark[4] & Yes & Yes & Yes \& No\cite{Nooijen_extensivity2005,Liu2026} & Yes  & Yes & Yes \& No\\
Convergent toward FCI & Yes & Yes & Yes & No\cite{LiPaldus2003}  & Yes & Yes \\
Black box\footnotemark[5] & No & No & No & No & Yes & No \\
\end{tabular}
\footnotetext[1]{A complete model space (CMS) is the set of determinants with all possible occupancy patterns of active spin-orbitals by a fixed number of electrons. A general model space (GMS) 
is any arbitrary set of determinants, which encompasses degenerate multireferences.
A special model space is defined by Refs.\ \onlinecite{Meissner1989,Meissner1990}, and includes all CMS and some GMS.}
\footnotetext[2]{Whether the Schr\"{o}dinger equation is left-multiplied by $e^{-\hat{T}}$ before projection.} 
\footnotetext[3]{Whether the C condition [Eq.\ (\ref{C-condition2})] is imposed. For a CMS, it is equivalent to the T condition [Eq.\ (\ref{Tcondition})] implied in the Jeziorski--Monkhorst\cite{Jeziorski1981} and Kucharski--Bartlett\cite{Kucharski1991} ans\"{a}tze.}
\footnotetext[4]{``Yes \& No'' means that the method can be size-extensive
or not, depending on the model space adopted.\cite{Meissner1989} See Appendix for more details.}
\footnotetext[5]{Whether the theory's algebraic working equations are invariant with the details of the multireference.}
\end{ruledtabular}
\end{table*}

\subsubsection{The Jeziorski--Monkhorst ansatz\label{sec:JeziorskiMonkhorst}}

Jeziorski and Monkhorst\cite{Jeziorski1981} pioneered a state-universal or Hilbert-space MRCC theory for a CMS.
A CMS is the one spanned by determinants with all possible patterns of occupancy of active orbitals by a fixed number of electrons. 
A degenerate multireference is generally not a CMS, and hence the Jeziorski--Monkhorst theory cannot serve as a degenerate CC theory, while
the present $\Delta$CC and QCC theories can be applied to any model space, including a CMS. 
The Jeziorski--Monkhorst theory has been further developed by Piecuch and coworkers.\cite{PiecuchPaldus1992,PiecuchPaldus1993,Paldus1993,Piecuch1994,PiecuchPaldus1994,LiPiecuch1994,PiecuchLi1994,PiecuchPaldus1995,KowalskiPRA2000,KowalskiIJQC2000,Kowalski2001,Kowalski2004}

Other MRCC theories include the state-specific Brillouin--Wigner theory of Huba\v{c}, Pittner, and coworkers,\cite{Hubac1994,Hubac1998,Pittner1999,Pittner2004,Pittner2005,PittnerPiecuch2009}
and state-specific theory of Mukherjee and coworkers.\cite{Mukherjee1998_1,Mukherjee1999_2,Mukherjee1999_3,Mukherjee2000_4,Mukherjee2004_5} 
They are not closely related to the present $\Delta$CC or QCC theories. Nor are valence-universal (Fock-space) theories.\cite{Mukherjee1975,Mukherjee1975_2,Mukherjee1977,Lindgren1978,Kutzelnigg1984,Haque1984,Haque1985,Haque1985_2,Nooijen1997_1,Nooijen1997_2,Nooijen1997,Petraco2002} 
For an excellent overview of these and other MRCC theories, see Evangelista {\it et al.}\cite{Evangelista2006}

For a CMS of size $M$, the Jeziorski--Monkhorst ansatz may be written as a projected Schr\"{o}dinger equation,
\begin{eqnarray}
\hat{P}_I e^{-\hat{T}_I}  \hat{H}_I e^{\hat{T}_I} |I\rangle = \hat{P}_I \sum_{J=1}^{M} e^{-\hat{T}_I} e^{\hat{T}_J} | J \rangle E_{JI}, \label{JM_ansatz}
\end{eqnarray}
where $|I\rangle$ and $|J\rangle$ are two of the reference determinants in the CMS and $\hat{P}_I$ is a projector to the external space
spanned by determinants reachable by $\hat{T}_I |I\rangle$. In a CMS, the demarcation between the internal and external spaces is the same for all references, and therefore, 
\begin{eqnarray}
\langle J | {\hat{T}_I} | I \rangle = 0. \label{Tcondition}
\end{eqnarray}
We call this the T condition in this article to distinguish it from the C condition, although they are mathematically the same in a CMS. 
The effective Hamiltonian matrix $\bm{E}$ is defined by its elements as
\begin{eqnarray}
E_{JI} &=& \langle J | e^{-\hat{T}_I} \hat{H}_I e^{\hat{T}_I} |I\rangle, \label{JM_ansatz2}
\end{eqnarray}
whose eigenvalues are the desired energies. 
In a CMS, since $\hat{T}_I^\dagger |J\rangle = 0$, $\langle J|$ and $\langle J|e^{-\hat{T}_I}$ span the same space, 
justifying the left-multiplication of $e^{-\hat{T}_I}$ in Eq.\ (\ref{JM_ansatz2}). 

The Jeziorski--Monkhorst theory is convergent toward FCI. It is also size-extensive, as proven by the originators.\cite{Jeziorski1981} 
However, when the number of active orbitals grows proportionally with system size, its application to an infinitely extended system seems
all but intractable. 

The Jeziorski--Monkhorst ansatz bears similarity with the $\Delta$CC ansatz, 
and their motivations and justifications are essentially the same.  However,
they differ in four important respects:\ (i) The Jeziorski--Monkhorst theory is limited to a CMS and cannot be used as a degenerate CC theory; (ii) The 
Jeziorski--Monkhorst theory left-multiplies $e^{-\hat{T}_I}$ with the Schr\"{o}dinger equation, while $\Delta$CC theory does not; 
(iii) The Jeziorski--Monkhorst theory invokes the T condition [Eq.\ (\ref{Tcondition})], while $\Delta$CC theory uses the C condition [Eq.\ (\ref{DCC_ansatz2})]. 
Although they are equivalent in a CMS, they differ substantively for a degenerate multireference; (iv) Numerous $\hat{P}_I$ operators in the $\Delta$CC 
ansatz [Eqs.\ (\ref{DCC_ansatz})--(\ref{S})] are absent in the Jeziorski--Monkhorst ansatz.

Owing to difference (iv), the Jeziorski--Monkhorst theory is not black-box, while $\Delta$CC theory is. 
These projectors in $\Delta$CC theory ensure that only excitations up to the truncation rank of $\hat{T}_I$  from each degenerate reference are taken into account,
whereas the Jeziorski--Monkhorst theory considers all effective Hamiltonian matrix elements with any relative excitation ranks. 
Consequently, the number and types of the terms in the formalism of the Jeziorski--Monkhorst theory vary depending 
on the details of a multireference, while they are invariant in $\Delta$CC theory. 

\subsubsection{The Li--Paldus ansatz\label{sec:LiPaldus}}

Li and Paldus\cite{PaldusLi2003,LiPaldus2003,LiPaldus2003_2,Paldus2004,LiPaldusIJQC2004} 
generalized the Jeziorski--Monkhorst ansatz to a GMS, which is also called an incomplete model space, 
referring to any arbitrary set of determinants.
A GMS can better withstand the problems of intruder states,\cite{Schucan1972,Paldus1993,Piecuch1994,shavitt} 
which plague a CMS. 
A degenerate multireference 
is a special case of GMS, and hence the Li--Paldus theory can serve as a degenerate CC theory. 

Letting $|I\rangle$ and $|J\rangle$ be two determinants in a GMS of size $M$, 
the Li--Paldus ansatz is written as 
\begin{eqnarray}
\hat{P}_I e^{-\hat{T}_I}  \hat{H}_I e^{\hat{T}_I} |I\rangle = \hat{P}_I \sum_{J=1}^{M} e^{-\hat{T}_I} e^{\hat{T}_J} | J \rangle E_{JI} \label{LP_ansatz}
\end{eqnarray}
with 
\begin{eqnarray}
E_{JI} &=& \langle J | \hat{H}_I e^{\hat{T}_I} |I\rangle, \label{LP_ansatz2}
\end{eqnarray}
whose eigenvalues are the desired energies. 
Equation (\ref{LP_ansatz}) is solved for $\hat{T}_I$ and $\bm{E}$ under the C condition,
\begin{eqnarray}
\langle J | e^{\hat{T}_I} | I \rangle = \delta_{JI}. \label{C-condition2}
\end{eqnarray}

The Li--Paldus and $\Delta$CC theories are more similar, but they differ in two respects:\ (i) The Li--Paldus theory
left-multiplies $e^{-\hat{T}_I}$ with the Schr\"{o}dinger equation, while $\Delta$CC theory does not; 
(ii) Many of the $\hat{P}_I$ operators of the $\Delta$CC ansatz are absent in the Li--Paldus ansatz. 
Owing to these differences, the Li--Paldus theory should produce different results from $\Delta$CC theory for the same (degenerate) multireference, 
unless the $\hat{T}_I$ operators are complete, whereupon both theories become equivalent to FCI (see Sec.\ \ref{sec:reverse}).  

As to difference (i), Jeziorski and Monkhorst ``derived''\cite{Jeziorski1981} the left-multiplication of $e^{-\hat{T}_I}$ in Eq.\ (\ref{JM_ansatz2})
by using $\hat{T}_I^\dagger |J\rangle = 0$, which is true only for a CMS. 
We, therefore, judged that the left-multiplication of $e^{-\hat{T}_I}$ be unjustifiable for a GMS and did not adopt it in the $\Delta$CC ansatz.
Furthermore, Eqs.\ (\ref{LP_ansatz})--(\ref{C-condition2}) seem to lack symmetry owing to $e^{-\hat{T}_I}$, which might as well be removed
to reach a more symmetric ansatz, which $\Delta$CC theory is.

The Li--Paldus theory was proven size-extensive by its developers,\cite{LiPaldus2003,LiPaldusIJQC2004} but the proof was 
challenged by Liu {\it et al.}\cite{Liu2026}
See also Lindgren,\cite{Lindgren1985_1,Lindgren1985_2,Lindgren1987} Meissner {\it et al.},\cite{Meissner1989} Mukherjee and coworkers,\cite{MukherjeeAQC1989,MukhopadhyayTCA1991} and Nooijen {\it et al.}\cite{Nooijen_extensivity2005}\ on this issue. 
The divergent conclusions seem to stem from the fact that the size-extensivity of a MRCC theory for a GMS is determined not so much by 
the ansatz but by the nature of the GMS adopted.\cite{Meissner1989} For instance, the Li--Paldus theory reduces to the Jeziorski--Monkhorst theory 
for a CMS and should be size-extensive therein.  
For the most compact GMS such as those for a few bond rearrangements, a metalloenzyme with a small active site, a one-dimensional 
metal, etc., the Li--Paldus theory may well be size-extensive also. 
Hence, it seems less meaningful to discuss size-extensivity or lack thereof of a multireference theory independently of the specification of a GMS. 
See Appendix for more on this issue. 

The Li--Paldus theory has been shown\cite{LiPaldus2003,PaldusLi2003} to be convergent toward FCI. 

\subsubsection{The Kucharski--Bartlett ansatz\label{sec:KucharskiBartlett}}

Kucharski and Bartlett\cite{Kucharski1991} introduced a variant of state-universal MRCC theory for a CMS, 
which differs from the Jeziorski--Monkhorst theory\cite{Jeziorski1981} by the absence and presence, respectively, of the left-multiplication of the $e^{-\hat{T}_I}$ operator. 
Its projected Schr\"{o}dinger equation, therefore, reads,
\begin{eqnarray}
\hat{P}_I   \hat{H}_I e^{\hat{T}_I} |I\rangle = \hat{P}_I \sum_{J=1}^{M} e^{\hat{T}_J} | J \rangle E_{JI}  \label{KB_ansatz}
\end{eqnarray}
with 
\begin{eqnarray}
E_{JI} = \langle J | \hat{H}_I e^{\hat{T}_I} |I\rangle, \label{KB_ansatz2}
\end{eqnarray}
where $|I\rangle$ and $|J\rangle$ refer to members of a CMS of size $M$.
Any excitation within the internal space is suppressed, i.e.,
\begin{eqnarray}
\langle J | \hat{T}_I | I \rangle = 0, \label{KB_Tcondition}
\end{eqnarray}
which is identified as the T condition. The energies are the eigenvalues of $\bm{E}$. 
It is size-extensive and convergent toward FCI. 

The Kucharski--Bartlett ansatz is similar to the $\Delta$CC ansatz as neither
elects to left-multiply $e^{-\hat{T}_I}$ with the Schr\"{o}dinger equation.
Some low-order approximations to the Kucharski--Bartlett theory are said to correspond to 
multireference MBPT (Ref.\ \onlinecite{Kucharski1989}) in much the same way that $\Delta$CC theory is related to $\Delta$MP theory (Sec.\ \ref{sec:HCPT}). This may also be due partly to the absence of $e^{-\hat{T}_I}$. 

However, they differ in three respects:\ (i) The Kucharski--Bartlett theory is limited to a CMS; (ii) $\Delta$CC theory uses the C condition, while the 
Kucharski--Bartlett theory invokes the T condition; (iii) Numerous $\hat{P}_I$ operators in $\Delta$CC ansatz are absent in the Kucharski--Bartlett ansatz.

Differences (i) and (ii) make the Kucharski--Bartlett theory inapplicable to a degenerate multireference.
Owing to difference (iii), the Kucharski--Bartlett theory does not have a single formalism applicable to all references and is not a black-box method. 

\subsubsection{The Meissner--Kucharski--Bartlett ansatz\label{sec:Meissner}}

The Kucharski--Bartlett theory\cite{Kucharski1991}  for a CMS (Sec.\ \ref{sec:KucharskiBartlett}) was extended  to a GMS
 by Meissner {\it et al.}\cite{Meissner1989,Meissner1990}
It still employs the T condition, but the left-multiplication of $e^{-\hat{T}_I}$ is revived, making the Meissner theory 
closer to the Li--Paldus theory and less so to 
$\Delta$CC theory. 
The same ansatz has been independently conceived by Mukhopadhyay and Mukherjee.\cite{Mukhopadhyay1989,Mukhopadhyay1991}
It is doubtful\cite{LiPaldus2003} that it is convergent toward FCI in the absence of the C condition. 

The Meissner theory is size-extensive with respect to extensions of both 
external and internal spaces (see Appendix for details). This remarkable property comes 
at the expense of limiting the model space to a special class of GMS,\cite{Meissner1989,Meissner1990} but not any general one. 
Whether the special class of GMS includes a general degenerate multireference is unclear, but is unlikely. Even if it did, owing to (i) the use 
of the T condition, (ii) the left-multiplication of $e^{-\hat{T}_I}$, and (iii) the lack of numerous $\hat{P}_I$ operators, the Meissner theory
differs from $\Delta$CC theory.




\subsubsection{The TD-CC ansatz\label{sec:TDCC}}

A particular pair of degenerate determinants, which forms an open-shell singlet state, belongs to the aforementioned special class of GMS 
of the Meissner theory\cite{Meissner1989,Meissner1990} (Sec.\ \ref{sec:Meissner}). Balkov\'{a} and coworkers\cite{Balkova1991JCP,Balkova1992CPL,Balkova1993JCP,Balkova1994JCP,Szalay1994JCP,Lutz2018JCP}
introduced TD-CC theory as a practical application of the Meissner theory for these specific degenerate determinants.
Its instance with singles and doubles, i.e., TD-CCSD, was adopted as a ``$\Delta$CCSD'' method for doubly degenerate references in the performance assessment 
study of Damour {\it et al.}\cite{Damour2024}

This TD-CC and present $\Delta$CC ans\"{a}tze both define size-extensive, black-box, degenerate CC theories, but they differ in two respects: (i) The former (TD-CC)
adopts a specific pair of degenerate determinants, while the latter ($\Delta$CC) can handle any and all degenerate
and nondegenerate references. 
Second, TD-CC theory is based on the Meissner ansatz,\cite{Meissner1989,Meissner1990}  which is distinguished from $\Delta$CC theory 
by left-multiplying $e^{-\hat{T}_I}$ and lacking the C condition and numerous projectors. 
It is doubtful\cite{LiPaldus2003} that the TD-CC series is convergent toward FCI.

\subsubsection{The QCC ansatz\label{sec:newSUMRCC}}

On the basis of the foregoing comparison of various MRCC theories, 
we have conceived a new one by removing $\hat{P}_J$ from Eq.\ (\ref{DCC_ansatz}) and at the same time $\hat{P}_I$ from Eq.\ (\ref{DCC_ansatz2}).
The ansatz is a hybrid of the Li--Paldus and Kucharski--Bartlett ans\"{a}tze, but is distinct from either. It can use any GMS including a CMS and degenerate multireference.
We call this theory quasidegenerate coupled-cluster (QCC) theory. 

QCC theory adopts the C condition and forgoes the left-multiplication of $e^{-\hat{T}_I}$.
Its projected Schr\"{o}dinger equation thus reads,
\begin{eqnarray}
\hat{P}_I   \hat{H}_I e^{\hat{T}_I} |I\rangle = \hat{P}_I \sum_{J=1}^{M} e^{\hat{T}_J} | J \rangle E_{JI} \label{newSUMRCCansatz1}
\end{eqnarray}
with 
\begin{eqnarray}
H_{JI} &=& \langle J | \hat{H}_I e^{\hat{T}_I} |I\rangle, \label{newSUMRCCansatz2} 
\end{eqnarray}
and the C condition,
\begin{eqnarray}
S_{JI} &=& \langle J | e^{\hat{T}_I} | I \rangle = \delta_{JI}, \label{C-condition3}
\end{eqnarray}
where $|I\rangle$ and $|J\rangle$ are any reference determinants in a GMS of size $M$. 
The energies are the eigenvalues of $\bm{E}=\bm{S}^{-1}\bm{H}$. 

We have elected not to left-multiply $e^{-\hat{T}_I}$ in Eq.\ (\ref{newSUMRCCansatz1}) because it would alter the equation mathematically without a 
compelling physical justification. 
The present ansatz 
is more symmetric than either of 
the Li--Paldus [Eqs.\ (\ref{LP_ansatz})--(\ref{C-condition2})] or Kucharski--Bartlett  [Eqs.\ (\ref{KB_ansatz})--(\ref{KB_Tcondition})] ansatz, although
for a CMS, QCC theory reduces to the Kucharski--Bartlett theory. 

These defining equations can also be reached by deleting some $\hat{P}_I$ operators from the $\Delta$CC ansatz in Sec.\ \ref{sec:DCC}.
Since size-extensivity seems to strongly depend on the model space (see Appendix), 
we can only speculate at this point that QCC theory is size-extensive for a CMS and degenerate multireference, but not for any GMS.
It is convergent toward FCI, as proven in Sec.\ \ref{sec:reverse}. 
The deletion of the projectors makes the theory non-black-box, but also more suitable for strong-correlation problems, as
the theory takes into account electron correlation within the internal space more fully. 
It should be noticed that $\Delta$CC theory is also applicable to any GMS.

We implement the QCC method in a determinant-based, general-order algorithm in Sec.\ \ref{sec:determinant} and examine its numerical performance in Sec.\ \ref{sec:Calc}.
In this article, we focus on $\Delta$CC theory (and QCC theory to a lesser extent) 
for degenerate multireferences, postponing a fuller investigation of QCC  and $\Delta$CC theories for quasidegenerate multireferences in harder strong-correlation problems.

\subsection{Relationship to EOM-CC theory\label{sec:EOMCC}}

There seems no simple relationship between $\Delta$CC and EOM-CC theories,\cite{Harris1977,Emrich1,Emrich2,Sekino1984,Geertsen1989,Comeau1993,Stanton1993,Kowalski2001EOMCCSDT,kallay_cc,HirataEOM2004,KrylovEOM2008,BartlettWIRE2012} as they are based on rather 
different design philosophies. In EOM-CC theory, the $I$th-state energy, $\tilde{E}_I$, is obtained by solving the following equation
for the linear operator $\hat{{\cal R}}_I$ (not to be confused with a resolvent operator):
\begin{eqnarray}
\hat{P}_0 \hat{H}e^{\hat{T}_0} \hat{{\cal R}}_I |0\rangle = \hat{P}_0 e^{\hat{T}_0}  \hat{{\cal R}}_I | 0\rangle \tilde{E}_I  \label{EOMCC1}
\end{eqnarray}
or equivalently,
\begin{eqnarray}
\hat{P}_0 \left( e^{-\hat{T}_0} \hat{H}e^{\hat{T}_0} \right) \hat{{\cal R}}_I |0\rangle = \hat{P}_0 \hat{{\cal R}}_I | 0\rangle \tilde{E}_I , \label{EOMCC2}
\end{eqnarray}
where $|0\rangle$ is typically, but not necessarily, the neutral, singlet, ground-state determinant, which is nondegenerate, and $\hat{P}_0$ 
is the projector onto the determinant space spanned by $(1+\hat{T}_0)|0\rangle$. 
The amplitudes of $\hat{T}_0$ are, in turn, pre-determined by
single-reference CC theory by solving the following equation for $\hat{T}_0$: 
\begin{eqnarray}
\hat{P}_0 \left( e^{-\hat{T}_0} \hat{H}e^{\hat{T}_0}\right)  |0\rangle = \hat{P}_0 | 0\rangle \tilde{E}_0 , \label{CC_groundstate}
\end{eqnarray}
which is equivalent to Eq.\ (\ref{CC_ansatz2}) for $I=0$.
Since $\hat{T}_0$ and $\hat{{\cal R}}_I$ commute, acting $\hat{{\cal R}}_I$ from the left on both sides of Eq.\ (\ref{CC_groundstate}) and subtracting it from Eq.\ (\ref{EOMCC2}),
we arrive at an equation-of-motion method\cite{Dunning1967,Rowe1968,Shibuya1970,Simons1973,Simons1976} with the CC effective Hamiltonian, $e^{-\hat{T}_0} \hat{H}e^{\hat{T}_0}$,
\begin{eqnarray}
\hat{P}_0 \left[ \left( e^{-\hat{T}_0} \hat{H}e^{\hat{T}_0}\right) , \hat{{\cal R}}_I \right] |0\rangle = \hat{P}_0 \hat{{\cal R}}_I | 0\rangle \tilde{\omega}_I , \label{EOMCC3}
\end{eqnarray}
where $\tilde{\omega}_I = \tilde{E}_I - \tilde{E}_0$. 

At this point, there is considerable latitude in choosing the form of $\hat{{\cal R}}_I$. 
When $\hat{{\cal R}}_I$ is an excitation operator of the same rank as $\hat{T}_0$, Eq.\ (\ref{EOMCC3}) is connected (and thus linked), postulating a size-extensive and intensive method for 
the total energy of and excitation energy to the $I$th excited state.\cite{Harris1977,Emrich1,Emrich2,Sekino1984,Geertsen1989,Comeau1993,Stanton1993,Kowalski2001EOMCCSDT,kallay_cc,HirataEOM2004} When $\hat{{\cal R}}_I$ is an ionization or electron-attachment operator, this ansatz defines ionization-potential 
(IP) or electron-attachment (EA) EOM-CC theory,\cite{Stanton_ip,Bartlett_ip,Stanton_ip2,MusialIP2003,Gour2005,kamiya_1,nooijen_eom,MusialEA2003,kamiya_2} respectively.
The $\hat{{\cal R}}_I$ can be a double-ionization operator, postulating the double-ionization-potential (DIP) EOM-CC theory.\cite{Sattelmeyer2003,Kus2011,Shen2013,Gururangan2025} 
The double-electron-attachment (DEA) EOM-CC theory\cite{Musial2011,Shen2013,Gulania2021} has also been developed. The $\hat{{\cal R}}_I$ can even incorporate a spin-flip operation, leading to various spin-flip EOM-CC theories,\cite{Krylov2001,Slipchenko2002,Levchenko2004,Manisha2024} which are shown to have superior performance.
To maintain the size-extensivity and intensivity, the rank of $\hat{{\cal R}}_I$ needs to be chosen carefully.\cite{HirataTCA}

The size-extensive and intensive ansatz of EOM-CC theory (i.e., the ranks of $\hat{T}_0$ and $\hat{{\cal R}}_I$ are the same) is naturally derived by the first-order time-dependent perturbation (linear response) theory applied
to single-reference CC theory.\cite{Monkhorst1977,Koch1990,Koch1990_2,Rico1993} In this picture, $e^{\hat{T}_0}$ describes the electron correlation in, typically, a neutral, singlet, nondegenerate ground state, whose wave function has
an exponential structure, while $\hat{{\cal R}}_I$ takes care of an excitation process, dominated by just a few destination determinants, which are often degenerate and automatically linearly combined in a spin- and spatial-symmetry-adapted manner. 

In $\Delta$CC theory, in contrast, massive electron rearrangements caused by excitations, ionizations, electron attachments, spin flips, etc.\ are described by 
a reference wave function, which is a single determinant or a linear combination
of degenerate determinants with appropriate electron occupancies. 
Electron correlation thereof then becomes a small perturbation captured accurately
by an exponential operator. 

Generally, $\Delta$CC and EOM-CC theories are complementary to each other as follows (see their numerical tests in Sec.\ \ref{sec:Calc}):

(1) {\it Accuracy:} $\Delta$CC theory treats each state on an equal footing and is applicable to all possible states in one and the same uniform formalism. It is only slightly dependent on the spin-orbitals that come from the orbital reference,\cite{Harris1977} but otherwise it does not rely on any particular state or the fidelity of its description. 
The accuracy of $\Delta$CC theory is much more uniform and primarily determined by how dominant the reference determinants are in the exact wave functions, i.e., 
the degree of strong correlation.
EOM-CC theory, on the other hand, relies on 
the accurate CC description of the nondegenerate ground state and reaches excited or other states by simulated one-photon absorption, ionization, etc.\ processes. As a result, EOM-CC theory cannot describe multi-electron processes as accurately as one-electron processes and does not even have roots with excitation ranks exceeding the truncation order of $\hat{{\cal R}}_I$. However, a method that overcomes this weakness has been developed by Mosquera.\cite{Mosquera2025}

(2) {\it Applicability:} $\Delta$CC theory can home in on any state out of order with any numbers of $\alpha$- and $\beta$-spin electrons, spin multiplicities, or spatial symmetries. 
It has roots for all possible states spanned by a basis set. 
In contrast, EOM-CC theory has a smaller number of roots, which are determined from the lower-energy side using a trial-vector algorithm.\cite{Bartlett1972,Bartlett1973,Davidson1975,Hirao1982} Although this algorithm is robust and efficient, it can miss high-spin states or interior states such as core excited or ionized states. 
(There are, however, ways to go around this shortcoming.\cite{Zuev2015,Peng2015}) 
Most applications of EOM-CC theory are spectral simulation of one-photon absorption, ionization, etc., and its selective superior performance for transitions with sizable one-photon transition probabilities
may be considered a practical advantage. EOM-CC theory also naturally defines oscillator strengths for the transitions from the ground state, which may prove more cumbersome with
$\Delta$CC theory, where the wave functions of two different references are not necessarily orthogonal to each other.

(3) {\it Computational Cost:} $\Delta$CC theory solves nonlinear $T$-amplitude equations for any state, and suffers from convergence difficulties more frequently than 
EOM-CC theory, whose algorithms are exceedingly stable. $\Delta$CC theory also encounters complex energies\cite{Zivkovic1978,Paldus1993,Piecuch1994} more often than EOM-CC theory. (The latter
can also have complex energies in principle, but almost never does in practice.) $\Delta$CC theory has the same cost scaling per state as EOM-CC theory of the same rank,
e.g., $O(n^4)$ for CCS, $O(n^6)$ for CCSD, etc., where $n$ is the number of spin-orbitals. However,
the prefactor of the cost function of $\Delta$CC theory is greater than that of EOM-CC theory at least
in the primitive optimal-scaling algorithm of the former implemented in this study.

(4) {\it Development Cost:} In the existing paradigm, one must separately develop the CC series for ground states, the EOM-CC series for excited states, 
the IP-EOM-CC series for ionized states, the EA-EOM-CC series for electron-attached states,
the DIP-EOM-CC series for doubly ionized states, and so on. For $\Delta$CC theory, in contrast,
 just one formulation and code at a given rank suffice in all possible cases. Streamlined formulations, advanced optimizations, and proven algorithms already available 
 for single-reference CC theory for ground states 
 may be harnessed to achieve efficient and robust $\Delta$CC implementations quickly for all types of states. 
 In this sense, $\Delta$CC theory may prove economical in terms of the total development cost. 
Recall that the cost of {\it ab initio} theory developments has become so great that expert systems (a type of artificial intelligence) have been almost mandatory for decades.\cite{Hirata06:TCAReview,tce} 

\subsection{Relationship to MBGF theory}

Several groups have argued that a coupled-cluster version of MBGF theory\cite{Linderberg65,Hedin,linderbergohrn,pickup,paldusAQC1975,cederbaumacp,simonsrev,schirmer1982,ohrnborn,oddershede,ortiz_aqc,Ohrn_review2010,OrtizWire,OrtizDyson,Ortiz2022,deltamp,Hirata2017}
is IP- and EA-EOM-CC theories.\cite{Gunnarsson1978,nooijen_gf1,nooijen_gf2,Meissner1993,Kowalski1,Kowalski2,Kowalski3} Here, we propose an alternative viewpoint in which an equally natural coupled-cluster extension of MBGF theory
is $\Delta$CC theory. 

The $\Delta$MP$n$ energy differences (as described in Sec.\ \ref{sec:HCPT}) between the nondegenerate neutral and ionized (or electron-attached) states 
are equal to the $n$th-order MBGF [MBGF($n$)] method
in the diagonal, frequency-independent approximation for $n \leq 3$.\cite{deltamp,pickup,szabo,Hirata2017} At $n=4$, $\Delta$MP$n$ and MBGF($n$) 
begin to diverge from each other with the difference being the so-called semireducible and linked-disconnected diagrams.\cite{deltamp,Hirata2017}
Nevertheless, as $n \to \infty$, the $\Delta$MP$n$ energy differences converge at exact IPs and EAs (unless they diverge) for both Koopmans and satellite states.
This is because the semireducible and linked-disconnected diagrams recuperate the effects of off-diagonal elements and frequency-dependence of the self-energy, respectively; 
the diagonal, frequency-independent ``approximation'' is not so much an approximation to MBGF but another converging ansatz. 

MBGF theory also converges at exactness for most Koopmans and some satellite states.\cite{deltamp,Hirata2017} 
Surprisingly, for numerous other states, MBGF theory converges at wrong limits.\cite{Hirata_PRA2024}
The cause of this nonconvergence has been identified\cite{Hirata_PRA2024} as the nonanalyticity of Green's functions in many frequency domains.
Therefore, Feynman--Dyson perturbative MBGF theory is fundamentally flawed, whereas $\Delta$MP theory is sound as the latter
directly expands energies and wave functions, which are analytic, instead of Green's functions, which are nonanalytic. 

One way of generalizing a perturbation theory to a coupled-cluster theory is to perform an infinite partial summation of diagrams. 
Summing all ladder and ring diagrams of MBPT leads to D-MBPT($\infty$)
of Bartlett and Shavitt,\cite{Bartlett1977,Bartlett1978} which is identified as linearized coupled-cluster doubles (LCCD).\cite{shavitt} 
Similar infinite summations of ladder and ring diagrams of MBGF have been performed under such names as the Tamm--Dancoff approximation,\cite{Linderberg67,Purvis1974,Schirmer1978,Walter1981,Hirata_PRA2024}
Brueckner--Hartree--Fock method,\cite{Day1967,Baldo1999} and T approximation.\cite{Kadanoff_book} They may be regarded as
linearized-coupled-cluster Green's functions.
Since $\Delta$CC theory is an infinite partial summation of $\Delta$MP diagrams (see Sec.\ \ref{sec:HCPT}), 
$\Delta$CC theory can also be considered a coupled-cluster Green's function in the diagonal, frequency-independent ansatz
for nondegenerate and degenerate references. It is convergent toward FCI for all states and superior to either Feynman--Dyson perturbative MBGF or $\Delta$MP theory. 


A performance comparison of the $\Delta$CC, $\Delta$MP,  IP- and EA-EOM-CC, and MBGF methods will be presented in Sec.\ \ref{sec:Calc}.

\subsection{Relationship to HF theory\label{sec:HF}}

According to the Thouless theorem,\cite{Thouless1960} a coupled-cluster-singles (CCS) wave function is a single Slater determinant,
\begin{eqnarray}
|\Psi\rangle = e^{\hat{T}_1}|\Phi\rangle,
\end{eqnarray}
where both $|\Psi\rangle$ and $|\Phi\rangle$ designate a single determinant composed of a respective set of orthogonal spin-orbitals,
and $\hat{T}_1$ is the one-electron excitation operator. 

One can define a generalized HF theory for a nondegenerate reference in at least two ways. One way is to 
vary spin-orbitals that constitute $|\Phi\rangle$ or, equivalently, to vary $\hat{T}_1$ for a fixed $|\Phi\rangle$, so that its expectation value of $\hat{H}$ is minimized, 
\begin{eqnarray}
\text{min}\,\frac{\langle \Phi | e^{\hat{T}_1^\dagger} \hat{H} e^{\hat{T}_1} | \Phi \rangle}{\langle \Phi | e^{\hat{T}_1^\dagger} e^{\hat{T}_1} | \Phi \rangle}.
\end{eqnarray}
This is identified as the variational coupled-cluster (VCC) method\cite{Arponen1983,Pal1983,Haque1986,BartlettNoga1988,Kutzelnigg1998,Xian2005,CooperKnowles2010,Robinson2011,conjecture,BlackKnowles2018} limited to singles (VCCS)
(see also Refs.\ \onlinecite{Harris1977,NakatsujiPseudoorbital1977,NakatsujiPseudoorbital1978}). 
It is well known that this ansatz cannot be easily extended to VCCSD and higher owing to its nonterminating
working equations.\cite{Arponen1983,Pal1983,Haque1986,BartlettNoga1988,Kutzelnigg1998,Xian2005,CooperKnowles2010,Robinson2011,conjecture,BlackKnowles2018} 
Therefore, we will not attempt to generalize this ansatz to a degenerate multireference except to show in the supplementary material 
that its nonterminating equations are still tractable at the VCCS level.

Another way to define a generalized HF theory is to demand that the CCS wave function satisfy the Schr\"{o}dinger equation within the determinant space spanned by $(1+\hat{T}_1)|\Phi\rangle$:
\begin{eqnarray}
\langle \Phi | \hat{H} e^{\hat{T}_1} | \Phi\rangle &=&  \langle \Phi |E e^{\hat{T}_1} | \Phi \rangle, \\
\langle \Phi^a_i | \hat{H} e^{\hat{T}_1} | \Phi \rangle &=&  \langle \Phi^a_i |E e^{\hat{T}_1} | \Phi \rangle, 
\end{eqnarray}
which is identified as the projection CCS method. This is distinct from the usual, variational HF theory unless the orbital reference is already a HF solution (in which case, $\hat{T}_1=0$
is a trivial solution). The projection CCS energy is bounded from below by the corresponding, variational HF energy.\cite{Harris1977} 
It has the advantage of being straightforwardly extensible to CCSD and higher. 
If we adopt this projection CCS as a generalized, projection HF theory, $\Delta$CCS is then its natural extension 
to a degenerate multireference. Its performance will be discussed in Sec.\ \ref{sec:Calc}.

The sum of the zeroth- and first-order energies [Eqs.\ (\ref{MP0}) and (\ref{MP1})] of single-reference MP theory 
recovers the HF energy expression. 
Therefore, the third way of defining a generalized HF theory is to identify it with the MP1 method. Then, 
 it may be argued that the $\Delta$MP1 method is another projection HF theory for degenerate and nondegenerate references. 
It is a ``projection'' HF theory because the orbitals in the reference determinants are predetermined and not varied. We shall examine its performance in Sec.\ \ref{sec:Calc}.


\subsection{Relationship to FCI theory\label{sec:reverse}}

The $T$-amplitudes of the single-reference {\it full} CC method can be restored uniquely and unambiguously
from the excitation amplitudes ($C$-amplitudes) of FCI. This process is called the cluster analysis, and its feasibility decides whether a CC theory forms a converging series 
toward FCI.  It has been established by Li and Paldus\cite{LiPaldus2003,PaldusLi2003} that the cluster analysis is indeed feasible for their MRCC theory, which is 
therefore exactly convergent partly thanks to the C condition (see below).

Since both {\it full} $\Delta$CC and {\it full} QCC methods are identical to {\it full} Li--Paldus MRCC method for the same GMS
(their truncated instances are different from one another), 
$\Delta$CC and QCC theories constitute converging series of approximations
toward FCI. 

Below, we document the cluster analysis of the full $\Delta$CC (full QCC) method.\cite{LiPaldus2003,PaldusLi2003}
Let $|\Psi_I\rangle$ be the normalized, exact (FCI) wave function for the $I$th state,
\begin{eqnarray}
|\Psi_I\rangle = \sum_{J}^{\text{int.}}  |J\rangle \,{C}_{JI} + \sum_{A}^{\text{ext.}}  |A \rangle \,C^\prime_{AI}, \label{FCI}
\end{eqnarray}
where $|J\rangle$ and $|A\rangle$ designate an internal and external determinant, respectively, and $\bm{C}$ and $\bm{C}^\prime$
are the FCI coefficients. Writing the inverse of the $M$-by-$M$ matrix $\bm{C}$ (where $M$ is the degree of degeneracy) as $\bm{C}^{-1}$, 
we have
\begin{eqnarray}
\sum_I^{\text{int.}} |\Psi_I\rangle \,(\bm{C}^{-1})_{IJ}  &=& |J\rangle + \sum_I^{\text{int.}}\sum_{A}^{\text{ext.}}  |A \rangle \,(\bm{C}^\prime)_{AI} (\bm{C}^{-1})_{IJ}  \nonumber\\
 &=& |J\rangle + \sum_A^{\text{ext.}} |A\rangle\langle A| \hat{C}_1 + \hat{C}_2 + \dots |J \rangle, \label{FCI2}
\end{eqnarray}
where $I$ and $A$ run over all internal (``int.'')\ and external (``ext.'')\ determinants, respectively, and 
$\hat{C}_n$ is the $n$-electron excitation operator, whose amplitudes are given by $\bm{C}^\prime\bm{C}^{-1}$. 
The matrix $\bm{C}$ is invertible if the FCI wave functions are correctly one-to-one mapped onto determinants, and this mapping is facilitated by trial-vector algorithms,\cite{Bartlett1972,Bartlett1973,Davidson1975,Hirao1982} in practice.

On the other hand, as per Eq.\ (\ref{wavefunction}), the full $\Delta$CC wave function is written as
\begin{eqnarray}
|\Psi_I\rangle &=& \sum_{J}^{\text{int.}}  e^{\hat{T}_J} |J\rangle \,{C}_{JI} \nonumber\\
&=&  \sum_{J}^{\text{int.}} |J\rangle \,{C}_{JI} + \sum_{J}^{\text{int.}} \left( \hat{T}_J + \hat{T}_J^2/2! +\dots \right) |J\rangle \,{C}_{JI}, 
\end{eqnarray}
in which $\hat{P}_J$ is absent because this is the {\it full} $\Delta$CC method. Comparing this equation with Eq.\ (\ref{FCI}), 
we see that the $\bm{C}$ of the former ($\Delta$CC) is the same as the $\bm{C}$ of the latter (FCI) {\it if and only if} the C condition [Eq.\ (\ref{DCC_ansatz2})] is imposed.

This last point will become clearer when the above equation is multiplied by $\bm{C}^{-1}$ from the right, leading to
\begin{eqnarray}
\sum_I^{\text{int.}} |\Psi_I\rangle \,(\bm{C}^{-1})_{IJ}  &=& \ |J\rangle 
+ \sum_{K\neq J}^{\text{int.}}  |K \rangle \langle K | \hat{T}_J + \hat{T}_J^2/2!  +\dots |J\rangle \nonumber\\
&& + \sum_{A}^{\text{ext.}}  |A \rangle \langle A | \hat{T}_J + \hat{T}_J^2/2!  +\dots |J\rangle  \label{DCCtoFCI} \\
&=& \ |J\rangle + \sum_{A}^{\text{ext.}}  |A \rangle \langle A | \hat{T}_J + \hat{T}_J^2/2!  +\dots |J\rangle, \nonumber\\
\end{eqnarray}
where the C condition, $\langle K | \hat{T}_J + \hat{T}_J^2/2! + \dots | J\rangle = \delta_{KJ}$, is invoked in the last equality. 
The final expression mirrors the last line of Eq.\ (\ref{FCI2}), implying
\begin{eqnarray}
\hat{C}_1 + \hat{C}_2 + \hat{C}_3 + \dots = \hat{T}_J + \hat{T}_J^2/2! + \hat{T}_J^3/3!  +\dots.
\end{eqnarray}
Decomposing $\hat{T}_J$ by excitation ranks, 
\begin{eqnarray}
\hat{T}_J &\equiv&  \hat{T}_1 + \hat{T}_2 + \hat{T}_3 + \dots, 
\end{eqnarray}
we can determine the exact $T$-amplitudes from FCI's $C$-amplitudes recursively,
\begin{eqnarray}
\hat{T}_1 &=&  \hat{C}_1, \\  
\hat{T}_2 &=&  \hat{C}_2 - \hat{T}_1^2/2!, \\  
\hat{T}_3 &=&  \hat{C}_3 - \hat{T}_2\hat{T}_1 - \hat{T}_1^3/3!,   
\end{eqnarray}
and so on. 

This proves the equivalence of the FCI method and the full $\Delta$CC or full QCC method and hence the exact convergence of the $\Delta$CC and QCC hierarchies.
The foregoing proof is originally due to Paldus and Li.\cite{PaldusLi2003} 

\section{Computer implementations\label{sec:implementation}}

\subsection{Determinant-based, general-order algorithm\label{sec:determinant}}

In a determinant- or string-based algorithm,\cite{knowles_fci,hirata_cc} a Slater determinant is computationally represented by
a pair of $\alpha$- and $\beta$-strings of bits, with the $n$th bit storing the $\alpha$- or $\beta$-spin electron occupation number of the $n$th orbital. 
A wave function, which is a linear combination of 
these determinants, can then be specified by an array of expansion coefficients indexed by these string pairs. 
The action of any quantum-mechanical operator such as $\hat{H}_I$, $\hat{T}_I$, $e^{\hat{T}_I}$, and $\hat{P}_I$ on a wave function is described computationally
by faithfully executing the second-quantized definition of each operator as a series of bit manipulations. 

In this way, virtually any {\it ab initio} electron-correlation
theory that has a well-defined determinantal ansatz can be implemented in a general-order algorithm by making relatively simple modifications to
 a string-based FCI code.\cite{knowles_fci} The computational cost is also that of a FCI calculation regardless of the order of the hierarchical theory, and its intended utility 
is to provide benchmark data for the smallest systems that help characterize the theory's behavior and performance or verify more efficient implementations. 
General-order CI,\cite{knowles_fci} CC,\cite{hirata_cc,kallay_cc,olsen_cc} EOM-CC,\cite{kallay_cc,hirata_eomcc,hirata_ipeomcc} combined CC and MBPT,\cite{Hirata_CCPT,Hirata_CCPT_erratum} MBPT,\cite{knowles_mp} $\Delta$MP,\cite{deltamp} and MBGF\cite{deltamp,Hirata2017} have been developed by this strategy. 
For pioneering symbolic derivations and implementations of MRCC theories, the reader is referred to Janssen and Schaefer,\cite{Janssen1991} Li and Paldus,\cite{LiPaldus1994} Abrams and Sherrill,\cite{Abrams2005} Lyakh {\it et al.},\cite{Lyakh2005} and Evangelista {\it et al.}\cite{Evangelista2006}

$\Delta$CC theory has a well-defined determinantal ansatz and can thus be implemented into a general-order, determinant-based algorithm largely reusing 
computational kernels of the determinant-based CC code\cite{hirata_cc} as part of the {\sc polymer} program.\cite{polymer} 
A general-order QCC method (Sec.\ \ref{sec:newSUMRCC}) can be realized simultaneously in the same code.\cite{polymer} 
The iterative procedure of the $n$th-order $\Delta$CC method for $M$-fold degenerate references is as follows:

(1) Generate initial guesses of the $T$-amplitudes for all degenerate references. (In our implementation, they are either zero or the exact values recovered from FCI as per Sec.\ \ref{sec:reverse}.)

(2) Form $e^{\hat{T}_I} |I\rangle$ for all degenerate references as linear combinations of determinants.\cite{hirata_cc} 

(3) Form $\hat{H}_I e^{\hat{T}_I} |I\rangle$ for all degenerate references as linear combinations of determinants.\cite{hirata_cc} (While the appearance of the normal-ordered $\hat{H}_I$ differs from one degenerate reference to another, its content is the same, and just one common subroutine 
that acts the {\it ab initio} $\hat{H}$ on a wave function suffices.)

(4) Build the $M$-by-$M$ matrices $\bm{H}$ and $\bm{S}$ of Eqs.\ (\ref{H}) and (\ref{S}) in the internal space. 
Their elements across two determinants differing by more than $n$ spin-orbitals are zeroed by 
the $\hat{P}_I$ operators in Eqs.\ (\ref{H}) and (\ref{S}). (In the QCC method, no elements need to be zeroed.) 

(5) Invert $\bm{S}$ by the LU decomposition, and then form $\bm{E} = \bm{S}^{-1}\bm{H}$. (Even though $\bm{S}=\bm{1}$ at convergence, this is not the case
during the iterative process and it must not be assumed that $\bm{E} = \bm{H}$.) Determine and print the eigenvalues of $\bm{E}$. (Complex eigenvalues are 
detected occasionally in cases of convergence difficulties.\cite{Zivkovic1978,Paldus1993,Piecuch1994})

(6) Compute the residuals, i.e., the differences between the left- and right-hand sides of Eq.\ (\ref{DCC_ansatz}) in the external space or those of Eq.\ (\ref{DCC_ansatz2}) in the internal space, both
 computable by the quantities obtained in steps (2) and (3). 
The $\hat{P}_J$ operator sets $e^{\hat{T}_J}|J\rangle$ to zero if the external determinant 
is more than $n$ spin-orbitals different from $|J\rangle$. (In the QCC method, this zeroing is suppressed.) 

If the norm of the residuals is less than a preset threshold, the convergence is achieved and the program halts; otherwise, update the $T$-amplitudes for all degenerate references and go back to step (2). 

The standard MBPT-style update, which adds the residuals (divided by the zeroth-order energy differences) to the most recent $T$-amplitudes in the  internal (external) space, 
works reasonably well, but not as well as the same
in the single-reference CC methods for the ground states. In cases of convergence difficulty, Pulay's direct inversion in the iterative subspace (DIIS)\cite{PulayDIIS} is found effective. 
In the full $\Delta$CC method, when the exact $T$-amplitudes are used as initial guesses, convergence is achieved instantly. 

\subsection{Algebraic, optimal-scaling, order-by-order algorithm\label{sec:tce}}

Optimal-scaling algorithms of $\Delta$CCS and $\Delta$CCSD are implemented based on
the algebraic equations given in Secs. \ref{sec:DCCS} and \ref{sec:DCCSD}. The equations have been derived and  
transformed  by {\sc tce}\cite{tce,tce2,nwchem}  into computational sequences consisting of unary matrix additions and binary matrix multiplications after the optimizations known as the strength reduction and factorization.
These optimizations define the so-called intermediates, lowering the size-dependence of costs of $\Delta$CCS and $\Delta$CCSD to the 
optimal $O(n^4)$ and $O(n^6)$, respectively, with the number of spin-orbitals ($n$), not counting the $O(n^5)$ integral-transformation step. 

Note that the $T$-amplitude equations in Secs.\ \ref{sec:DCCS} and \ref{sec:DCCSD} 
are left disconnected or even unlinked, and the cancellation of the latter occurs numerically. This algorithm choice has been made because 
the presence of these unlinked terms has no impact on the size-dependence of cost, while a complete removal of disconnected terms is neither possible nor necessary. 
(These disconnected terms are linked, and do not violate size-extensivity. 
See Sec.\ \ref{sec:size} and Appendix.)

The computational sequences are then converted by debug-mode {\sc tce} to the {\sc fortran90} subroutines as part of the {\sc polymer} program\cite{polymer} 
that evaluate the energy and $T$-amplitude equations.  
In this pilot implementation, our primary concerns have been the correctness of the formulas and computer codes, and hence the codes do not exploit spin, spatial, or index-permutation symmetries. 
They are also for serial executions only. Hence, while their size-dependence of cost is optimal, their wall-clock-time execution speeds are slow, leaving much room for improvement. 

Overall, the iterative procedure of the algebraic $\Delta$CCS and $\Delta$CCSD algorithms is essentially 
the same as the determinant-based algorithm outlined above since the disconnected and unlinked terms are left intact to be canceled numerically. 
Only and major differences exist in the way in which the matrix elements and residuals are evaluated in the correctly scaling subroutines synthesized by {\sc tce}.  
Below, we document the iterative procedure of the algebraic algorithm, taking the $\Delta$CCSD method for $M$-fold degenerate references as an example.
(The non-black-box QCC method cannot be implemented in this manner.)

\begin{figure}
\includegraphics[scale=0.5]{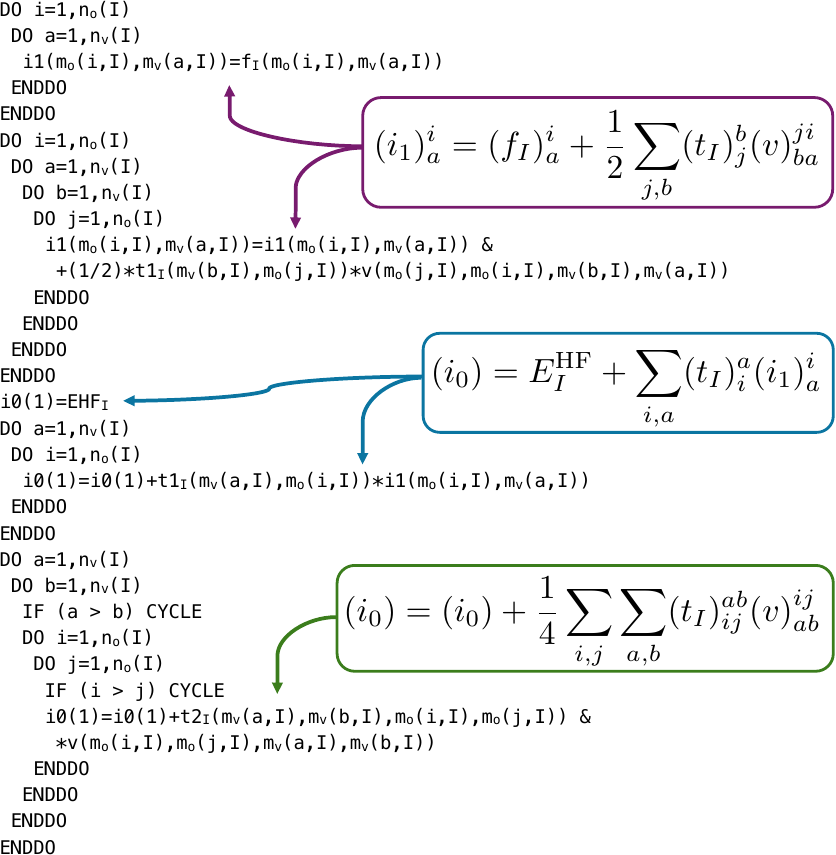}
\caption{The {\sc fortran90} pseudocode for the diagonal $\Delta$CCSD energy for the $I$th degenerate reference determinant [Eq.\ (\ref{DCCSD3})]. The element of array \texttt{i0(1)} returns the energy, array \texttt{i1} stores an intermediate formed by the strength reduction and factorization, arrays \texttt{t1}$_\texttt{I}$ and \texttt{t2}$_\texttt{I}$ are, respectively, the $T_1$- and $T_2$-amplitudes, variable \texttt{EHF}$_\texttt{I}$ and array \texttt{f}$_\texttt{I}$ are the $I$-dependent HF energy and Fock matrix, and array \texttt{v} is the
anti-symmetrized two-electron integrals. Variables \texttt{n}$_\texttt{o}$\texttt{(I)} and \texttt{n}$_\texttt{v}$\texttt{(I)} are the numbers of occupied and virtual spin-orbitals in the $I$th determinant, while 
arrays \texttt{m}$_\texttt{o}$\texttt{(i,I)} and \texttt{m}$_\texttt{v}$\texttt{(a,I)} store the lists of occupied and virtual spin-orbitals of the $I$th determinant. The actual code is synthesized by {\sc tce}.}
\label{fig:code}
\end{figure}

(1) For each ($I$th) degenerate reference, store the numbers [$n_\text{o}(I)$ and $n_\text{v}(I)$] and lists [$m_\text{o}(i,I)$ for $i = 1, \dots, n_\text{o}(I)$ and $m_\text{v}(a,I)$ for $a = 1, \dots, n_\text{v}(I)$]  of occupied and virtual spin-orbitals. 

(2) For each ($I$th and $J$th) degenerate reference pair, evaluate and store the (nonzero) parities, i.e.,  $\langle I^{ab}_{ij} | J^{cd}_{kl} \rangle$, $\langle I^{ab}_{ij} | J^{c}_{k} \rangle$, $\langle I^{a}_{i} | J^{cd}_{kl} \rangle$, and $\langle I^{a}_{i} | J^{c}_{k} \rangle$. See Eqs.\ (\ref{DCCSD7}) and (\ref{DCCSD8}).

(3) For each ($I$th) degenerate reference, evaluate and store the $I$-dependent HF energy, $E^\text{HF}_I$, and Fock matrix elements, $(f_I)^p_q$, by Eqs.\ (\ref{EHF}) and (\ref{Fock}).

(4) Generate initial guesses of the $T$-amplitudes for all degenerate references. (They are zero in the present implementation.)

(5) For each ($I$th) degenerate reference, evaluate the diagonal $\Delta$CCSD energy $\langle I | \hat{H}_I e^{\hat{T}_I}|I\rangle$ [Eq.\ (\ref{DCCSD3})] 
using the {\sc tce}-synthesized subroutine with $E^\text{HF}_I$ and $(f_I)^p_q$ formed in step (3). 

(Figure \ref{fig:code} illustrates the structure of the pilot code synthesized by {\sc tce} for the diagonal $\Delta$CCSD energy. 
While {\sc tce} correctly identifies a common factor as the intermediate $(i_1)^i_a$ and its synthesized code has the correct cost scaling, 
the nested loops over gapped lists of occupied or virtual spin-orbitals pose an efficiency bottleneck.)

(6) For each ($I$th) degenerate reference, form the left-hand sides of the $T$-amplitude equations, i.e., Eqs.\ (\ref{DCCSD1}) and (\ref{DCCSD2}), using 
the {\sc tce}-synthesized subroutines. Populate the off-diagonal elements of $\bm{H}$ with them, while its diagonal elements have been computed in step (5).   

(7) For each ($I$th) degenerate reference, form the exponential wave functions, i.e., Eqs.\ (\ref{DCCSD3}) and (\ref{DCCSD4}), using 
the {\sc tce}-synthesized subroutines. Populate the off-diagonal elements of $\bm{S}$ with them, while its diagonal elements are unity. 
Invert $\bm{S}$ by the LU decomposition, and form $\bm{E} = \bm{S}^{-1} \bm{H}$. Determine and print the eigenvalues of $\bm{E}$.

(8) For each ($I$th) degenerate reference, evaluate the right-hand sides of the $T$-amplitude equations, i.e., Eqs.\ (\ref{DCCSD7}) and (\ref{DCCSD8}), using 
parities determined in step (2) and the exponential wave functions and $\bm{E}$ formed in step (7). Compute the residuals. 
 In the external space, they are the differences between the left- and right-hand sides of the $T$-amplitude equations, i.e., Eq.\ (\ref{DCCSD1}) minus Eq.\ (\ref{DCCSD7}) or Eq.\ (\ref{DCCSD2}) minus Eq.\ (\ref{DCCSD8}); in the internal space, they are $\bm{S}-\bm{1}$. 

If the norm of the residuals is less than a preset threshold, the convergence is achieved. Otherwise, update the $T$-amplitudes for all degenerate references in the same way 
as in the determinant-based algorithm, and go back to step (5).

\section{Numerical Tests\label{sec:Calc}}

The $\Delta$CC and QCC methods through the FCI limits have been applied to the same group of small molecules previously used in the 
general-order CC,\cite{hirata_cc} EOM-CC,\cite{hirata_eomcc} and IP- and EA-EOM-CC benchmarks.\cite{hirata_ipeomcc} They are compared with
the general-order CI and EOM-CC methods through the FCI limits as well as with the general-order 
$\Delta$MP (Ref.\ \onlinecite{deltamp}) and MBGF methods\cite{Hirata2017,Hirata_PRA2024} through the nineteenth order. 
The CI and CC methods are characterized by the truncation order ($k$) of the cluster (and linear) excitation operators, specified by 
suffixes:\ S ($k=1$), SD ($2$), SDT ($3$),
SDTQ ($4$), SDTQP ($5$), SDTQPH ($6$), SDTQPHS ($7$), and SDTQPHSO ($8$).\cite{note_acronym} See Refs.\ \onlinecite{hirata_cc,hirata_eomcc,hirata_ipeomcc,Hirata_PRA2024}
for the geometries, basis sets, whether the frozen-core and/or frozen-virtual approximations are invoked, and other details.

In all cases, the orbital reference is the determinant with an equal number of $\alpha$- and $\beta$-spin electrons obeying the aufbau principle, whose orbitals are determined by
the spin-restricted HF method. The zeroth-order energy ($E_{II}^{(0)}$) of the $I$th determinant, by which the degeneracy is judged, is the sum of the occupied HF orbital energies
of each orbital reference.

When evaluating the performance, the computational cost and its size-dependence should be factored in. Although 
the costs of the determinant-based algorithms scale exponentially with the number of spin-orbitals ($n$) regardless 
of the method's rank, the intrinsic size-dependencies of the costs are as follows:\ $\Delta$CCS scales as $O(n^4)$, $\Delta$CCSD as $O(n^6)$, $\Delta$CCSDT as $O(n^8)$, or 
the $k$th-order $\Delta$CC as $O(n^{2k+2})$. The CI, EOM-CC, and QCC methods scale identically as the $\Delta$CC method of the same rank. 
The $\Delta$MP2, $\Delta$MP3, and $\Delta$MP4 methods scale as $O(n^4)$, $O(n^6)$,
and $O(n^7)$, respectively, if the algorithms are thoroughly optimized.\cite{shavitt} An MBGF method should have 
the same scaling as the $\Delta$MP method of the same perturbation order.
In the above, the $O(n^5)$ integral-transformation step is excluded  from the cost.

Calculations were carried out with the {\sc polymer} program.\cite{polymer} The numerical data are available in the supplementary material.

\subsection{Excited states}

\subsubsection{CH$^+$\label{sec:CH+}}

\begin{figure}
\includegraphics[scale=0.65]{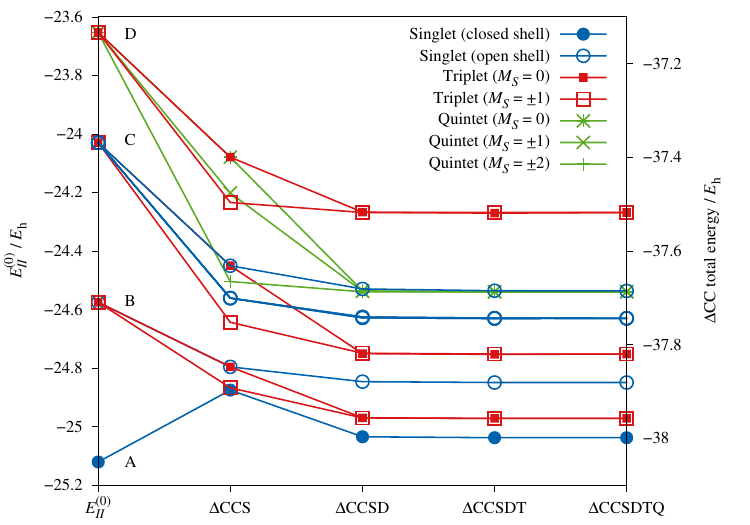}
\caption{The total energies of the ground and excited states of CH$^+$ ($r_\text{CH}$ = 1.131\,\AA) obtained by the $\Delta$CC methods and 6-31G** basis set with 
the highest virtual and lowest occupied orbitals kept frozen.\cite{hirata_eomcc}
The determinant references with
the numbers of the $\alpha$- and $\beta$-spin electrons being $(3,3)$, $(4,2)$, or $(5,1)$ are used, leading to roots corresponding to 
the total magnetic spin quantum numbers of $M_S = 0, 1$, or $2$, respectively. The zeroth-order energies ($E_{II}^{(0)}$) of the reference determinants refer to
the left axis, while the rest of the data to the right axis. The references are labeled ``A'' through ``D.''}
\label{fig:CH+_U}
\end{figure}

\begin{figure}
\includegraphics[scale=0.5]{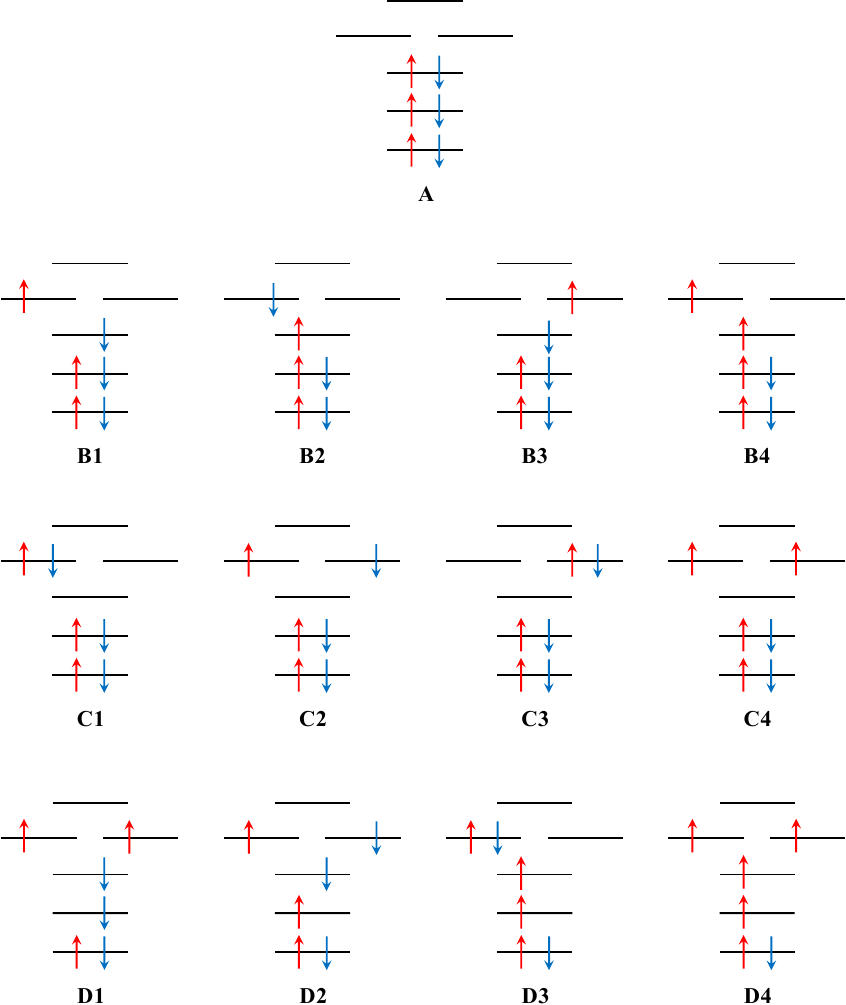}
\caption{Occupancy diagrams (not exhaustive) of some low-lying determinant references of CH$^+$. 
Labels ``A'' through ``D'' correspond to $E_{II}^{(0)}$ of Fig.\ \ref{fig:CH+_U}.}
\label{fig:CHplus}
\end{figure}

The CH$^+$ molecule has six electrons, of which four are correlated in the frozen-core approximation adopted here. The $\Delta$CC energies of several low-lying states 
obtained with (degenerate) determinant references with the number of the $\alpha$- and $\beta$-spin electrons being $(3,3)$, $(4,2)$, or $(5,1)$ are shown in Fig.\ \ref{fig:CH+_U}.

The ground-state wave function has the dominant single determinant labeled ``A'' in Fig.\ \ref{fig:CH+_U}, which is used as the orbital reference. 
Its occupancy diagram shown in Fig.\ \ref{fig:CHplus} obeys the aufbau principle. Therefore, for the ground state, $\Delta$CC theory is equal to single-reference CC theory with the spin-restricted HF reference,
which is rapidly convergent toward the exact (FCI) result reached by CCSDTQ (blue, filled circles).\cite{hirata_cc} In fact, CCSD is already nearly exact. 

Since the lowest-unoccupied molecular orbital (LUMO) is doubly degenerate, the second lowest-lying determinants already have a high degree of degeneracy.  
They are four-fold degenerate in the (3,3) sector or eight-fold degenerate 
in the combined (3,3), (4,2), and (2,4) sectors. Some of these degenerate references are shown in Fig.\ \ref{fig:CHplus} with label ``B.'' Occupancy diagrams B1, B2, and B3 are three of the four 
degenerate references in the (3,3) sector, whereas diagram B4 is one of the two in the (4,2) sector. For the lowest excited states already, the use of $\Delta$CC (not single-reference CC) theory
is essential, although determinants of different sectors do not couple with one another, and can thus be handled by separate $\Delta$CC calculations. 

As seen in Fig.\ \ref{fig:CH+_U}, the eight-fold degeneracy of references B is lifted as they split 
into a degenerate pair of triplets (red squares) and doubly degenerate open-shell singlets (blue, open circles). 
The triplets can be reached from the reference determinants in the (3,3) sector, and they have the total magnetic spin quantum number ($M_S$) of zero (red, filled squares) and may be 
called ``low spin.'' 
Those that come from the references in the (4,2) or (2,4) sector (red, open squares) have $M_S= 1$ or $-1$, respectively, and may be called ``high spin.''

In accordance with Hund's rule, the triplets lie lower than the singlets. Both states' energies are practically exact at the $\Delta$CCSD level. They maintain  the expected degeneracy accurately
 even when the roots come from different sectors. 

TD-CCSD (Ref.\ \onlinecite{Balkova1991JCP,Balkova1992CPL,Balkova1993JCP,Balkova1994JCP,Szalay1994JCP,Lutz2018JCP}) is designed to 
describe open-shell singlets such as the one encountered here (blue, open circles). This state's reference determinants are, however, nominally four-fold degenerate in the (3,3) sector,
exceeding the two-determinant limit of the TD-CC ansatz.  Since two out of the four-fold degeneracy originates from the degeneracy of the LUMOs, which are orthogonal to each other,
TD-CCSD should still be able to handle these states. 
However, the $\Delta$CCSD method, which is distinct from TD-CCSD and more general, can treat them equally well without any special considerations needed for the degree 
or nature of the degeneracy of the references. 

Only at the $\Delta$CCS level is the degeneracy 
of the triplets nonphysically lifted. Their energies are also less converged. The loss of degeneracy is due to the fact that configurations B1 and B2 in Fig.\ \ref{fig:CH+_U}
differ by two spin-orbitals and their coupling is not accounted for by  $\Delta$CCS. The high-spin ($M_S=\pm1$)
state (red, open square) is more accurately described by $\Delta$CCS than the low-spin ($M_S= 0$) state (red, filled square). This is because the high-spin configuration B4 dominates in
the ground-state wave function in the (4,2) sector, which can therefore be described well even by single-reference CC theory. Its degenerate counterpart in the (3,3) sector, in contrast, is an excited state whose wave function
has equal contributions from the low-spin configurations B1 and B2, which couple with each other. Hence, this is a manifestation of the well-known observation that 
a high-spin state 
tends to be dominated by a single determinant, and thus constitutes a convenient reference. This insight underlies the ingenious spin-flip methods of Krylov and coworkers.\cite{Krylov2001,Krylov2001_2,Slipchenko2002,Shao2003,Levchenko2004,Manisha2024}  $\Delta$CC theory too can take advantage of this insight by simply 
adopting a high-spin reference, free of any additional formulation or coding work. 

References labeled ``C'' are four-fold degenerate in the (3,3) sector, nondegenerate in each of the (4,2) and (2,4) sectors, and altogether six-fold degenerate.
Upon inclusion of electron correlation, they split into doubly degenerate open-shell singlets and a nondegenerate open-shell singlet (blue, open circles) as well as triplets (red squares).
The same observations as above for the states labeled by B apply:\ For all states, the $\Delta$CCSD energies are nearly exact. A nonphysical lifting of degeneracy of the triplets occurs at the $\Delta$CCS level, 
and their energies from the high-spin references are more accurate.

References labeled ``D'' are ten-fold degenerate in the (3,3) sector, six-fold generate in each of the (4,2) and (2,4) sectors, and nondegenerate in each of the (5,1) and (1,5) sectors; altogether 24-fold degenerate.
Only the triplets (red squares) and quintets (green crosses) spawned from them are plotted in Fig.\ \ref{fig:CH+_U}. Once again, their states' degeneracies are spuriously 
lifted at the $\Delta$CCS level, and the greater the $|M_S|$, the more accurate their energies. The $\Delta$CCS energies of the highest-spin references 
are nearly as accurate as the $\Delta$CCSD energies. This is understandable because the quintet state is the ground state in the (5,1) sector, whose wave function is dominated 
by the single determinant D4 of Fig.\ \ref{fig:CHplus}. This determinant has so few $\beta$-spin occupied orbitals and so few $\alpha$-spin virtual orbitals that there is expected to be little electron correlation. 
$\Delta$CCS only needs to describe the orbital relaxation from  reference A to D4, which it does accurately. 

\begin{figure}
\includegraphics[scale=0.65]{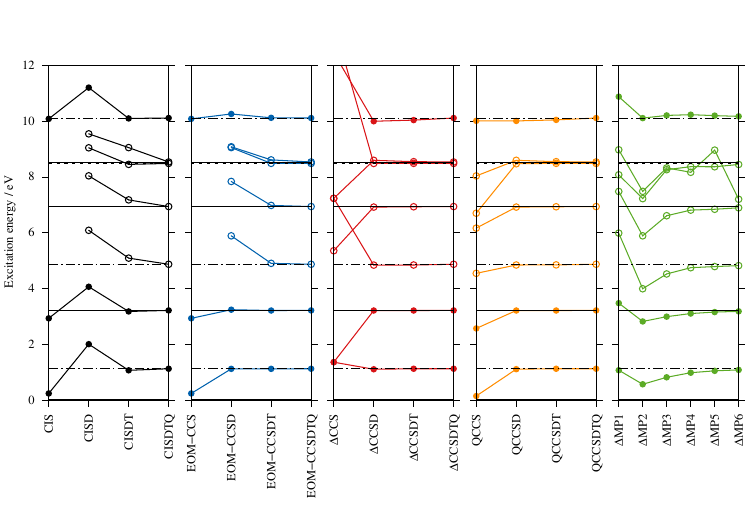}
\caption{Same as Fig.\ \ref{fig:CH+_U}, but for the vertical excitation energies of CH$^+$ computed by the CI, EOM-CC, $\Delta$CC, QCC, and $\Delta$MP methods
using low-spin ($M_S=0$) reference determinants.\cite{hirata_eomcc}
The solid, dotted-dashed, and dashed horizontal lines indicate the exact (FCI) values for singlet,  triplet, and quintet states, respectively. Filled circles correspond to predominantly one-electron excitations
 and open circles to predominantly two-electron excitations.}
\label{fig:CH+_excited}
\end{figure}

Figure \ref{fig:CH+_excited} compares the vertical excitation energies of the same molecule\cite{hirata_eomcc} obtained by the CI, EOM-CC, $\Delta$CC, and QCC methods through the FCI limits  as well as by the $\Delta$MP method through the sixth order. 
The references are low-spin determinants. 
Filled circles plot the excitations to states with predominantly one-electron character, while the open circles 
correspond to those with predominantly two-electron character. The ground-state wave function is 92.2\% zero-electron character. See Table 2 of Ref.\ \onlinecite{hirata_eomcc} for the precise percentage 
weights of the singles and doubles in their FCI wave functions.

Going from CISD to EOM-CCSD, the errors for the one-electron excitations (filled circles) 
decrease from 0.8--0.9 eV to invisibly small ($<$ 0.03 eV). For the two-electron excitations (open circles), CIS and EOM-CCS, which are one and the same method, lack the corresponding
roots, while CISD and EOM-CCSD do have roots, but they suffer from substantial errors of approximately 1 eV. EOM-CCSDT is nearly exact for both types of excitations (errors less than 0.04 eV). 
In general, the $k$th-order EOM-CC is always 
superior to the $k$th-order CI, and is quantitative for excitations of up to $(k-1)$-electron character. The accuracy is independent of the spin multiplicities of the initial or final states. 
These performance characteristics are well known.\cite{hirata_eomcc,shavitt,bartlettRMP,BartlettWIRE2012}

Going from EOM-CC to $\Delta$CC, we make the following observations:\ Unlike EOM-CCS, $\Delta$CCS has roots for all states regardless of their excitation character. However, its results
are rather erratic, posing a considerable challenge when assigning energies to states especially for highly degenerate references. Errors in the excitation energies are substantial and do not correlate
with the excitation character. Hence, the practical utility of $\Delta$CCS seems rather limited, but there are exceptions (see below). 

Once we raise the rank to $\Delta$CCSD, the theory's accuracy and thus utility improve dramatically. It has roots for all states and the 
excitation energies are nearly exact for both one- and two-electron excitations (errors less than 0.03 eV), as foreshadowed by Fig.\ \ref{fig:CH+_U}. 
Unlike EOM-CCSD, which has errors approaching 1 eV for two-electron excitations (and lacks roots for three-electron excitations), $\Delta$CCSD 
displays uniform high accuracy for any state insofar as
the state's wave function is dominated by a linear combination of degenerate determinants. 

Only when a wave function has a noticeable contribution from an energetically nearby determinant (the situation that may be variously referred to as strong correlation, nondynamical correlation, quasidegeneracy, 
 etc.)\ does a higher-order $\Delta$CC method become necessary to maintain accuracy.
Hence, the performance expectation of $\Delta$CC theory is essentially the same as that of single-reference CC theory for nondegenerate ground states, which is well documented.\cite{shavitt,bartlettRMP,BartlettWIRE2012}

The results of the $\Delta$CC and QCC calculations are interchangeable except for those of the lowest-rank members:\ QCCS has roots for all excitations and its excitation
energies are better than the $\Delta$CCS roots, though not necessarily superior to the CIS or EOM-CCS roots when the latter exist. 
The similarity in performance of higher-rank members originates from the same in their ans\"{a}tze, and a closer
inspection shows that QCC is slightly more accurate than $\Delta$CC of the same rank at an unnoticeably elevated cost. 
This slight accuracy boost of QCC comes at the cost of losing the black-box nature of the method. 
$\Delta$CC and QCC theories are both applicable to a GMS, and their comparable performance for bond breaking will be investigated in the future. 

The $\Delta$MP$n$ results are plotted in the rightmost subgraph. They are convergent toward the respective, correct, exact (FCI)  limits unless they are divergent. 
The $\Delta$MP1 method has roots for two-electron excitations and is distinctly superior to $\Delta$CCS, but not necessarily to CIS, EOM-CCS, or QCCS.
The $\Delta$MP2 method is poor, apparently giving unbalanced descriptions of the initial and final states. The $\Delta$MP3 method and onward tend to correct these errors  in a monotonic, but slowly convergent manner. Since $\Delta$MP3 is a $O(n^6)$ procedure, it should be compared with CISD, EOM-CCSD, and $\Delta$CCSD, which are
also the $O(n^6)$ methods. The performance of these methods is in the order:\ $\Delta$CCSD $>$ EOM-CCSD $>$ $\Delta$MP3 $>$ CISD. 
When an emphasis is placed on two-electron excitations, the order changes to:\ $\Delta$CCSD $\gg$ $\Delta$MP3 $>$ EOM-CCSD $\approx$ CISD. 

Lastly, some remarks on the convergence of the iterative solution algorithms of the $\Delta$CC $T$-amplitude equations (Sec.\ \ref{sec:implementation}) may be in order.
Without DIIS, $\Delta$CC calculations tend to fail for many references. With DIIS, the $\Delta$CC calculations for degenerate references B and C of Fig.\ \ref{fig:CH+_U} converge at a similar rate as 
the single-reference CC calculations for the ground state (A). For the ten-fold degenerate references D, the convergence difficulty has been encountered even with DIIS, but not uniformly; the 
convergence is slow for some roots, while rapid for others. Convergence difficulties are sometimes accompanied by complex energies.\cite{Zivkovic1978,Paldus1993,Piecuch1994}
Overall, however, $\Delta$CC theory has proven to be a  generally well-behaved and practically viable method.

\subsubsection{CH$_2$}

\begin{figure}
\includegraphics[scale=0.65]{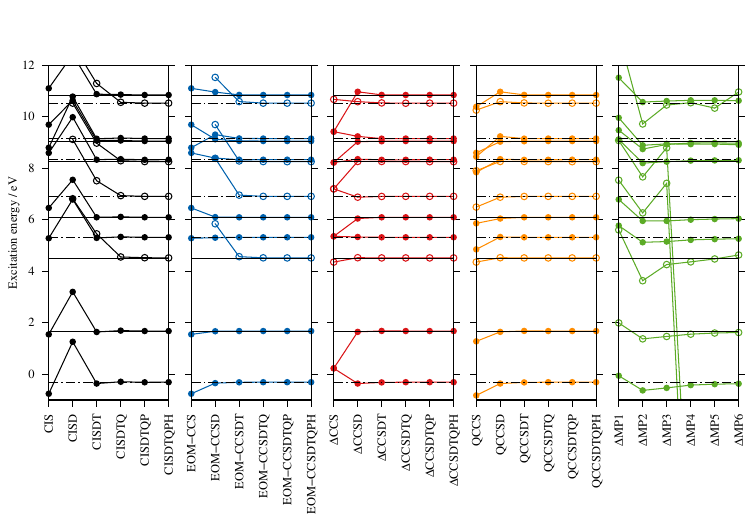}
\caption{The vertical excitation energies of CH$_2$ ($r_\text{CH} = 1.102\,\AA$ and $a_\text{HCH}=104.7^\circ$) (relative to its singlet ground state) 
calculated by the CI, EOM-CC, $\Delta$CC, QCC, and $\Delta$MP 
methods and 6-31G* basis set with the highest virtual and lowest occupied orbitals kept frozen in the low-spin reference determinants.\cite{hirata_eomcc}
The solid and dotted-dashed horizontal lines indicate the exact (FCI) values for singlet and triplet states, respectively. Filled circles correspond to predominantly one-electron excitations
 and open circles to predominantly two-electron excitations.}
\label{fig:CH2_excited}
\end{figure}

The vertical excitation energies of the CH$_2$ molecule,\cite{hirata_eomcc} a twelve-electron system in the frozen core and virtual approximation, are computed by $\Delta$CC and other methods, using
the singlet HF reference (note that the true ground state is triplet). 
The results are shown in Fig.\ \ref{fig:CH2_excited}. They reinforce the general performance trends observed for CH$^+$ in Sec.\ \ref{sec:CH+}. 

CIS $=$ EOM-CCS lack two-electron excitation roots (open circles), but they are impressively accurate for one-electron excitations (filled circles). CISD overestimates excessively all shown 
excitation energies (errors of approximately $1.5$ eV). 
EOM-CCSD does not have this problem and is nearly exact 
for one-electron excitations. However, for two-electron excitations, it has errors sometimes in excess of 1 eV. 

$\Delta$CCS has roots for all excited states, but its excitation energies do not seem useful (and are
challenging to assign for highly degenerate references). $\Delta$CCSD is as accurate as EOM-CCSD for one-electron excitations and distinctly superior to the latter for two-electron excitations.
$\Delta$CCSD is practically converged at the FCI limits for all shown excitation energies, and is, therefore, an attractive alternative to EOM-CCSD.

The QCC methods behave nearly identically as the $\Delta$CC methods except at the singles-only level. 
The fact that black-box $\Delta$CC theory produces essentially the same accurate results (except at the $\Delta$CCS level) as the slightly more general, but non-black-box QCC theory
underscores the practical utility of the former.

$\Delta$MP1 has roots for all excitations, whose numerical values seem reasonable if not superior to CIS = EOM-CCS. $\Delta$MP2 tends to underestimate all excitation energies, which are corrected systematically by $\Delta$MP3 and onward
except for at least two sets of two-electron excitations where the perturbation series oscillate wildly. In fact, it is almost a rule than an exception that perturbation series that initially appear to converge turn 
to diverge eventually,\cite{olsen_mp} making it crucial to perform  infinite diagram summation. This is realized by $\Delta$CC theory. 

\subsection{Ionized states}

\subsubsection{BH\label{sec:BH}}

\begin{figure}
\includegraphics[scale=0.65]{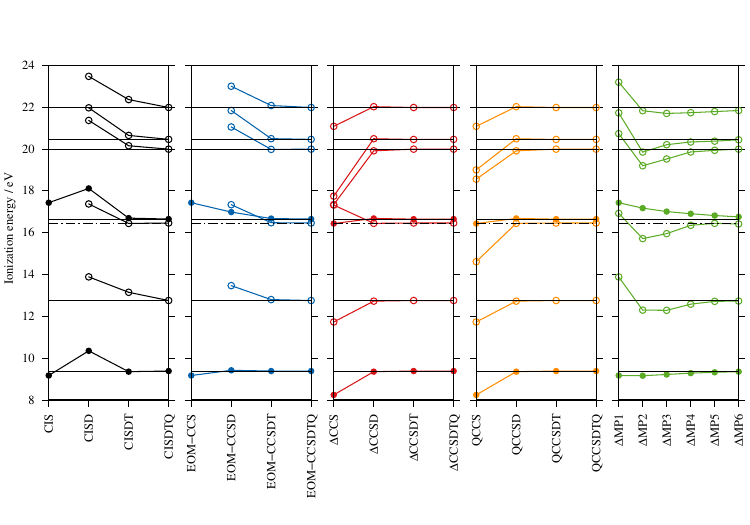}
\caption{The vertical ionization energies of BH ($r_\text{BH} = 1.232\,\AA$) calculated by the CI, EOM-CC, $\Delta$CC, QCC, and $\Delta$MP 
methods and 6-31G** basis set with the highest virtual and lowest occupied orbitals kept frozen in the low-spin reference determinants.\cite{hirata_ipeomcc}
The solid and dotted-dashed horizontal lines indicate the exact (FCI) values for doublet and quartet states, respectively. Filled circles correspond to Koopmans
ionization potentials and open circles to satellite ionization transitions.}
\label{fig:BH_ionized}
\end{figure}

The BH molecule is a four-electron system in the frozen core and virtual approximation. Its vertical ionization energies have been computed by the hole-particle CI, IP-EOM-CC, $\Delta$CC,
and QCC methods through the FCI limits as well as by the $\Delta$MP method up to the sixth order. The results are shown in Fig.\ \ref{fig:BH_ionized}.

Here, CIS = IP-EOM-CCS use the 1h (one hole) operator
as the linear ionization operator, CISD and IP-EOM-CCSD employ the 1h and 2h-1p (two-hole-one-particle) operators, CISDT and IP-EOM-CCSDT the 1h, 2h-1p, and 3h-2p operators. These combinations
ensure the size-extensivity and intensivity of the methods\cite{HirataTCA} and are the standard choices. For the $\Delta$CC and $\Delta$MP methods, such considerations are unnecessary, and the same formalisms and codes handle any states with any numbers of $\alpha$- and $\beta$-spin electrons; they only need a reference determinant as input. See the last paragraph of Sec.\ \ref{sec:EOMCC} for more on this point.

There are two Koopmans ionization potentials (IPs) (filled circles) shown in Fig.\ \ref{fig:BH_ionized}:\ The first IP at 9.4 eV and the second IP at 16.6 eV.
The singles weights of the corresponding ionized wave functions are, respectively, 93\% and 76\% only.\cite{hirata_ipeomcc} 
The rest of the IPs are satellites (open circles) and their corresponding ionized wave functions have 94\% or more two-electron character.

The CIS $=$ IP-EOM-CCS $=$ $\Delta$MP1 roots exist only for the Koopmans IPs and their results correspond to the HF orbital energies (the signs reversed) as per the Koopmans theorem.
The agreement is excellent for the first IP owing to the well-known systematic cancellation of correlation and orbital relaxation errors.\cite{szabo} 
The agreement is less good for the second IP likely because of its 
noticeable two-electron character. The 2h-1p CISD method is poor with errors in excess of 1 eV for all shown IPs. IP-EOM-CCSD erases these large error nearly completely 
for the first IP, but it brings about only modest improvements for the satellite IPs and second Koopmans IP, whose destination states have, respectively, dominant and appreciable two-electron character. 

$\Delta$CC theory displays remarkable performance for all IPs. Their results are practically converged at the $\Delta$CCSD level (errors less than 0.08 eV). The same goes to the results of the
QCC calculations beyond singles.

The $\Delta$MP series seems convergent for all IPs. We shall explore higher-order $\Delta$MP methods in what follows.

\begin{figure}
\includegraphics[scale=0.65]{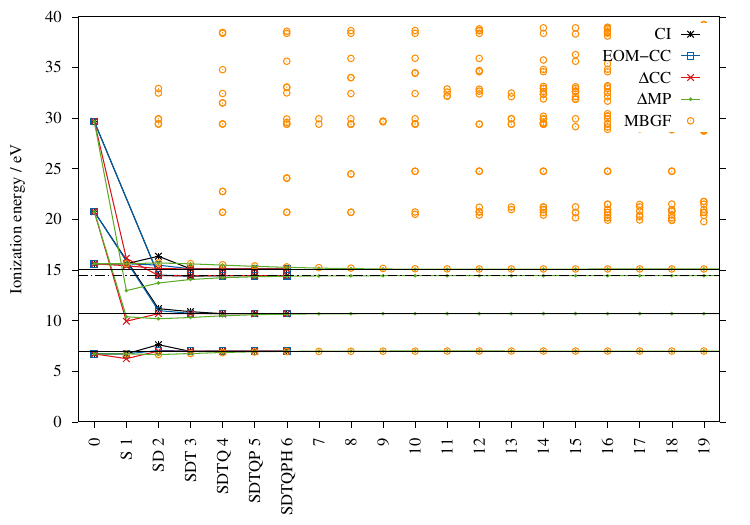}
\caption{The vertical ionization energies of BH ($r_\text{BH} = 1.232\,\AA$) using the minimal (STO-3G) basis set with all electrons correlated
in the low-spin reference determinants.\cite{Hirata_PRA2024}
The horizontal axis gives the truncation order of excitation operators or the perturbation order. 
The CI, EOM-CC, $\Delta$CC, and $\Delta$MP data are shown for the two lowest Koopmans ionization potentials and two doubly degenerate satellite ionization transitions in between.
The solid and dotted-dashed horizontal lines indicate the corresponding exact (FCI) data for doublet and quartet states, respectively.
The MBGF data plot all roots of the inverse Dyson equation.} 
\label{fig:BH_STO3G_all}
\end{figure}

\begin{figure}
\includegraphics[scale=0.65]{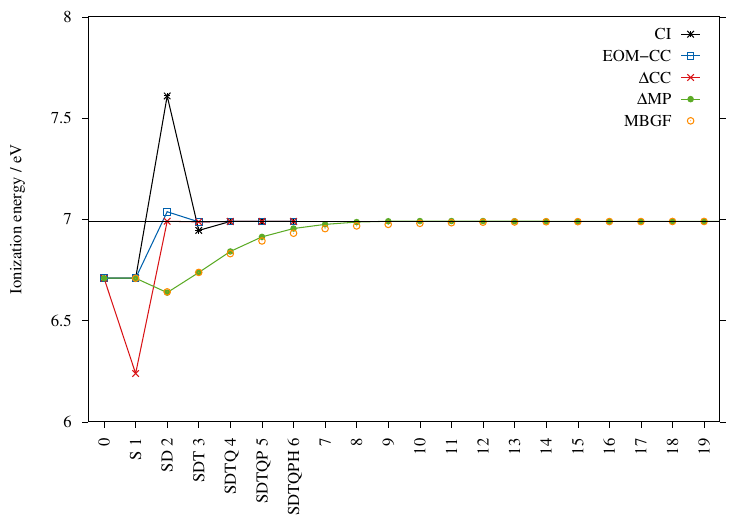}
\caption{The enlargement of Fig.\ \ref{fig:BH_STO3G_all} around the first Koopmans state. }
\label{fig:BH_STO3G_root1}
\end{figure}

\begin{figure}
\includegraphics[scale=0.65]{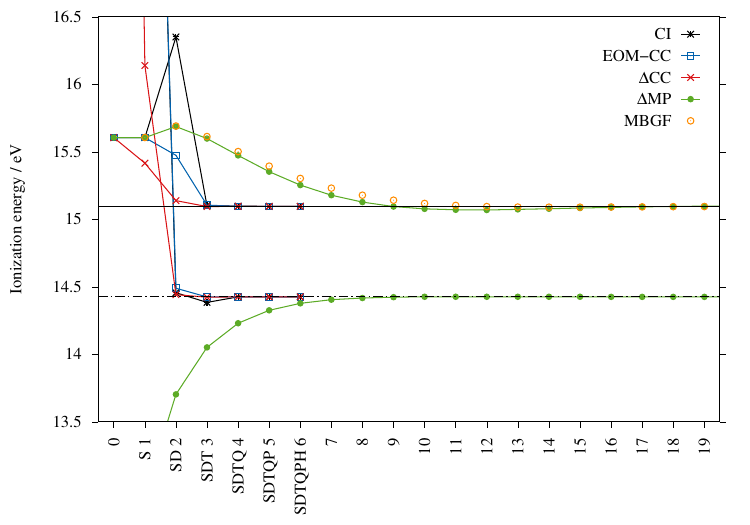}
\caption{The enlargement of Fig.\ \ref{fig:BH_STO3G_all} around the second Koopmans state at 15.1 eV. The MBGF roots for the nearby satellite state at 14.4 eV are nonconvergent\cite{Hirata_PRA2024} 
and outside this graph.}
\label{fig:BH_STO3G_root3}
\end{figure}

\begin{figure}
\includegraphics[scale=0.65]{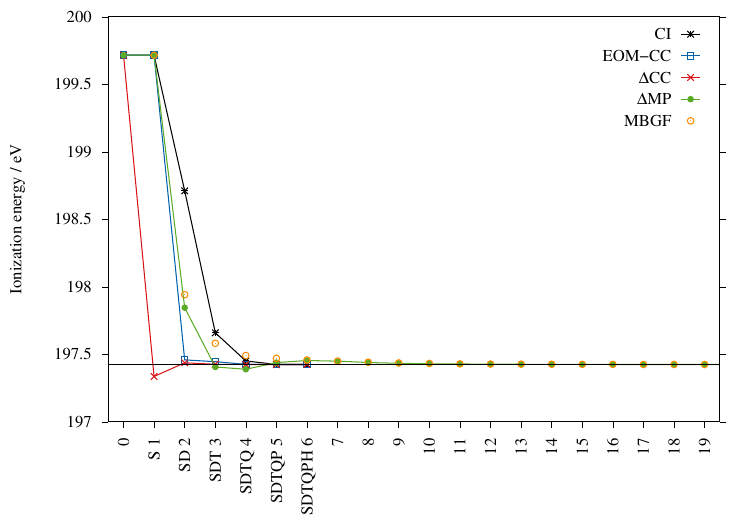}
\caption{The enlargement of Fig.\ \ref{fig:BH_STO3G_all} around the third Koopmans (or core-ionized) state.}
\label{fig:BH_STO3G_core}
\end{figure}

Figures \ref{fig:BH_STO3G_all}--\ref{fig:BH_STO3G_core} extend the analysis on the BH molecule's IPs
to higher-order $\Delta$MP and MBGF calculations 
through the nineteenth order. 
(This is an all-electron calculation with the minimal basis set and differs from the preceding one.)

It may be recalled\cite{deltamp,Hirata2017,Hirata_PRA2024} that the $\Delta$MP$n$ method is equivalent to the MBGF($n$) method in the diagonal, frequency-independent approximation
for $n \leq 3$.  The $\Delta$MP series is convergent at the FCI limits (unless divergent). 
The MBGF series  converges at the FCI limits for some Koopmans states, but for most other states, it does not.\cite{Hirata_PRA2024}

Figure \ref{fig:BH_STO3G_all} confirms these. For the two Koopmans  and two satellite states considered,  with increasing $n$, the $\Delta$MP$n$ results (green dots) 
quickly snap onto the horizontal lines of the FCI data. In contrast, the plots of the IPs obtained by higher-order MBGF (orange, open circles) 
fall upon the FCI lines of the two Koopmans states only, but not of the two satellite states. The roots for the latter converge at wrong limits around  20 eV, 
exposing a severe, fundamental flaw of Feynman--Dyson MBGF.\cite{Hirata_PRA2024} 

Figure \ref{fig:BH_STO3G_root1} zooms in on the first Koopmans IP around 7 eV. For this root, both the $\Delta$MP$n$ and MBGF($n$) methods 
converge at the FCI limit upon increasing $n$. The rate of convergence is slightly faster with $\Delta$MP than with MBGF, despite the frequent characterization of the former as an approximation to the latter.\cite{szabo,deltamp,Hirata2017,Hirata_PRA2024} CIS = EOM-CCS yield the same value  as $\Delta$MP1 or MBGF(1), obeying the Koopmans theorem.
$\Delta$CCS is distinctly worse than the other four methods. However, $\Delta$CCSD is nearly exact, slightly more accurate than IP-EOM-CCSD and distinctly more so
than 2h-1p CISD, $\Delta$MP3, or MBGF(3), which are all $O(n^6)$ procedures. IP-EOM-CCSDT catches up with $\Delta$CCSDT, but the 3h-2p CISDT, $\Delta$MP5, and  MBGF(5)
methods  still fall short. 


Figure \ref{fig:BH_STO3G_root3} focuses on the second Koopmans IP around 15.1 eV and its nearby satellite ionization transition at 14.4 eV. For the Koopmans IP, the performance of the methods 
is in the order:\ $\Delta$CC $>$ IP-EOM-CC $>$ CI  as well as  $\Delta$CC $>$ $\Delta$MP $\approx$ MBGF.
For the satellite transition, which is a two-electron process, 
CIS = IP-EOM-CCS do not even have a root. CISD is accurate, which is accidental because CISDT has a greater error. IP-EOM-CCSD has a slightly visible error, but IP-EOM-CCSDT
is essentially converged. In contrast, $\Delta$CCSD is nearly exact. 
The $\Delta$MP series is monotonically convergent, and its rate is not particularly slower than those for the Koopmans IPs. 
On the other hand, as discussed above, the corresponding MBGF roots are nowhere near the exact value and outside this graph. 

In Fig.\ \ref{fig:BH_STO3G_core}, we turn to the core ionization, i.e., the third Koopmans IP. 
There has been much attention to such ionizations occurring in the extreme ultraviolet range,\cite{VuraWeis2025} as
they can report element-specific dynamical information about spin and oxidation states, particularly, of transition metals in a molecule.
The Koopmans approximation, i.e., CIS = IP-EOM-CCS = $\Delta$MP1 = MBGF(1), is poor, as orbital relaxation effects are large.
$\Delta$CCS works remarkably well in this case. It may be because the one-electron excitation operator takes into account the relaxation effects, while 
correlation effects are small because core orbitals are so far removed from valence orbitals. Whether this is merely accidental remains to be checked 
by a more systematic study. 
IP-EOM-CCSD and $\Delta$CCSD are essentially converged, while CISD is much less accurate. $\Delta$MP and MBGF series both converge at 
the FCI limits, with $\Delta$MP3 being near exact. 

The $\Delta$CC methods can home in on these core ionizations (and excitations) by simply specifying the core-hole determinants as their references. 
This is in contrast with the trial-vector algorithms\cite{Bartlett1972,Bartlett1973,Davidson1975,Hirao1982} of the CI and EOM-CC methods, which are ill-suited for such interior eigenvalues,\cite{Zuev2015}
although efficient and reliable modifications to them are now available.\cite{Zuev2015,Peng2015} 
A more fundamental issue surrounding core transitions has to do with their Feshbach resonances with an ionization continuum, causing Auger decay on the one hand and making 
Hermitian, Gaussian-basis-set quantum mechanics problematic on the other.\cite{Vidal2019} Here, too, a pioneering method of core-valence separation by Cederbaum {\it et al.}\cite{Cederbaum1980}\ has been adopted in
the EOM-CC methods\cite{Coriani2015,Vidal2019,Skomorowski2021_1,Skomorowski2021_2} with accurate simulation results. A proper treatment of 
core transitions by $\Delta$CC theory may also require the same type of provisions to account for such resonances with a continuum.

\subsubsection{C$_2$}

\begin{figure}
\includegraphics[scale=0.65]{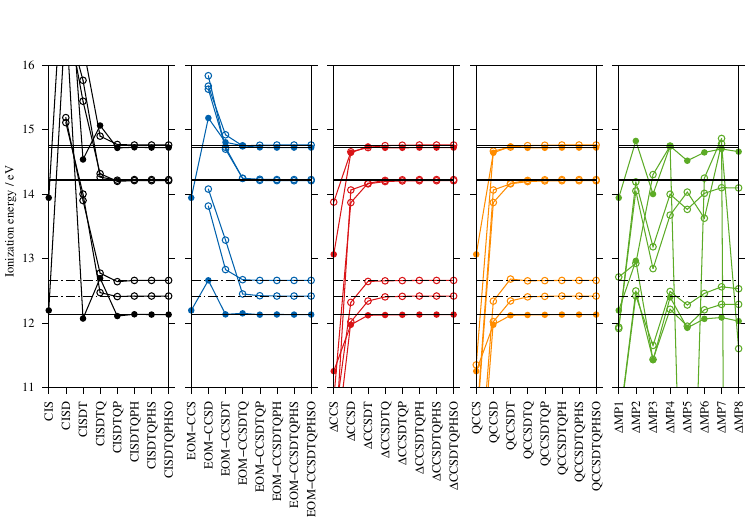}
\caption{The vertical ionization energies of C$_2$ ($r_\text{CC} = 1.262\,\AA$) calculated by the CI, EOM-CC, $\Delta$CC, QCC, and $\Delta$MP 
methods and 6-31G basis set with the two highest virtual and two lowest occupied orbitals kept frozen in the low-spin reference determinants.\cite{hirata_ipeomcc}
The solid and dotted-dashed horizontal lines indicate the exact (FCI) values for doublet and quartet states, respectively. Filled circles correspond to Koopmans 
states  and open circles to satellite states.}
\label{fig:C2_ionized}
\end{figure}

The C$_2$ molecule's ground-state wave function has a large multireference character. Consequently, even its Koopmans ionization transitions
have appreciable three-electron character.\cite{hirata_ipeomcc} The CI series does not converge until we reach CISDTQP, 
and even IP-EOM-CC theory struggles until we invoke IP-EOM-CCSDT 
for Koopmans IPs (Ref.\ \onlinecite{hirata_ipeomcc}) or IP-EOM-CCSDTQ for satellite ionization transitions. 
In contrast, $\Delta$CCSD is already usefully accurate for the first and second Koopmans IPs with errors of 0.16 and 0.08 eV, respectively 
(as compared with 0.53 and 0.46 eV at IP-EOM-CCSD or 4.68 and 3.50 eV at 2h-1p CISD). $\Delta$CCSDT is near exact for these states (errors less than 0.02 eV).

For satellite states, $\Delta$CCSDT is quantitative with errors less than 0.08 eV and $\Delta$CCSDTQ is essentially exact.
In comparison, IP-EOM-CCSDT has errors up to 0.87 eV. 

QCC theory displays similar behavior and performance as $\Delta$CC theory. For this strong-correlation
problem, the difference seems smaller (and, in fact, $\Delta$CCS works slightly better than QCCS).

The $\Delta$MP series are wildly oscillatory. Their plots provide a contrast to the rapid and much smoother convergence of the $\Delta$CC plots, underscoring
the importance and power of the infinite diagram summation in the latter.

\subsection{Electron-attached states}

\subsubsection{CH$^+$}

\begin{figure}
\includegraphics[scale=0.65]{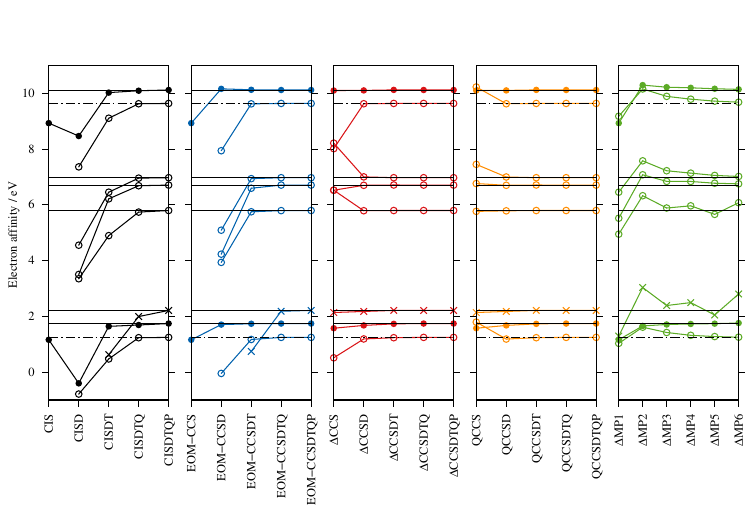}
\caption{The vertical electron affinities of CH$^+$ ($r_\text{CH} = 1.120\,\AA$) calculated by the CI, EOM-CC, $\Delta$CC, QCC, and $\Delta$MP 
methods and 6-31G* basis set with the highest virtual and lowest occupied orbitals kept frozen in the low-spin reference determinants.\cite{hirata_ipeomcc}
The solid and dotted-dashed horizontal lines indicate the exact (FCI) values for doublet and quartet states, respectively. Filled circles correspond to Koopmans 
states, open circles to 2p-1h satellite states, and crosses to the 3p-2h satellite state.}
\label{fig:CH_affinity}
\end{figure}

Figure \ref{fig:CH_affinity} plots the vertical electron affinities (EAs) of CH$^+$ computed by the hole-particle CI, EA-EOM-CC, $\Delta$CC, QCC,
 and $\Delta$MP methods. There are two Koopmans EAs (filled circles) in the shown energy range. The rest are satellite electron-attachment transitions, and 
 one of them is the 3p-2h type (crosses), while the other four are the 2p-1h type (open circles). 
 
For the Koopmans EAs, CIS = EA-EOM-CCS = $\Delta$MP1 yield the HF orbital energy (the sign reversed) as per the Koopmans theorem.
The results are rather poor because unlike in IPs, correlation and orbital-relaxation errors in EAs do not cancel with each other, but rather they pile up.\cite{szabo} 
CISD has errors approaching 2 eV, while EA-EOM-CCSD is nearly converged. $\Delta$CCS and QCCS are impressively accurate and $\Delta$CCSD
and QCCSD are essentially exact. $\Delta$MP2 is also accurate. These may suggest that there is little correlation but considerable orbital relaxation
upon electron attachment at least in this small molecule. 

CIS fails to locate the roots corresponding to the 2p-1h satellite states. CIS and CISD also lack the root for the 3p-2h state. 
EA-EOM-CCSD does have the 2p-1h roots, but their EAs are in error by nearly 2 eV and are not useful. EA-EOM-CCSDT nearly completely corrects these errors, but
it is still erroneous by 1.5 eV for the 3p-2h satellite electron attachment, and it takes EA-EOM-CCSDTQ for its prediction to be quantitative (error of 0.03 eV). These performance trends are well known.\cite{hirata_ipeomcc,shavitt,bartlettRMP,BartlettWIRE2012}

In contrast, $\Delta$CC and QCC series are practically converged at the CCSD level. For the 3p-2h state, which poses a considerable difficulty for EA-EOM-CC,
even $\Delta$CCS and QCCS are quantitative (errors of 0.07 eV) and almost as accurate as EA-EOM-CCSDTQ. This is because the state's wave function 
is dominated by doubly degenerate 3p-2h determinants and is straightforward to describe by $\Delta$CC theory. The $\Delta$MP series are also generally rapidly convergent except for 
a few signs of oscillatory divergence. For this particular problem, the order of performance is $\Delta$CC $>$ $\Delta$MP $>$ EA-EOM-CC $>$ CI.

\subsection{Summary of the numerical tests}

\begin{figure}
\includegraphics[scale=0.65]{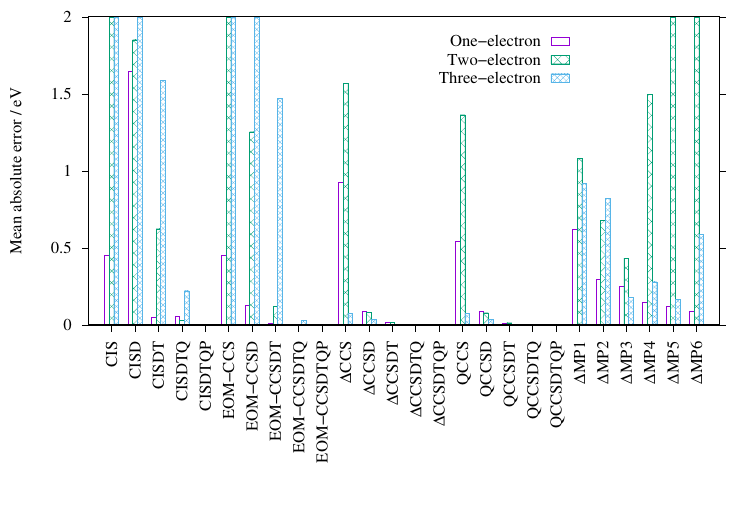}
\caption{The mean absolute errors of the CI, EOM-CC, $\Delta$CC, QCC, and $\Delta$MP methods 
from the FCI method for predominantly one-, two-, and three-electron excitation, ionization, or electron-attachment transitions.
An error of infinity is recorded when the corresponding root does not exist.}
\label{fig:Errors}
\end{figure}

The foregoing benchmark calculations' results can be summarized in Fig.\ \ref{fig:Errors} in the form of comparison of their mean absolute errors from the FCI data.
The minimal-basis-set calculations on the BH molecule are excluded, and an error of infinity is recorded when the corresponding root does not exist. 
The order of performance is QCC $\approx$ $\Delta$CC $>$ EOM-CC $>$ CI as well as
QCC $\approx$ $\Delta$CC $>$ $\Delta$MP $>$ MBGF corrected for their cost scaling. 
The last method is not included in Fig.\ \ref{fig:Errors}, but the order is based on the established fact that 
for most Koopmans roots, $\Delta$MP $\approx$ MBGF, while for most satellite roots, the $\Delta$MP series are convergent toward the correct FCI limits (unless divergent), but the MBGF series 
are generally not.\cite{Hirata_PRA2024} QCC and $\Delta$CC theories' results are interchangeable beyond the CCS level, and in this sense,
the rather strong restrictions (in the form of projectors) imposed on the $\Delta$CC ansatz by its black-box requirement have the minimal impact on the performance. 
The practical utility of $\Delta$CCS is limited for general states and transitions, but there are a few pieces of evidence suggesting that it is accurate for core ionizations and high-spin states,
where there are large orbital-relaxation and small correlation effects, and possibly  for electron affinities also.   

In short, $\Delta$CC theory is a viable alternative to the powerful and versatile EOM-CC theory with complementary pros and cons; it may even become 
a replacement of the latter in many applications. 

\section{Conclusions and outlook}

A systematically converging, size-extensive, and black-box {\it ab initio} CC theory is introduced, which applies
 the time-honored exponential ansatz of single-reference CC theory to degenerate
as well as nondegenerate reference determinants with any numbers and occupancies of $\alpha$- and $\beta$-spin electrons.  
This degenerate CC ($\Delta$CC) theory is a natural CC extension of $\Delta$MP theory, the former being an infinite diagram summation
of the latter.

It can be applied to excited, ionized, and electron-attached states with any number of electrons excited, removed, added, and/or spin-flipped. 
For a given truncation order of the cluster excitation operator, all of these situations can be handled uniformly by one and the same set of formalisms implemented into a single
computer program.  Nowhere in the $\Delta$CC formalisms do we need an ionization, double-ionization, electron-attachment, spin-flip, etc.\ operator; instead, 
(degenerate) reference determinants are  responsible for describing electron rearrangements, leaving the exponential operator to concentrate on capturing
electron correlation. For instance, $\Delta$CC theory reaps the same benefits as the spin-flip methods of Krylov and coworkers\cite{Krylov2001,Slipchenko2002,Levchenko2004,Manisha2024}
by simply adopting high-spin determinants as the reference.

For all excitation, ionization, and electron-attachment energies of one-, two-, and three-electron types considered in this study, 
 $\Delta$CC including singles and doubles ($\Delta$CCSD) achieves the mean absolute errors of 
0.08, 0.09, and 0.03 eV, respectively. 
This is compared with non-size-extensive CISD, which has the large errors of 1.6, 1.9, and $\infty$ eV, with ``$\infty$'' meaning that there are no roots. 
The widely used EOM-CCSD method displays much smaller errors of 0.13, 1.3, and $\infty$ eV, but its performance falls short of $\Delta$CCSD. 
The order of performance is, therefore, $\Delta$CC $>$ EOM-CC $>$ CI.

The third-order $\Delta$MP method ($\Delta$MP3) has the same $O(n^6)$ scaling as $\Delta$CCSD, CISD, and EOM-CCSD and has the corresponding errors of 0.25, 0.43, and 0.18 eV, respectively,
for one-, two-, and three-electron transitions, which are adequate.
However, $\Delta$MP4 has the errors of 0.15, 1.5, and 0.28 eV, showing  signs of divergence. $\Delta$MP theory is, therefore, hard to use in a predictive calculation
that requires a converging series of approximations. 
Feynman--Dyson MBGF theory is even worse, guaranteed to converge at wrong limits for most satellite ionized or electron-attachment transitions.\cite{Hirata_PRA2024}
The order of performance is $\Delta$CC $>$ $\Delta$MP $>$ MBGF, underscoring the importance of an infinite diagram summation in $\Delta$CC theory.

The $\Delta$CCS method can be viewed as a projection HF theory for degenerate and nondegenerate references. In the general context, however, its practical utility seems limited 
because its accuracy is low. However, there are pieces of evidence and  theoretical arguments suggesting that it is quantitative
for core ionizations, high-spin states, and possibly electron affinities, and thus can be an extremely fast method of choice for such cases. However, a more systematic, larger-scale benchmark study is needed to reach a definitive 
conclusion.

$\Delta$CC theory is a promising alternative to the popular, powerful, versatile EOM-CC theory with complementary pros and cons; it 
may even be a replacement of the latter in many applications. EOM-CC theory may be ideally suited for simulations of one-photon absorption/emission spectra and various
one-particle properties. Its trial-vector algorithms are robust and efficient. On the other hand, $\Delta$CC theory is considerably more accurate
for a wide range of electronic configurations and transitions. 
It should be able to reuse highly optimized formalisms and algorithms of the well-established single-reference CC methods for the ground states. 
Our pilot implementations have revealed only slightly more frequent convergence difficulties of the iterative solution algorithm of the amplitude equations.

$\Delta$CC theory is closely related to, but distinct from Li and Paldus's and Kucharski and Bartlett's MRCC theories. A new MRCC (QCC) ansatz has been introduced, which
$\Delta$CC theory is directly related to, but is still distinct from. It is size-extensive for a degenerate multireference, exactly convergent, and applicable to 
any general --- complete or incomplete --- model spaces.
While both $\Delta$CC and QCC theories are applicable to a GMS, the latter takes a full account of electron correlation within the internal space at the expense of losing the black-box character. 
For weakly correlated problems studied here, the results of the $\Delta$CC and QCC calculations are interchangeable beyond singles. 
It is important to quantify the performance of $\Delta$CC and QCC theories for bond breaking and other strong-correlation problems.

Looking ahead, we find it important to develop an efficient, optimal-scaling algorithm of the $\Delta$CC methods through a sufficiently high order (the present
$\Delta$CCS and $\Delta$CCSD implementations are optimal scaling, but inefficient).
Formulas, algorithms, and computer codes should be developed for their analytical derivatives, 
transition probabilities between two $\Delta$CC states, noniterative inclusions of higher-order connected excitation operators such as CCSD(T)\cite{Raghavachari1989,Bartlett1990} and CCSD(2),\cite{Hirata2004}
explicitly correlated ans\"{a}tze, crystals under periodic boundary conditions, etc. 

Having roots for all reference determinants, $\Delta$CC theory 
can serve as a basis for a finite-temperature generalization of CC theory,\cite{sanyal,mandal,mandal2,WhiteChan2,Harsha1,Harsha2,Harsha3} which should be an infinite diagram summation of finite-temperature MBPT.\cite{Jha2019,HirataJha2019,Jha2020,HirataJha2020,HirataPRA2021,HirataJCP2021,HirataCPL2022} 
Such a theory may integrate quantum-mechanical CC theory and classical-mechanical 
Ursell--Mayer cumulant theory,\cite{Ursell,Mayer,Montroll1941} providing {\it ab initio} descriptions 
of gases and liquids of electrons,\cite{Cizek1969,Freeman1977,Bishop1978,Callahan2022,Butler2024} atoms,\cite{Ursell,Mayer,Montroll1941,bloch,Brout1959,Morita1961} and nucleons\cite{Coester1958,CoesterKummel1960,Bethe1964,Day1967,Kummel1978,Baldo1999} at zero and nonzero temperatures and of their phase transitions
and critical pheonomena.\cite{Hirata_JPCL2025}

\appendix*
\section{Size-extensivity\label{sec:linked}}

Size-extensivity\cite{HirataTCA,HirataARPC,Nooijen_extensivity2005} is among the most contentious issues in {\it ab initio} electron-correlation theory, in particular, in
relation to multireference theories.\cite{Nooijen_extensivity2005} 

A size-extensive electron-correlation theory is the one that yields a correlation energy (relative to a size-extensive mean-field
reference energy) that has the correct volume ($V$) dependence of $O(V)$ for an infinite system in a finite computational time. 
(The last requirement excludes FCI and the multireference theories limited to a CMS from the category of meaningfully size-extensive theories.)

In a multireference theory, there are two potential sources of the loss of size-extensivity according to Meissner {\it et al.}:\cite{Meissner1989}\ The effective 
Hamiltonian and its diagonalization (see also Refs.\ \onlinecite{Lindgren1985_1,Lindgren1985_2,Lindgren1987,MukherjeeAQC1989,MukhopadhyayTCA1991,Nooijen_extensivity2005}). The former pertains to 
the size-extensive recuperation of a correlation energy from the external space, whereas the latter has to do with the same from the internal 
space. Clearly, not only does the ansatz but also the model space and its volume dependence play a crucial role
in ensuring the size-extensivity with respect to both external- and internal-space extensions.

We will first document an analytic (i.e., nondiagrammatic) proof of the size-extensivity for $\Delta$CC theory, which encompasses the same for $\Delta$MP theory. We then consider 
diagrammatic criteria of size-extensivity of a general, degenerate theory. Unlike those for a single-reference theory, which may  be precisely
expressed 
in terms of diagrammatic linkedness, the criteria for a degenerate theory seem ambiguous and limited, varying with the details of
the degenerate multireference in a complex manner. 

\subsection{Size-extensivity of $\Delta$MP and $\Delta$CC theories (analytic) \label{app:analytic}}

Here, we present an analytic proof of the size-extensivity of $\Delta$MP and $\Delta$CC theories in three steps:\ (1) 
Prove the size-extensivity of the perturbation corrections to the grand potential $\Omega^{(n)}$ 
of finite-temperature MBPT;\cite{HirataJCP2021} (2) Prove the size-extensivity of the perturbation corrections to the energy $\tilde E_I^{(n)}$ of $\Delta$MP theory;\cite{hirschfelder,Lindgren1974,HoseKaldor1979,Kucharski1989} (3) Prove 
the size-extensivity of the energy $\tilde E_I$ of $\Delta$CC theory. (Subscript $I$ labels states, while a tilde designates an eigenvalue of the respective effective Hamiltonian matrix.)

Let us first document the well-known analytic proof of the size-extensivity of single-reference RSPT. In the latter, 
the $n$th-order perturbation correction $E^{(n)}$ to the energy $E$ is the $n$th derivative 
of the $E$ defined exactly with a perturbation-scaled Hamiltonian $\hat{H}=\hat{H}_0 + \lambda \hat{V}$,
\begin{eqnarray}
E^{(n)} &\equiv& \left. \frac{1}{n!}\frac{\partial^n E(\lambda)}{\partial \lambda^n}\right|_{\lambda=0}, \label{RSPT}
\end{eqnarray}
where $\lambda$ is the dimensionless perturbation strength. Since the exact $E(\lambda)$ always 
has the correct volume dependence of $O(V)$ regardless of the value of $\lambda$ ($0 \leq \lambda \leq 1$), 
the perturbation correction $E^{(n)}$ is also $O(V)$ 
at any order (see page 156 of Shavitt and Bartlett\cite{shavitt}). Hence, single-reference RSPT can be proven to be size-extensive, independently of 
the linked-diagram theorem.\cite{shavitt,GellmannLow1951,Brueckner1955,Goldstone1957,Hugenholtz1957,Frantz,Manne} 

This proof is not valid for Brillouin--Wigner perturbation theory (BWPT) because its perturbation corrections do not conform to the derivative definition of 
Eq.\ (\ref{RSPT}). In fact, BWPT is generally not size-extensive.

This strategy does not work, either, for degenerate RSPT at least directly. 
This is because the theory determines perturbation corrections to energies for $M$ degenerate references simultaneously,
\begin{eqnarray}
E_I^{(n)} &\equiv& \left. \frac{1}{n!}\frac{\partial^n E_I(\lambda)}{\partial \lambda^n}\right|_{\lambda=0}, \label{degRSPT}
\end{eqnarray}
for $1 \leq I \leq M$. When $M$ increases with volume, the whole 
notion of the volume dependence of the individual $E_I$ loses validity.

However, this strategy can still be utilized indirectly to prove the size-extensivity of degenerate RSPT, which encompasses $\Delta$MP theory.
We first apply it to finite-temperature MBPT,\cite{HirataJCP2021} 
whose thermodynamic functions sum over  the $\Delta$MP energies. For instance, the $n$th-order 
perturbation correction to the grand potential $\Omega^{(n)}$ in the grand canonical ensemble\cite{HirataJCP2021} is defined as
\begin{eqnarray}
\Omega^{(n)} &\equiv& \left. \frac{1}{n!}\frac{\partial^n \Omega(\lambda)}{\partial \lambda^n}\right|_{\lambda=0}, \label{Omega}
\end{eqnarray}
which is clearly $O(V)$ because the exact $\Omega$ is $O(V)$ at all values of $\lambda$ ($0 \leq \lambda \leq 1$). On the other hand,
$\Omega$ and the $I$th state energy $E_I$ are related to each other by
\begin{eqnarray}
\Omega &\equiv& -\frac{1}{\beta} \ln \sum_I e^{-\beta(E_I - \mu N_I)}, \label{Omega22} 
\end{eqnarray}
where $\beta = (k_\text{B}T)^{-1}$ is the inverse temperature, $\mu$ is the chemical potential, $N_I$ is the number of electrons in the $I$th state,
and the summation is taken over all states $I$ with any number of electrons.

The perturbation corrections to $\Omega$ are related to the perturbation corrections to the $I$th state energy $\tilde E_I^{(n)}$ by\cite{HirataJCP2021} 
\begin{eqnarray}
&& \Omega^{(n)} = \langle D_I^{(n)} \rangle \nonumber\\
&&+ \frac{(-\beta)}{2!} \sum_{i=1}^{n-1} \left( \langle D_I^{(i)}D_I^{(n-i)} \rangle - \Omega^{(i)}\Omega^{(n-i)}  \right) \nonumber\\
&&+ \frac{(-\beta)^2}{3!} \sum_{i=1}^{n-2} \sum_{j=1}^{n-i-1} \left( \langle D_I^{(i)}D_I^{(j)} D_I^{(n-i-j)}\rangle - \Omega^{(i)}\Omega^{(j)}\Omega^{(n-i-j)}  \right) \nonumber\\
&& + \dots + \frac{(-\beta)^{n-1}}{n!}  \left\{ \langle (D_I^{(1)})^n\rangle - (\Omega^{(1)})^n  \right\}, \label{Omega_n}
\end{eqnarray}
where $D_I^{(n)} = \tilde E_I^{(n)} - \mu^{(n)} N_I$ and $\langle X_I \rangle$ denotes the zeroth-order thermal average of $X_I$ over all states $I$,
\begin{eqnarray}
\langle X_I \rangle \equiv \frac{\sum_I X_I e^{-\beta (E_I^{(0)} - \mu^{(0)} N_I )}}{\sum_I e^{-\beta (E_I^{(0)} - \mu^{(0)} N_I )}}.
\end{eqnarray}
Here, $\tilde E_I^{(n)}$ is the $I$th eigenvalue of the effective Hamiltonian matrix of $\Delta$MP theory (not of MP theory) 
because many zeroth-order references are exactly degenerate. 

Let us assume for a moment that $\Delta$MP theory is {\it not} size-extensive. Recalling that Eq.\ (\ref{Omega_n})
is true for any perturbation order $n$ at any temperature $T$, we see that its first term 
immediately renders $\Omega^{(n)}$ nonextensive, contradicting
the size-extensivity of finite-temperature MBPT just established. It thus proves that the initial assumption is false and that $\Delta$MP theory is size-extensive. 
It also imples that systematic cancellations of 
nonextensive second and higher powers of energies take place in all subsequent terms of Eq.\ (\ref{Omega_n}). This has indeed been borne out,
not within each sum, but holistically across all sums.\cite{HirataJCP2021}

Since $\tilde E_I$ of $\Delta$CC theory is an infinite partial summation of the $\Delta$MP energies $\tilde E^{(n)}_I$ (see Sec.\ \ref{sec:HCPT} and Appendix \ref{app:DCC}), if the latter are $O(V)$, the former is also $O(V)$.
This concludes the analytic proof of the size-extensivity of $\Delta$CC theory.
Numerical tests (not shown) support this conclusion.

\subsection{Energy diagrams of MP theory}

\begin{figure}
\includegraphics[scale=0.6]{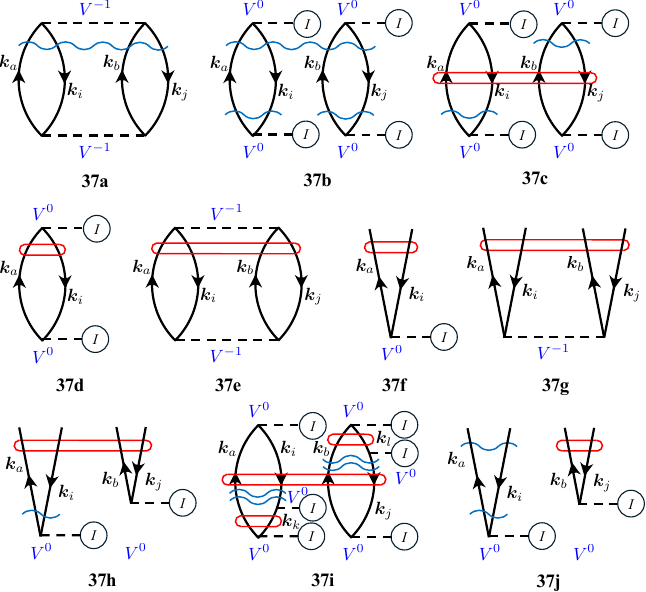}
\caption{Diagrams of MP theory, $\Delta$MP theory, or finite-temperature MBPT for a $d$-dimensional HEG. 
Lines are labeled by wave vectors, and the volume dependence of vertexes are indicated. 
}
\label{fig:extensivity}
\end{figure}

Let us first review the well-established diagrammatic criteria of size-extensivity of a single-reference theory:\ When its 
diagrammatic equations are linked, the theory is size-extensive (an unlinked diagram is a disconnected diagram at least one of whose disconnected parts is closed).\cite{shavitt,HirataTCA,HirataARPC,Nooijen_extensivity2005} 
We take MP theory (an instance of single-reference RSPT)  as an example. 

We illustrate the $k$-vector counting method of determining the volume dependence of a diagram's value. Specifically, we show 
how a linked, closed (thus connected) diagram expresses an extensive energy. Let us take the second-order correlation correction to the energy of an infinite
crystal of $d$-dimensional homogeneous electron gas (HEG) as an example. One of its energy diagrams is drawn as diagram  {\bf 37a} of Fig.\ \ref{fig:extensivity}.
Each two-electron integral vertex has the volume dependence of $O(V^{-1})$,\cite{HirataTCA,HirataARPC}  whereas each line (indexed by wave vector $\bm{k}$) 
can take one of the $O(V)$ values. Of the four $\bm{k}$ vectors, however, only three are linearly independent because
of the momentum conservation law,
\begin{eqnarray}
\bm{k}_i + \bm{k}_j \equiv \bm{k}_a + \bm{k}_b, \label{mom}
\end{eqnarray} 
modulo reciprocal lattice vectors. The summation over these three independent $\bm{k}$ vectors contributes $O(V^3)$ to the 
volume dependence. Together, the correlation energy scales correctly as $O(V^{-2}V^3) = O(V)$.
Here, the resolvent line (blue wiggly line), which denotes a division by an $O(V^0)$ orbital energy difference, has no impact on the volume dependence. 

In contrast, unlinked (closed, disconnected) diagram {\bf 37b} has the nonphysical volume dependence of $O(V^2)$:\ Each one-electron integral vertex
is an $O(V^0)$ quantity.\cite{HirataTCA,HirataARPC} Of the four $\bm{k}$ vectors, only two are independent because they obey two momentum conservation laws,
\begin{eqnarray}
\bm{k}_i \equiv \bm{k}_a, \quad \bm{k}_j \equiv \bm{k}_b,
\end{eqnarray} 
modulo reciprocal lattice vectors, making the diagram scale as $O(V^0V^2) = O(V^2)$. 

The linked-diagram theorem\cite{shavitt,GellmannLow1951,Brueckner1955,Goldstone1957,Hugenholtz1957,Frantz,Manne}  proved
that in MP theory, all unlinked diagrams systematically cancel out, leaving fully linked energy and amplitude (wave-function) equations. 


\subsection{Diagonal energy diagrams of $\Delta$MP theory\label{app:multiref}}

Next, let us consider a hypothetical, diagonal energy diagram {\bf 37c} in the context of $\Delta$MP theory. 
It is the same as the unlinked diagram {\bf 37b} except that an actual resolvent line (blue wiggly line) is replaced 
by a fictitious resolvent line (red oblong), imposing the degeneracy condition,
\begin{eqnarray}
|\bm{k}_i|^2 + |\bm{k}_j|^2 = |\bm{k}_a|^2 + |\bm{k}_b|^2,
\end{eqnarray} 
which would affect the $k$-vector counting logic. 
This condition reduces the volume dependence of the value of diagram {\bf 37c} by a factor of $V^{1/d}$, bringing its scaling down to $O(V^{2-1/d})$.  
Hence, diagram {\bf 37c}, despite being unlinked, scales correctly as $O(V)$ in one-dimensional HEG, but still incorrectly in 
two- and three-dimensional HEG. 

In short, the extensivity of a closed, connected diagram is predicated on the fact that all $k$ vectors 
are allowed to take any values 
under the only constraints of momentum conservation laws (whose violation makes the corresponding vertex's value zero).\cite{shavitt,HirataTCA,HirataARPC} Since a diagram with
a fictitious resolvent line is a small fraction of the isomorphic diagram without such lines, its volume dependence varies depending on the nature of a degenerate
multireference.

(The size-extensivity  proof\cite{LiPaldus2003} of the Li--Paldus MRCC theory 
seems to be based on the implicit assumption that diagrams can be connected by a model-space vertex, which in our case
corresponds to a fictitious resolvent line. See, e.g., Figs.\ 1 and 2 of Ref.\ \onlinecite{LiPaldus2003}. 
This argument seems valid if and only if the GMS is intensive,  
strongly attenuating the volume dependence of every diagram containing a model-space vertex. 
Since the user of the Li--Paldus theory has complete freedom to construct a GMS, the theory may be considered 
size-extensive insofar as the GMS is chosen carefully.\cite{Meissner1989}  For any arbitrary GMS, however, the Li--Paldus theory is highly unlikely to be size-extensive, 
as pointed out by Nooijen {\it et al.}\cite{Nooijen_extensivity2005}\ and by Liu {\it et al.}\cite{Liu2026} A size-extensive multireference theory for any GMS 
may not exist in the first place!\cite{conjecture})

In $\Delta$MP theory,  we do not have such freedom and 
a degenerate multireference grows as $O(V^{2/3})$ in  likely the most highly degenerate system of three-dimensional HEG; we 
cannot assume that a fictitious resolvent line has the ability to connect otherwise disconnected parts of a diagram. Diagram {\bf 37c} 
should thus be deemed disconnected (unlinked), destroying the size-extensivity. 

Therefore, a diagonal energy diagram of $\Delta$MP theory must be connected (linked) for it to be size-extensive. Generally, a size-extensive degenerate theory
is characterized by a connected diagonal energy equation. 

\subsection{Off-diagonal energy diagrams of $\Delta$MP theory\label{app:diag}}

Meissner {\it et al.}\cite{Meissner1989}\ argued 
that there are two sources of the loss of size-extensivity in a multireference theory:\ The effective Hamiltonian and its subsequent
diagonalization.\cite{Lindgren1985_1,Lindgren1985_2,Lindgren1987,MukherjeeAQC1989,MukhopadhyayTCA1991,Nooijen_extensivity2005} 
It is through the diagonalization that the off-diagonal elements enter the state energies.
Let us, therefore, focus on the diagonalization step to try to determine the correct volume dependence of the off-diagonal energies. 

In $\Delta$MP (or $\Delta$CC) theory, 
the purpose of the diagonalization of $\bm{E}^{(n)}$ is not so much to capture a correlation energy within the internal space in a size-extensive manner, but merely to
determine the correct linear combinations of degenerate references that converge at the FCI wave functions. It is also 
exceedingly difficult to obtain an extensive correlation energy from diagonalization,\cite{march,conjecture} not to mention that it is unclear 
what the correct volume dependence of a correlation energy in the internal space should be.

On the other hand, owing to the trace invariance,
\begin{eqnarray}
\text{Tr}\, \bm{E}^{(n)} &=& \sum_{I=1}^{M} \tilde{E}_I^{(n)}, \label{traces}
\end{eqnarray}
where $\tilde{E}_I^{(n)}$ is the $I$th eigenvalue of $\bm{E}^{(n)}$, one can immediately see that the {\it average} of the eigenvalues 
within an internal space always scales correctly as $O(V)$ assuming that
the diagonal energies are connected (Appendix \ref{app:multiref}). 
In an extreme, but common case where the degeneracy of references is not lifted upon inclusion of electron correlation, 
$\Delta$MP theory is already size-extensive as all off-diagonal elements of $\bm{E}^{(n)}$ are zero. 
(This situation should be distinguished from a truncated CI theory, whose diagonal energies are not extensive in the first place 
because the $C$-amplitudes that enter their formulas come from unlinked equations.) 

The effects of nonzero off-diagonal elements are, therefore, to lift the degeneracy and to give some breadth
to the distribution of the eigenvalues. What is then the volume dependence of this breadth? Defining it by the variance,
\begin{eqnarray}
\text{Var}\left[\tilde{E}_I^{(n)}\right]  &\equiv& E\Big[ \left(\tilde{E}_I^{(n)}\right)^2 \Big] - \left( E\left[ \tilde{E}_I^{(n)} \right] \right)^2, \label{variances}
\end{eqnarray}
where $E[ X_I ]$ denotes the average of $X_I$ over all states $I$, we see that it should be an $O(V)$ quantity because it is proportional to
the heat capacity, which is extensive, as per the fluctuation-dissipation theorem.\cite{Reichl}  

This is also supported by finite-temperature MBPT,\cite{HirataJCP2021,Hirata_PRA2024}  which supplies  formulas for the quantities related to the variance.
For example, the second-order grand potential $\Omega^{(2)}$ contains a factor,
\begin{eqnarray}
\langle (\tilde{E}_I^{(1)})^2 \rangle - \langle \tilde{E}_I^{(1)} \rangle^2 
&=& \sum_{p,q}^{\text{denom.}=0} |f^p_q(T)|^2 f_p^- f_q^+  \nonumber\\
&& + \frac{1}{4} \sum_{p,q,r,s}^{\text{denom.}=0} |v^{pq}_{rs}|^2 f_p^- f_q^- f_r^+ f_s^+, \nonumber\\ \label{Omega2}
\end{eqnarray}
where $f^p_q(T)$ is the finite-temperature Fock matrix element, and 
$f_p^-$ and $f_p^+$ are the Fermi--Dirac occupancy and vacancy functions, respectively, for the $p$th orbital.
The summations are 
restricted (``denom.$=0$'') to those cases whose fictitious denominators are zero ($\epsilon_q - \epsilon_p = 0$ in the first term and $\epsilon_r + \epsilon_s - \epsilon_p - \epsilon_q = 0$
in the second term). Equation (\ref{Omega2}) is a part of $\Omega^{(2)}$, and is therefore an $O(V)$ quantity. Since the variance is the high-temperature limit of 
Eq.\ (\ref{Omega2}), it also scales as $O(V)$.

The diagrammatic representations of the two terms of Eq.\ (\ref{Omega2}) are drawn as  {\bf 37d} and {\bf 37e}, whose ``denom.$=0$'' conditions are represented  
by the same red oblong denoting a fictitious resolvent line. Other than the fact that the up- and down-going lines run over all orbitals
and accompany the $f_p^-$ and $f_p^+$ factors, these finite-temperature diagrams are connected, having 
the same correct volume dependence of $O(V)$ as their zero-temperature counterparts. 
In short, the variance should scale as $O(V)$ and its diagrams must be connected. 
(They are called anomalous diagrams.\cite{kohn,HirataJCP2021}) 

What is then the volume dependence 
of the off-diagonal elements of $\bm{E}^{(n)}$ that is consistent with the $O(V)$ dependence of the variance?
Using the trace invariance of a cyclic product of matrices, we can rewrite Eq.\ (\ref{Omega2}) as
\begin{eqnarray}
\langle (\tilde{E}_I^{(1)})^2 \rangle - \langle \tilde{E}_I^{(1)} \rangle^2 
&=&  \langle E_{IJ}^{(1)} E_{JI}^{(1)}  \rangle, \label{EJI}
\end{eqnarray}
where the summation is implied over all $J$th states ($J\neq I$) that are degenerate with the $I$th state at the zeroth order;
the thermal average is then taken over all states $I$. Comparing this with Eq.\ (\ref{Omega2}), we see that the connectedness of diagram {\bf 37d}
is ensured by the connectedness of diagram {\bf 37f}, which represents $E_{JI}^{(1)}$ with $|J\rangle = |I^a_i\rangle$. Likewise, the connectedness of 
diagram {\bf 37e} is guaranteed by the connectedness of diagram {\bf 37g}, expressing $E_{JI}^{(1)}$ with $|J\rangle = |I^{ab}_{ij}\rangle$.

Generally, perturbation corrections to grand potential  consist of cumulants of thermal averages\cite{HirataJCP2021}  such as 
\begin{eqnarray}
&& \langle {E}_{IJ}^{(2)} {E}_{JI}^{(1)} \rangle - \langle {E}_{I}^{(2)} \rangle\langle {E}_{I}^{(1)} \rangle , \\
&& \langle {E}_{IJ}^{(1)} {E}_{JK}^{(1)}  {E}_{KI}^{(1)} \rangle - 3 \langle {E}_{I}^{(1)} \rangle\langle {E}_{IJ}^{(1)}{E}_{JI}^{(1)} \rangle 
- \langle {E}_{I}^{(1)} \rangle\langle {E}_{I}^{(1)} \rangle\langle {E}_{I}^{(1)} \rangle, \nonumber\\
\end{eqnarray}
where summations over $J$ and $K$ are implied. It is easy to see that these thermodynamic functions and thus
 the average and variance of the eigenvalues of $\bm{E}^{(n)}$ in $\Delta$MP theory
have the correct volume dependence when all off-diagonal elements of $\bm{E}^{(n)}$ are connected.
Conversely, if a disconnected diagram such as {\bf 37h}
persists, an unlinked diagram such as {\bf 37c} ensues, destroying the size-extensivity. 

Hence, the diagrammatic criterion for the off-diagonal energies of a size-extensive degenerate theory is that they be connected. 
(The connectedness must not be confused with extensivity; an off-diagonal ``energy'' is not extensive. 
An energy diagram with two dangling lines, e.g., diagram {\bf 37f}, 
scales as $O(V^0)$, whereas one with four dangling lines such as  {\bf 37g} is an $O(V^{-1})$ quantity. See Refs.\ \onlinecite{HirataTCA,HirataARPC}.)

However, this diagrammatic criterion may be too strict; the smaller volume of a degenerate multireference should temper the volume dependence of 
a disconnected diagram with one or more fictitious resolvent lines. For example, a two-part closed disconnected diagram
with three fictitious resolvent lines scales correctly as $O(V)$ even in three-dimensional HEG and should be allowed to 
remain in a size-extensive theory. Such an example is given as diagram {\bf 37i}. It is a sixth-order variance diagram $\langle E_{IJ}^{(3)}E_{JI}^{(3)} \rangle$ 
formed by contracting the off-diagonal energy diagram {\bf 48d} of Appendix \ref{app:DMP} with its Hermitian conjugate. 
It is disconnected, but scales as $O(V^{2-3/d})$ in $d$-dimensional HEG, with each fictitious resolvent line responsible for a factor of $V^{-1/d}$. 
This is the correct scaling of $O(V)$  (or lower) for the variance   even in three-dimensional HEG, suggesting that the disconnected, off-diagonal energy diagram {\bf 48d} may 
be permissible in $\Delta$MP theory. 

The foregoing analysis suggests that the diagrammatic size-extensivity criteria for a multireference theory are not as instructive or definitive as they are for a single-reference theory. 
The connectedness of the off-diagonal energies is one of the sufficient conditions and not a necessary one.

\subsection{The broadened definition of linkedness\label{app:Sandars}}

While diagram {\bf 37h} is an off-diagonal energy, isomorphic diagram {\bf 37j} represents a wave function,
whose disconnected part on the right, in turn, expresses the off-diagonal energy $E_{JI}^{(1)}$ with $|J\rangle = |I^b_j\rangle$. 
A disconnected wave-function diagram whose disconnected part denotes an energy will always spawn 
a disconnected energy diagram. Therefore, we consider it unlinked in
the broadened definition of linkedness.\cite{Sandars1969,shavitt,Kucharski1989,HoseKaldor1979}
In other words, an apparently open disconnected part is considered closed if it is intersected by a fictitious resolvent line at the top. 
Diagram {\bf 37j} is deemed unlinked because its disconnected part on the right is closed by the fictitious resolvent line at the top. 


\subsection{Size-extensivity of $\Delta$MP theory (diagrammatic) \label{app:DMP}}

To summarize the foregoing analysis, 
a size-extensive degenerate theory is defined by the connected (i.e., linked) equations for both diagonal and off-diagonal energies and by the linked amplitude equations
in the broadened definition of the linkedness.\cite{Sandars1969,HoseKaldor1979,Kucharski1989,shavitt} 
Here, the linkedness of the amplitude equation is only a necessary condition for the connectedness (linkedness) of the energies, and disconnected energy diagrams may 
emerge from a linked amplitude equation. Therefore, both energy and amplitude equations need to be checked for linkedness at each order. 
When these conditions are met, the energy matrix $\bm{E}$ has the eigenvalues that scale correctly with volume, and
the theory is size-extensive with respect to both external- and internal-space extensions. They are sufficient conditions, and a degenerate theory that does not strictly 
obey them can still be size-extensive. 

Here, we illustrate the size-extensivity of $\Delta$MP theory on the basis of these diagrammatic cconditions up to the third perturbation order, retaining only the one-electron part 
of the Hamiltonian to avoid proliferation of diagrams. Our argument follows
that of Jeziorski and Monkhorst,\cite{Jeziorski1981} which challenged the size-extensivity proof by Hose and Kaldor,\cite{HoseKaldor1979} if not their conclusion.
What follows is not a proof, but an illustration.

In this simplified $\Delta$MP theory, the $n$th-order perturbation corrections to energy and wave function are still given recursively by
\begin{eqnarray}
E^{(n)}_{JI} &=& \langle J^{(0)} | \hat{V}_I | I^{(n-1)} \rangle,\label{HCPTrecursion1again} \\
|I^{(n)}\rangle &=& \hat{R}_I \Big( \hat{V}_I |I^{(n-1)} \rangle - \sum_{i=1}^{n-1} \sum_{J=1}^M |J^{(n-i)}\rangle  E^{(i)}_{JI} \Big) \label{HCPTrecursion2again}
\end{eqnarray}
with 
\begin{eqnarray}
\hat{V} _I= E_{II}^{(1)} + \sum_{p,q} \left( (f_I)^p_q - \epsilon_p \delta_{pq} \right) \left\{ \hat{p}^\dagger \hat{q} \right\}.
\end{eqnarray}

\begin{figure}
\includegraphics[scale=0.6]{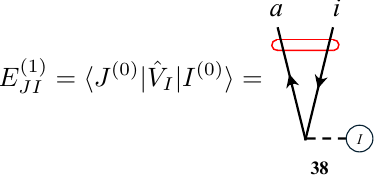}
\caption{The diagram for the first-order correction to the off-diagonal energy $E_{JI}^{(1)}$ with the one-electron perturbation  $\hat{V}_I$. It is connected. The fictitious resolvent line (red oblong) 
demands $\epsilon_i - \epsilon_a = 0$, where $|J\rangle = |I^a_i\rangle$.}
\label{fig:EJI1}
\end{figure}

\begin{figure}
\includegraphics[scale=0.6]{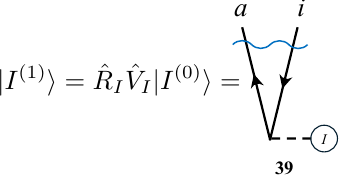}
\caption{The diagram for the first-order correction to wave function with the one-electron perturbation  $\hat{V}_I$. It is connected. The resolvent line (blue wiggly line) 
represents $(\epsilon_i - \epsilon_a)^{-1}$.}
\label{fig:I1}
\end{figure}

The first-order correction to the diagonal energy, $E_{II}^{(1)}$, is extensive. 
The first-order correction to the off-diagonal energy, $E_{JI}^{(1)} = \langle J | \hat{V}_I | I \rangle$, is given diagrammatically in Fig.\ \ref{fig:EJI1}. It is connected.
The first-order wave function, $|I^{(1)}\rangle = \hat{R}_I \hat{V}_I|I\rangle$, is drawn in Fig.\ \ref{fig:I1}. It is also connected (and thus linked).

\begin{figure}
\includegraphics[scale=0.6]{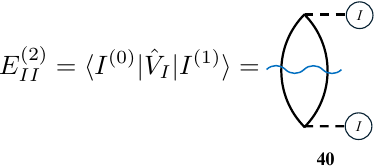}
\caption{The skeleton diagram for the second-order correction to the diagonal energy $E_{II}^{(2)}$ with the one-electron perturbation  $\hat{V}_I$. It is connected. }
\label{fig:EII2}
\end{figure}

\begin{figure}
\includegraphics[scale=0.6]{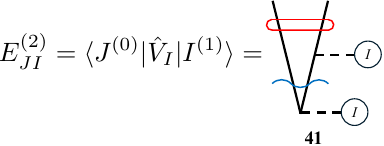}
\caption{The skeleton diagram for the second-order correction to the off-diagonal energy $E_{JI}^{(2)}$ with the one-electron perturbation  $\hat{V}_I$, where $|J\rangle = |I^a_i\rangle$
and $\epsilon_i - \epsilon_a = 0$. It is connected. }
\label{fig:EJI2_singles}
\end{figure}

\begin{figure}
\includegraphics[scale=0.6]{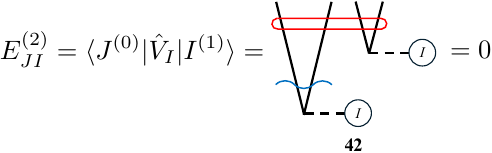}
\caption{The skeleton diagram for the second-order correction to the off-diagonal energy $E_{JI}^{(2)}$ with the one-electron perturbation  $\hat{V}_I$,  where $|J\rangle = |I^{ab}_{ij}\rangle$
and $\epsilon_i + \epsilon_j - \epsilon_a - \epsilon_b = 0$. It is disconnected, but shown to vanish by the factorization theorem.}
\label{fig:EJI2_doubles}
\end{figure}

The second-order correction to the diagonal energy, $E_{II}^{(2)} = \langle I | \hat{V}_I | I^{(1)} \rangle = \langle I | \hat{V}_I \hat{R}_I \hat{V}_I | I \rangle$, 
is drawn in Fig.\ \ref{fig:EII2}. It is connected. 
The second-order off-diagonal energies, $E_{JI}^{(2)} = \langle J| \hat{V}_I | I^{(1)} \rangle = \langle J | \hat{V}_I \hat{R}_I \hat{V}_I | I \rangle$,
are depicted in Figs.\ \ref{fig:EJI2_singles} and \ref{fig:EJI2_doubles} for $|J\rangle = |I^a_i\rangle$ and $|J\rangle = |I^{ab}_{ij}\rangle$, respectively.
Diagram {\bf 41} is connected, but diagram {\bf 42} is disconnected and should not persist. In fact, the factorization theorem of Frantz and Mills\cite{Frantz,HoseKaldor1979}
(see also Hose and Kaldor\cite{HoseKaldor1979})
can be invoked to show that the latter is zero because the denominators cancel out as
\begin{eqnarray}
\sum_{a,b,i,j}^{\substack{\epsilon_i  - \epsilon_a + \epsilon_j - \epsilon_b = 0 \\ \epsilon_i - \epsilon_a \neq 0}} \frac{1}{\epsilon_i - \epsilon_a} &=& 
 \sum_{a,b,i,j}^{\substack{\epsilon_i - \epsilon_a + \epsilon_j - \epsilon_b = 0 \\ \epsilon_i - \epsilon_a \neq 0  \\ \epsilon_j - \epsilon_b \neq 0}} \left( \frac{1/2}{\epsilon_i - \epsilon_a} + \frac{1/2}{\epsilon_j - \epsilon_b} \right) \nonumber\\
&=& \sum_{a,b,i,j}^{\substack{\epsilon_i - \epsilon_a + \epsilon_j - \epsilon_b = 0 \\ \epsilon_i - \epsilon_a \neq 0  \\ \epsilon_j - \epsilon_b \neq 0}} \frac{(\epsilon_i - \epsilon_a + \epsilon_j - \epsilon_b) }{2(\epsilon_i - \epsilon_a)(\epsilon_j - \epsilon_b)} \nonumber\\
&=&0.
\end{eqnarray}
Therefore, the second-order energy equation is connected.

\begin{figure}
\includegraphics[scale=0.6]{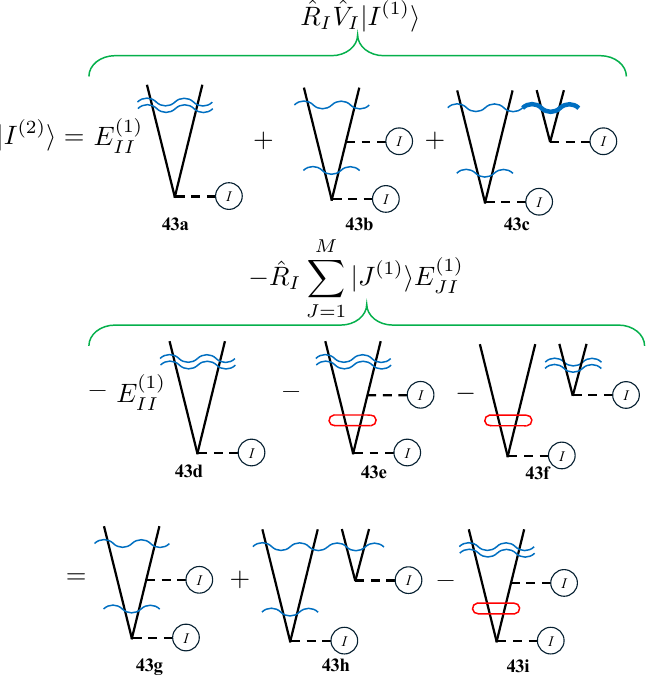}
\caption{Skeleton diagrams for the second-order correction to wave function with the one-electron perturbation  $\hat{V}_I$.
Two types of the resolvent line are distinguished by the bold blue wiggly line and thin blue wiggly line.
The thin part's partial denominator spanning a disconnected part cannot be zero, but the bold part's partial denominator can be.
The final results are linked. See the subsequent two figures and corresponding text for the explanations of the equalities.}
\label{fig:I2}
\end{figure}

\begin{figure}
\includegraphics[scale=0.6]{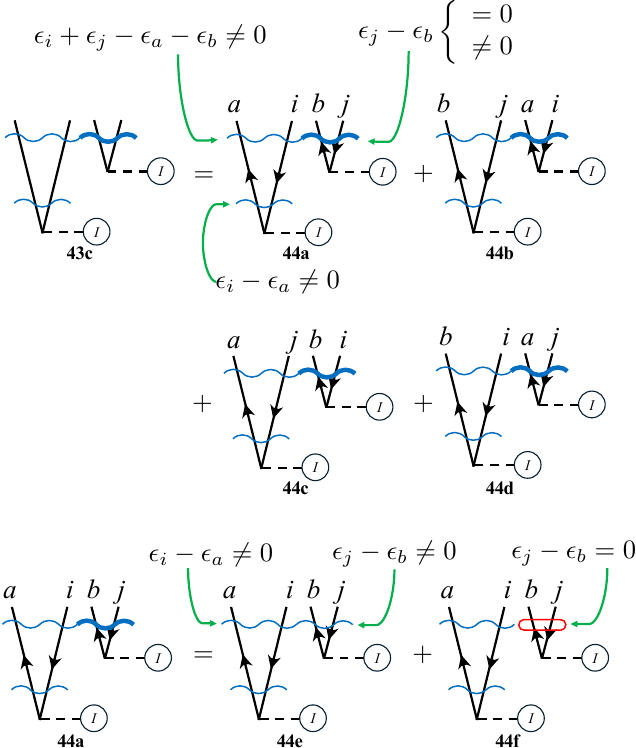}
\caption{Four individual diagrams collectively represented by skeleton diagram {\bf 43c}. Each of them (e.g., {\bf 44a}) is further split 
into linked ({\bf 44e}) and unlinked ({\bf 44f}) diagrams. }
\label{fig:I2_2}
\end{figure}

\begin{figure}
\includegraphics[scale=0.6]{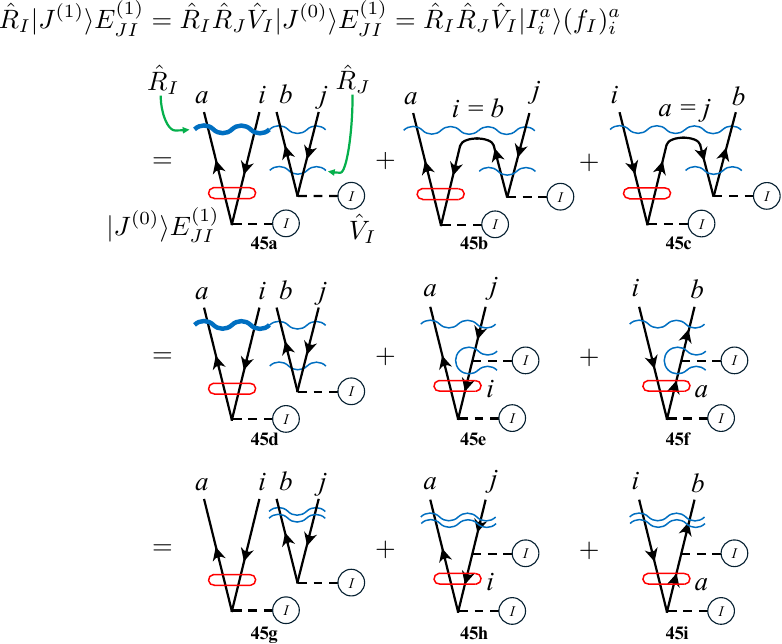}
\caption{A transformation of the diagrams in the renormalization term of $|I^{(2)}\rangle$, where $|J\rangle = |I^a_i\rangle$ and $\epsilon_i - \epsilon_a = 0$.
Every $f_I$ vertex denotes the Fock operator of the $I$th state because these diagrams and underlying normal ordering are relative to the $|I\rangle$ vacuum.}
\label{fig:I2_3}
\end{figure}

The skeleton diagrams of the second-order wave function, 
\begin{eqnarray}
|I^{(2)}\rangle &=& \hat{R}_I\hat{V}_I|I^{(1)}\rangle - \hat{R}_I\sum_J |J^{(1)}\rangle E_{JI}^{(1)} 
, \label{I2}
\end{eqnarray} 
are drawn in Fig.\ \ref{fig:I2}. The first three diagrams come from the principal term, whereas the subsequent three 
from the renormalization term. The negative sign on the renormalization term is not absorbed 
by diagrams in this article to highlight their partial cancellation with the principal term, contrary to the convention.\cite{Kucharski1989,HirataJCP2021} 

There are two subtle, but important differences
in the diagrams between MP and $\Delta$MP theories. One is that the denominator associated with a {\it partial} resolvent line spanning a disconnected part
may be positive, negative, or zero in $\Delta$MP theory. This may hide an unlinked contribution
within a seemingly linked diagram\cite{Kucharski1989} and also interfere with the application of the factorization theorem.\cite{HoseKaldor1979} Starting with this figure, we use a bold blue wiggly line 
to indicate the portion of the resolvent line spanning a disconnected part whose associated partial denominator can be zero, while a thin blue wiggly line contraindicates. 
The other is that not all time orderings of vertexes occur in $\Delta$MP theory. When folded diagrams are straightened into 
folded-resolvent diagrams\cite{Kucharski1989} by adopting the $|I\rangle$ vacuum,\cite{HoseKaldor1979} 
the vertexes in $E_{JI}^{(n)}$ always appear earlier (lower). This will also interfere with the application of the factorization theorem. 

Going back to Fig.\ \ref{fig:I2}, diagrams {\bf 43a} and {\bf 43d} are unlinked, and they cancel each other out. They are the only unlinked
diagrams in MP theory, but in $\Delta$MP theory, diagram {\bf 43f} is also unlinked because the disconnected part on the left 
is closed by the fictitious resolvent line. 
Diagram {\bf 43c} also hides an unlinked contribution because the bold part of the resolvent line allows the corresponding partial
denominator to be zero. This unlinked contribution is easily identified as equal to diagram {\bf 43f}, and hence they cancel each other.
Together, $|I^{(2)}\rangle$ consists of three diagrams {\bf 43g}, {\bf 43h}, and {\bf 43i}, all of which are linked.

The details of how diagram {\bf 43c} is split into linked and unlinked diagrams are 
given graphically in Fig.\ \ref{fig:I2_2}. Skeleton diagram {\bf 43c} stands for four individual diagrams
whose dangling line indices ($i,j,a,b$) are anti-symmetrized, giving rise to all time orderings of vertexes. In diagram {\bf 44a}, the lower thin resolvent line
denotes $(\epsilon_i - \epsilon_a)^{-1}$ and implies $\epsilon_i - \epsilon_a \neq 0$. The upper resolvent line
stands for $(\epsilon_i + \epsilon_j - \epsilon_a - \epsilon_b)^{-1}$ and requires $\epsilon_i + \epsilon_j - \epsilon_a - \epsilon_b \neq 0$.
The thin part of the upper resolvent line demands $\epsilon_i - \epsilon_a \neq 0$, whereas the bold part permits
$\epsilon_j - \epsilon_b = 0$.
Therefore, diagram {\bf 44a} consists of linked diagram {\bf 44e} ($\epsilon_j - \epsilon_b \neq 0)$  and unlinked diagram {\bf 44f}
($\epsilon_j - \epsilon_b = 0$), the latter canceling diagram {\bf 43f}. 

Figure \ref{fig:I2_3} illustrates how diagrams {\bf 43e} and {\bf 43f} emerge from the renormalization term, i.e., the second term
of Eq.\ (\ref{I2}). For $|J\rangle = |I^a_i\rangle$ with $\epsilon_i - \epsilon_a = 0$, we can form one unlinked diagram {\bf 45a}
and two folded diagrams {\bf 45b} and {\bf 45c}. In each case, $b$ and $j$ label a virtual and occupied orbital, respectively, in $|J\rangle$, 
and $\hat{R}_J$ introduces a factor of $(\epsilon_j - \epsilon_b)^{-1}$. In diagram {\bf 45b}, orbital $b$ is identified as orbital $i$, which is also a virtual orbital in $|J\rangle = |I^a_i\rangle$.
In diagram {\bf 45c}, orbital $j$ turns out to be orbital $a$, which is occupied in $|J\rangle$. We can transform the folded diagrams
into folded-resolvent diagrams\cite{Kucharski1989} ({\bf 45e} and {\bf 45f}) by lifting the top vertexes and straightening the lines at the expense of warping the resolvent lines. In this way, 
the line directions indicate the hole/particle distinction relative to the $|I\rangle$ vacuum  and the vertexes and fictitious resolvent lines originating from $E_{JI}^{(n)}$
tend to occur at the bottom. We can furthermore straighten the warped resolvent lines, arriving at the desired diagrams ({\bf 45h} and {\bf 45i}). 
Not all time orderings of vertexes arise when a fictitious resolvent line is present. 

\begin{figure}
\includegraphics[scale=0.6]{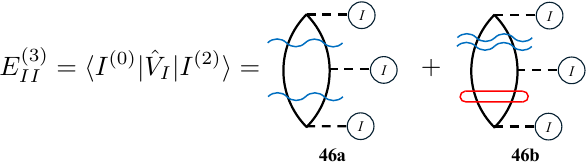}
\caption{Skeleton diagrams for the third-order correction to the diagonal energy $E_{II}^{(3)}$ with the one-electron perturbation  $\hat{V}_I$. They are connected.}
\label{fig:EII3}
\end{figure}

\begin{figure}
\includegraphics[scale=0.6]{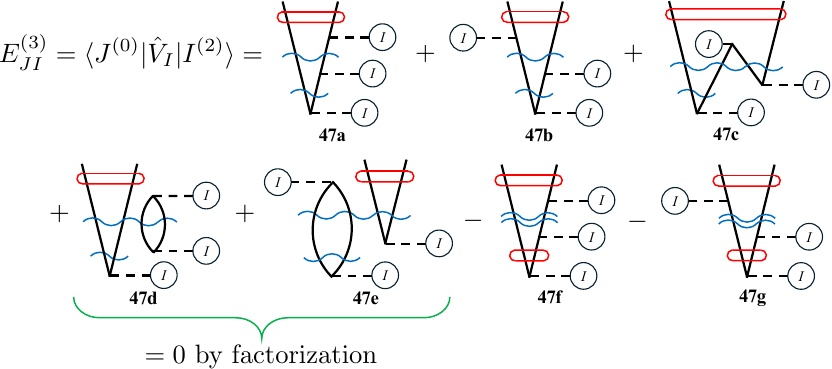}
\caption{Skeleton diagrams for the third-order correction to the off-diagonal energy $E_{JI}^{(3)}$ with the one-electron perturbation  $\hat{V}_I$, where $|J\rangle = |I^a_i\rangle$
and $\epsilon_i - \epsilon_a = 0$. The disconnected diagrams sum to zero by the factorization theorem, leaving only the connected diagrams.}
\label{fig:EJI3_singles}
\end{figure}

\begin{figure}
\includegraphics[scale=0.6]{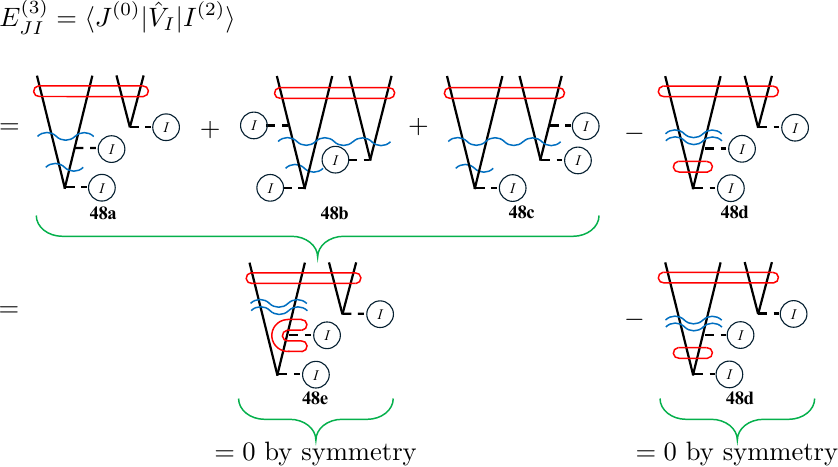}
\caption{Skeleton diagrams for the third-order correction to the off-diagonal energy $E_{JI}^{(3)}$ with the one-electron perturbation  $\hat{V}_I$,  where $|J\rangle = |I^{ab}_{ij}\rangle$
and $\epsilon_i + \epsilon_j - \epsilon_a - \epsilon_b = 0$. They are disconnected, but sum to zero by symmetry.}
\label{fig:EJI3_doubles}
\end{figure}

\begin{figure}
\includegraphics[scale=0.6]{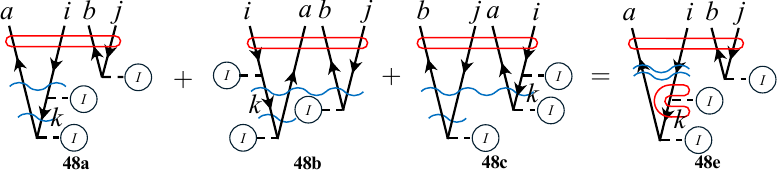}
\caption{An explanation for the diagrammatic identity (${\bf 48a}+{\bf 48b}+{\bf 48c}={\bf 48e}$) in Fig.\ \ref{fig:EJI3_doubles}.}
\label{fig:EJI3_doubles_2}
\end{figure}

Let us consider the third order. Figure \ref{fig:EII3} shows two skeleton diagrams for the third-order diagonal energy $E_{II}^{(3)}$, both of which are connected.
Diagram {\bf 46b} is isomorphic with the renormalization diagram of finite-temperature MBPT,\cite{HirataJCP2021} in which a resolvent line is shifted upward and piles up.

Figure \ref{fig:EJI3_singles} lists skeleton diagrams for the third-order off-diagonal energy $E_{JI}^{(3)}$, where $|J\rangle = |I^a_i\rangle$ and $\epsilon_i - \epsilon_a = 0$. 
They are all connected except for diagrams {\bf 47d} and {\bf 47e} that are disconnected and unlinked. However, they are individually zero, as indicated in the figure, 
because the thin resolvent line requiring $\epsilon_i - \epsilon_a \neq 0$ and the fictitious resolvent line demanding $\epsilon_i - \epsilon_a = 0$ are incompatible. 
(This is equivalent to applying the factorization theorem to them.)
Hence, $E_{JI}^{(3)}$ is connected in the single-excitation space.

Skeleton diagrams for the same third-order off-diagonal energy $E_{JI}^{(3)}$ but in the double-excitation space, i.e., with $|J\rangle = |I^{ab}_{ij}\rangle$ and $\epsilon_i + \epsilon_j - \epsilon_a - \epsilon_b = 0$ are listed in Fig.\ \ref{fig:EJI3_doubles}. They are reduced to two diagrams {\bf 48e} and {\bf 48d}, which are disconnected and unlinked.
They are, however, expected to either vanish if the degeneracy is caused by symmetry or persist but do not destroy the size-extensivity because of 
the multiple fictitious resolvent lines. The persistence of these disconnected diagrams seems to
contradict the proof of size-extensivity of degenerate RSPT by Hose and Kaldor,\cite{HoseKaldor1979} as first pointed out by Jeziorski and Monkhorst,\cite{Jeziorski1981}
although both groups seem to agree on the size-extensivity of the theory.

The first three diagrams ({\bf 48a}, {\bf 48b}, and {\bf 48c}) are consolidated into the nonzero diagram {\bf 48e}; the factorization theorem 
works only partially, even though all possible time orderings of vertexes appear. This is because in $\Delta$MP theory, 
denominator factors can be positive or negative, interfering with the factorization theorem. 
This diagrammatic identity is more fully explained in Fig.\ \ref{fig:EJI3_doubles_2}, in which the lines 
are now labeled. The three diagrams share the same numerator. Recalling the meaning of thin resolvent lines, the denominators are summed as
\begin{widetext}
\begin{eqnarray}
&& \sum_{i,j,k,a,b}^{\substack{\epsilon_i  - \epsilon_a + \epsilon_j- \epsilon_b = 0 \\ \epsilon_k - \epsilon_a \neq 0 \,;\, \epsilon_i - \epsilon_a \neq 0}}\frac{1}{\epsilon_k - \epsilon_a}\frac{1}{\epsilon_i - \epsilon_a} + \sum_{i,j,k,a,b}^{\substack{\epsilon_i  - \epsilon_a + \epsilon_j- \epsilon_b = 0 \\ \epsilon_k - \epsilon_a + \epsilon_j - \epsilon_b \neq 0 \\ \epsilon_k - \epsilon_a \neq 0  \,;\, \epsilon_j - \epsilon_b \neq 0}}\frac{1}{\epsilon_k - \epsilon_a+ \epsilon_j - \epsilon_b}\frac{1}{\epsilon_k - \epsilon_a} + \sum_{i,j,k,a,b}^{\substack{\epsilon_i  - \epsilon_a + \epsilon_j- \epsilon_b= 0 \\ \epsilon_k - \epsilon_a + \epsilon_j - \epsilon_b \neq 0 \\ \epsilon_k - \epsilon_a \neq 0 \,;\, \epsilon_j - \epsilon_b \neq 0}}\frac{1}{\epsilon_k - \epsilon_a+ \epsilon_j - \epsilon_b}\frac{1}{\epsilon_j - \epsilon_b} \nonumber\\
&& = \sum_{i,j,k,a,b}^{\substack{\epsilon_i  - \epsilon_a + \epsilon_j- \epsilon_b = 0 \\ \epsilon_k  = \epsilon_i \,;\, \epsilon_k - \epsilon_a \neq 0 \\ \epsilon_i - \epsilon_a \neq 0  \,;\, \epsilon_j - \epsilon_b \neq 0}}\frac{1}{\epsilon_k - \epsilon_a}\frac{1}{\epsilon_i - \epsilon_a} 
+ \sum_{i,j,k,a,b}^{\substack{\epsilon_i  - \epsilon_a + \epsilon_j- \epsilon_b = 0 \\ \epsilon_k  \neq \epsilon_i \,;\, \epsilon_k - \epsilon_a \neq 0 \\ \epsilon_i - \epsilon_a \neq 0 \,;\, \epsilon_j - \epsilon_b \neq 0}} \left( \frac{1}{\epsilon_k - \epsilon_a}\frac{1}{\epsilon_i - \epsilon_a} +  \frac{1}{\epsilon_k - \epsilon_a+ \epsilon_j - \epsilon_b}\frac{1}{\epsilon_k - \epsilon_a}  + \frac{1}{\epsilon_k - \epsilon_a+ \epsilon_j - \epsilon_b}\frac{1}{\epsilon_j - \epsilon_b} \right) 
\nonumber\\&& 
= \sum_{i,j,k,a,b}^{\substack{\epsilon_i  - \epsilon_a + \epsilon_j- \epsilon_b = 0 \\ \epsilon_k  = \epsilon_i \\ \epsilon_i - \epsilon_a \neq 0 \,;\, \epsilon_j - \epsilon_b \neq 0}}\frac{1}{(\epsilon_i - \epsilon_a)^2},
\end{eqnarray}
\end{widetext}
where the second term in the middle line is zero by virtue of the factorization theorem with $\epsilon_i  - \epsilon_a + \epsilon_j- \epsilon_b = 0$.
The incomplete cancellation of the three diagrams is another discrepancy from the proof of Hose and Kaldor.\cite{HoseKaldor1979} 

\begin{figure}
\includegraphics[scale=0.6]{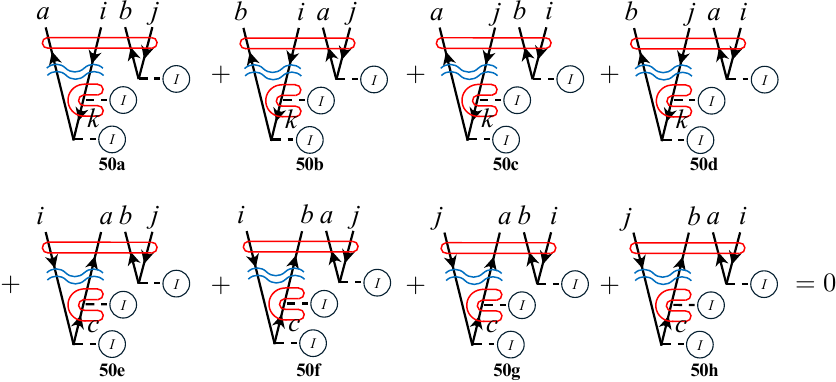}
\caption{A possible explanation as to why diagram {\bf 48e} in Fig.\ \ref{fig:EJI3_doubles} is zero by the symmetry 
responsible for causing the degeneracy. The same explanation applies to diagram {\bf 48d} being zero.}
\label{fig:EJI3_doubles_3}
\end{figure}

\begin{figure}
\includegraphics[scale=0.6]{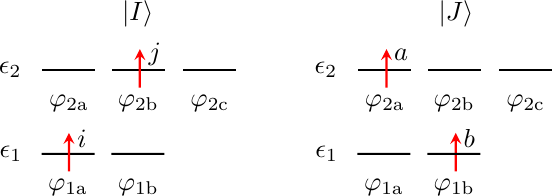}
\caption{Two degenerate reference determinants $|I\rangle$ and $|J\rangle$ that are a two-orbital difference from each other, i.e.,
$|J\rangle = |I^{ab}_{ij}\rangle$ with $\epsilon_i + \epsilon_j - \epsilon_a - \epsilon_b = 0$.
Here, $i$, $j$, $a$, and $b$ refer to $\varphi_\text{1a}$, $\varphi_\text{2b}$, $\varphi_\text{2a}$, and $\varphi_\text{1b}$, respectively,
and $\epsilon_i = \epsilon_b = \epsilon_1$ and $\epsilon_j = \epsilon_a = \epsilon_2$.}
\label{fig:occupancy}
\end{figure}

Figure \ref{fig:EJI3_doubles_3} provides an explanation as to how the disconnected, unlinked diagram {\bf 48e} sums up to zero
if the degeneracy is caused by symmetry. (The same explanation applies to the disappearance of diagram {\bf 48d}.)
Figure \ref{fig:occupancy} suggests two representative occupancy patterns of degenerate reference determinants $|I\rangle$ and $|J\rangle$, which are a two-orbital
difference from each other, i.e., $|J\rangle = |I^{ab}_{ij}\rangle$ with $\epsilon_i + \epsilon_j - \epsilon_a - \epsilon_b = 0$. 
The lines of the diagrams in Fig.\ \ref{fig:EJI3_doubles_3} indicate the hole/particle distinctions 
relative to the $|I\rangle$ vacuum, and $i$ and $j$ refer to $\varphi_{\text{1a}}$
and $\varphi_{\text{2b}}$, respectively, whereas $a$ and $b$ correspond to $\varphi_{\text{2a}}$
and $\varphi_{\text{1b}}$, respectively. 

Then, diagrams {\bf 50a}--{\bf 50d} are evaluated as
\begin{eqnarray}
\text{\bf 50a} &=& (-1)^{2+3} \sum_{i,j,k,a,b}^{\substack{\epsilon_i + \epsilon_j - \epsilon_a -\epsilon_b = 0 \\ \epsilon_i - \epsilon_a \neq 0\,;\, \epsilon_i = \epsilon_k}}
\frac{(f_I)^k_i (f_I)^a_k (f_I)^b_j}{(\epsilon_i - \epsilon_a)^2} \nonumber\\
&=& - \frac{(f_I)^{\text{1a}}_{\text{1a}} (f_I)^{\text{2a}}_{\text{1a}} (f_I)^{\text{1b}}_{\text{2b}}}{(\epsilon_1 - \epsilon_2)^2}, \\
\text{\bf 50b} &=& 0, \\
\text{\bf 50c} &=& 0, \\
\text{\bf 50d} &=& - \frac{(f_I)^{\text{2b}}_{\text{2b}} (f_I)^{\text{1b}}_{\text{2b}} (f_I)^{\text{2a}}_{\text{1a}}}{(\epsilon_2 - \epsilon_1)^2},
\end{eqnarray}
where the parity of $(-1)^{2+3}$ comes from the presence of two (fictitious) loops and three hole lines. The summation over $k$ is limited to 
occupied orbitals in $|I\rangle$ that have the same energy as $\epsilon_i$  in diagram {\bf 50a} or $\epsilon_j$  in diagram {\bf 50d}.
Diagrams {\bf 50b} and {\bf 50c} are zero because $\epsilon_i - \epsilon_b = 0$ and $\epsilon_j - \epsilon_a  = 0$, respectively.

Likewise, diagrams {\bf 50e}--{\bf 50h} are evaluated as
\begin{eqnarray}
\text{\bf 50e} &=& (-1)^{2+2} \sum_{i,j,a,b,c}^{\substack{\epsilon_i + \epsilon_j - \epsilon_a -\epsilon_b = 0 \\ \epsilon_i - \epsilon_a \neq 0\,;\, \epsilon_a = \epsilon_c}}
\frac{(f_I)^a_c (f_I)^c_i (f_I)^b_j}{(\epsilon_i - \epsilon_a)^2} \nonumber\\
&=&  \frac{(f_I)^{\text{2a}}_{\text{2a}} (f_I)^{\text{2a}}_{\text{1a}} (f_I)^{\text{1b}}_{\text{2b}}}{(\epsilon_1 - \epsilon_2)^2}
+ \frac{(f_I)^{\text{2a}}_{\text{2c}} (f_I)^{\text{2c}}_{\text{1a}} (f_I)^{\text{1b}}_{\text{2b}}}{(\epsilon_1 - \epsilon_2)^2}, \\
\text{\bf 50f} &=& 0, \\
\text{\bf 50g} &=& 0, \\
\text{\bf 50h} &=& \frac{(f_I)^{\text{1b}}_{\text{1b}} (f_I)^{\text{1b}}_{\text{2b}} (f_I)^{\text{2a}}_{\text{1a}}}{(\epsilon_2 - \epsilon_1)^2},
\end{eqnarray}
where $c$ runs over all virtual orbitals in $|J\rangle$ with $\epsilon_c = \epsilon_a$ in diagram {\bf 50e}, which are $\varphi_{\text{2a}}$ and $\varphi_{\text{2c}}$;
in diagram {\bf 50h}, $\epsilon_c = \epsilon_b$ and hence $c$ is limited to $\varphi_{\text{1b}}$. 

If the degeneracy is caused by symmetry, we expect 
\begin{eqnarray}
(f_I)^\text{1a}_\text{1a} &=& (f_I)^\text{1b}_\text{1b}, \label{identity1}\\
(f_I)^\text{2a}_\text{2a} &=& (f_I)^\text{2b}_\text{2b}, \label{identity2}\\
(f_I)^\text{2a}_\text{2c} &=& 0. \label{identity3}
\end{eqnarray}
The first two identities are self-evident. The last one can be inferred as follows: We choose as degenerate orbitals the eigenfunctions 
of a proper ($\hat{C}_n$) or improper ($\hat{S}_n$) rotation operator responsible for causing the degeneracy.
All eigenvalues are distinct ($e^{ik\pi/n}$ with an integer $k$). Since the Fock operator commutes 
with all symmetry operators, say, $[\hat{C}_n,\hat{f}_I] = 0$, we have $\langle \varphi_\text{2a} | \hat{C}_n \hat{f}_I | \varphi_\text{2c} \rangle 
- \langle \varphi_\text{2a} | \hat{f}_I \hat{C}_n | \varphi_\text{2c} \rangle$ = 0. (Note that the orbitals are not necessarily
eigenfunctions of $\hat{f}_I$.) 
Since the eigenvalues of $\hat{C}_n$  for $\varphi_\text{2a}$ 
and $\varphi_\text{2c}$ are different, we have $\langle \varphi_\text{2a} | \hat{f}_I | \varphi_\text{2c} \rangle = (f_I)^\text{2a}_\text{2c}  = 0$.

Then, we infer that diagrams {\bf 50a} through {\bf 50h} cancel each other and sum to zero, 
confirming that diagram {\bf 48e} is zero by symmetry. 
The same logic dictates that the whole set of skeleton diagram {\bf 48d} also vanishes as a sum.
Together, $E_{JI}^{(3)}$  in the double-excitation space is zero and does not introduce a disconnected diagram insofar as 
the degeneracy is caused by symmetry. The foregoing argument is a derivative of the one put forward by Jeziorski and Monkhorst.\cite{Jeziorski1981}

Conversely, $\Delta$MP theory {\it does} seem to suffer from disconnected and thus unlinked diagrams for accidental 
degeneracies. This seems incongruous to 
the analytic proof of the size-extensivity of $\Delta$MP theory in Appendix \ref{app:analytic}, which is independent of the origin of degeneracy.  
Therefore, it seems plausible that another mechanism that nullifies diagrams {\bf 48d} and {\bf 48e} may be at play. For instance, 
disconnected diagrams of the variance (or anomalous diagrams) 
occur for the first time at the sixth order as $\langle E_{IJ}^{(3)}E_{JI} ^{(3)}\rangle$ by contracting
two of these disconnected diagrams {\bf 48d} and {\bf 48e}. An example has  already been given as diagram {\bf 37i}.
It contains three fictitious resolvent lines, together reducing its volume dependence from the nonphysical
$O(V^2)$ to the correct $O(V)$, even for likely the largest degenerate multireference of three-dimensional HEG. 
Therefore, diagrams {\bf 48d} and {\bf 48e} may be allowed in a size-extensive degenerate theory, and this explanation may be more in line
with the analytic proof in Appendix  \ref{app:analytic}. 

\begin{figure}
\includegraphics[scale=0.6]{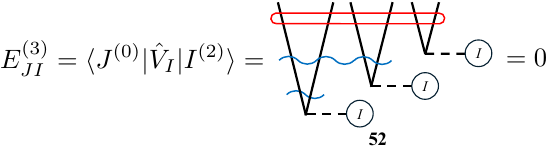}
\caption{The skeleton diagram for the third-order correction to the off-diagonal energy $E_{JI}^{(3)}$ with the one-electron perturbation  $\hat{V}_I$,  where $|J\rangle = |I^{abc}_{ijk}\rangle$
and $\epsilon_i + \epsilon_j + \epsilon_k - \epsilon_a- \epsilon_b - \epsilon_c = 0$. It is disconnected, but zero by the factorization theorem.}
\label{fig:EJI3_triples}
\end{figure}

Figure \ref{fig:EJI3_triples} shows the only diagram of $E_{JI}^{(3)}$  in the triple-excitation space, i.e., $|J\rangle = |I^{abc}_{ijk}\rangle$
with $\epsilon_i + \epsilon_j + \epsilon_k - \epsilon_a- \epsilon_b - \epsilon_c = 0$. It is disconnected, but can be shown to vanish
by virtue of the factorization theorem. 
Hence, the $E_{JI}^{(3)}$  is zero in the triple-excitation space and does not cause disconnectedness.

Together, the whole $E_{JI}^{(3)}$ is connected.

\begin{figure}
\includegraphics[scale=0.6]{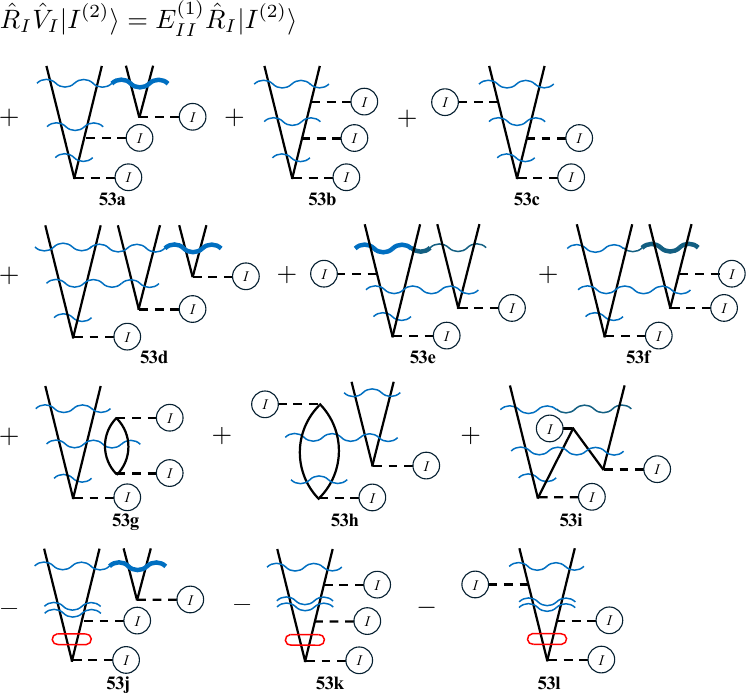}
\caption{Skeleton diagrams for the principal part of the third-order wave function with the one-electron perturbation  $\hat{V}_I$.}
\label{fig:RVI2}
\end{figure}

\begin{figure}
\includegraphics[scale=0.6]{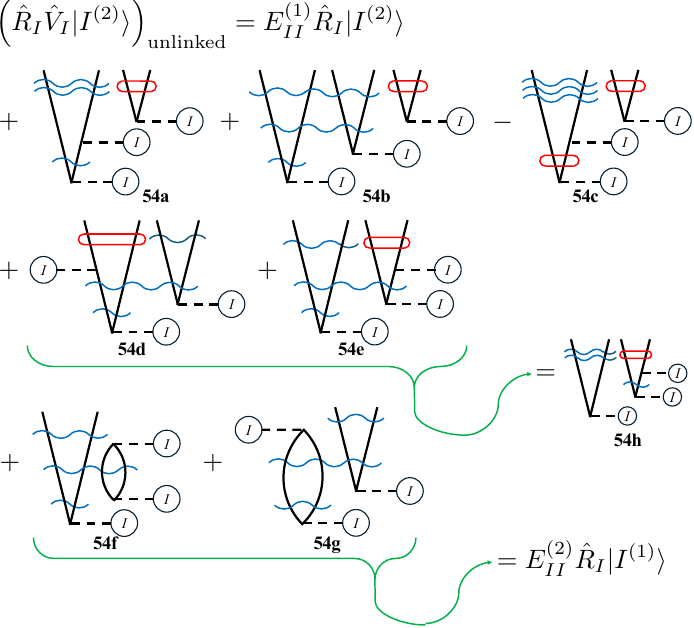}
\caption{The unlinked contributions to the principal part of the third-order wave function. The first and last four terms are canceled by
the unlinked contributions in Figs.\ \ref{fig:RJ2EJI1_unlinked} and \ref{fig:RJ1EJI2_unlinked}, respectively.}
\label{fig:RVI2_unlinked}
\end{figure}

\begin{figure}
\includegraphics[scale=0.6]{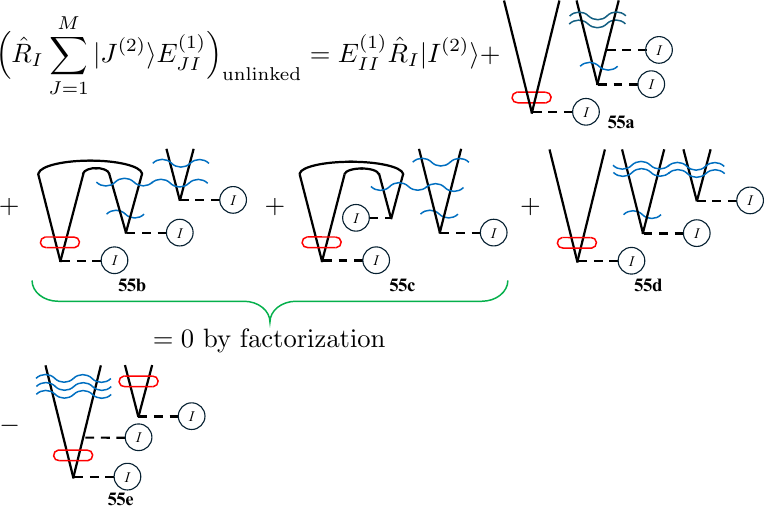}
\caption{The unlinked contributions to  the first renormalization term of the third-order wave function, i.e., the second term in the right-hand side of Eq.\ (\ref{I3}).
They cancel the first four terms of Fig.\ \ref{fig:RVI2_unlinked}.}
\label{fig:RJ2EJI1_unlinked}
\end{figure}

\begin{figure}
\includegraphics[scale=0.6]{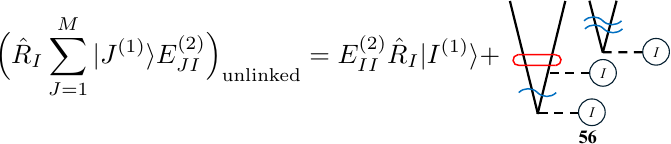}
\caption{The unlinked contributions to the second renormalization term of the third-order wave function, i.e., the last term in the right-hand side of Eq.\ (\ref{I3}).
They cancel the last four terms of Fig.\ \ref{fig:RVI2_unlinked}.}
\label{fig:RJ1EJI2_unlinked}
\end{figure}

The third-order correction to the wave function is given by
\begin{eqnarray}
|I^{(3)}\rangle = \hat{R}_I \hat{V}_I| I^{(2)}\rangle - \hat{R}_I \sum_{J=1}^M |J^{(2)}\rangle E_{JI}^{(1)}  - \hat{R}_I \sum_{J=1}^M |J^{(1)}\rangle E_{JI}^{(2)}. \nonumber\\ \label{I3}
\end{eqnarray}
Figure \ref{fig:RVI2} lists skeleton diagrams of the principal part, i.e., the first term in the right-hand side. 
Recalling that the thick span of a resolvent line (blue wiggly line) indicates that the associated partial denominator
can be zero, the unlinked contributions to the principal part consist of diagrams in Fig.\ \ref{fig:RVI2_unlinked}. Some of them
are consolidated by virtue of the factorization theorem, as indicated in the figure. 

All unlinked contributions to the renormalization terms,
i.e., the second and third terms in the right-hand side of Eq.\ (\ref{I3}), are drawn in Figs.\ \ref{fig:RJ2EJI1_unlinked} and \ref{fig:RJ1EJI2_unlinked}. 
Folded diagrams {\bf 55b} and {\bf 55c} are zero because the thin resolvent lines and fictitious resolvent lines impose incompatible conditions 
and thus annihilate them (which is equivalent to invoking the factorization theorem). 
It can be seen that all unlinked diagrams in the principal part (Fig.\ \ref{fig:RVI2_unlinked}) are systematically 
canceled out by the same in the renormalization parts (Figs.\ \ref{fig:RJ2EJI1_unlinked} and \ref{fig:RJ1EJI2_unlinked}). Therefore, $|I^{(3)}\rangle$ is linked, although 
it contains numerous disconnected diagrams.

This concludes the illustration of the connectedness of $E_{JI}^{(n)}$ and the linkedness of $|I^{(n)}\rangle$ for $1 \leq n \leq 3$, and thus the size-extensivity
of $\Delta$MP theory up to the third order under the simplifying assumption of the one-electron $\hat{V}_I$. 
It is believed, but not proven, that this continues to be the case
with the full Hamiltonian at higher perturbation orders. 
This is consistent with the size-extensivity of finite-temperature MBPT, which sums over $\Delta$MP energies.

\subsection{Size-extensivity of $\Delta$CC theory (diagrammatic) \label{app:DCC}}

Since $\Delta$CC theory is an infinite partial summation of the $\Delta$MP diagrams, the size-extensivity of $\Delta$MP theory implies
the same of $\Delta$CC theory. The strategy introduced by Jeziorski and Monkhorst\cite{Jeziorski1981} and adopted later by others\cite{Meissner1989} 
of demonstrating the size-extensivity of multireference CC theories can be used here, which shows the connectedness of energy equations and the linkedness of amplitude (wave-function) equations in their several low-order perturbation approximations, followed by mathematical induction. However, again, the following analysis is not a proof, but an illustration.

\begin{figure}
\includegraphics[scale=0.6]{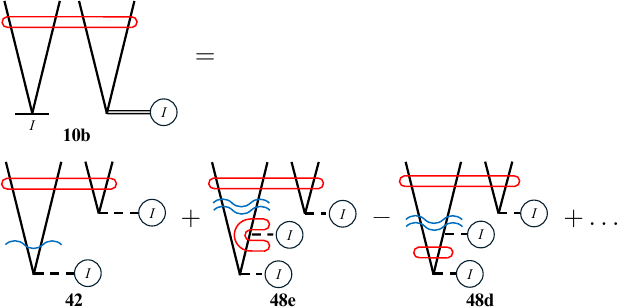}
\caption{The disconnected diagram in the off-diagonal energy $E_{JI}$ of $\Delta$CCSD theory, where $|J\rangle = |I^{ab}_{ij}\rangle$
and $\epsilon_i + \epsilon_j - \epsilon_a - \epsilon_b = 0$, as an infinite partial summation of the $\Delta$MP diagrams, which are zero.}
\label{fig:CC_EJI}
\end{figure}

Let us consider $\Delta$CCSD theory. All of its diagonal energy diagrams are connected (Fig.\ \ref{fig:HII}).

The diagrams of the off-diagonal energy $E_{JI}$ in the single-excitation space
are also connected (Fig.\ \ref{fig:HJI_singles_int}). 

The off-diagonal energy $E_{JI}$ in the double-excitation space contains an apparently disconnected
diagram {\bf 10b} (Fig.\ \ref{fig:HJI_doubles_int}). It is reproduced in Fig.\ \ref{fig:CC_EJI} with its perturbation expansion.  In the second-order perturbation approximation, this disconnected diagram is 
equal to diagram {\bf 42} and is zero (see Fig.\ \ref{fig:EJI2_doubles}). In the third-order perturbation approximation, it is the sum of diagrams {\bf 48a} through {\bf 48d}, which is either
zero by symmetry or size-extensive due to the multiple degeneracy conditions. 

The left-hand side of the $T_1$-amplitude equation (corresponding to the principal part) 
 contains only one unlinked diagram {\bf 2b} (Fig.\ \ref{fig:HJI_singles_ext}). As  discussed 
in Sec.\ \ref{sec:DCCSD}, it is canceled by the same unlinked diagram {\bf 11a} (Fig.\ \ref{fig:RHS_singles}) in the right-hand side (corresponding to the renormalization part).
In the right-hand side, an additional unlinked diagram emerges as a result of folding, which is diagram {\bf 12i} (Fig.\ \ref{fig:RHS_singles_folded1}). 
However, as indicated in Fig.\ \ref{fig:RHS_singles_unlinked}, this unlinked diagram is zero by virtue of the C condition. 

\begin{figure}
\includegraphics[scale=0.6]{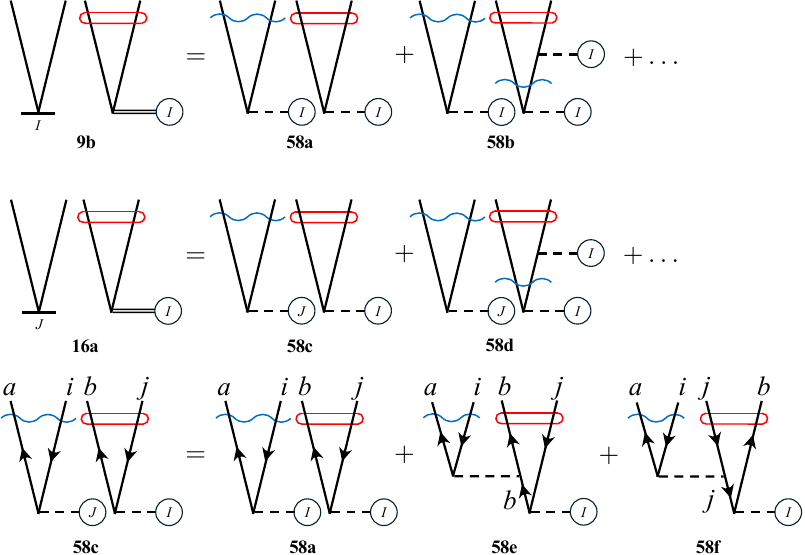}
\caption{The unlinked diagram {\bf 9b} in the left-hand side and the unlinked diagram {\bf 16a} in the right-hand side of the $T_2$-amplitude equation of $\Delta$CCSD as
infinite partial summations of the $\Delta$MP diagrams.
Their unlinked parts cancel each other exactly because diagrams {\bf 58a} and {\bf 58c} differ merely by connected diagrams {\bf 58e} and {\bf 58f}.}
\label{fig:CC_T2}
\end{figure}

The left-hand side of the $T_2$-amplitude equation in Fig.\ \ref{fig:HJI_doubles_ext}
has the disconnected diagram {\bf 7b} and unlinked product {\bf 7c}. 
The latter cancels the same unlinked product {\bf 15a} in the right-hand side (Fig.\ \ref{fig:RHS_doubles}), but the former persists. 
This disconnected diagram {\bf 7b} can be divided into linked-disconnected and genuinely unlinked contributions, and 
only the latter is illegal in a size-extensive theory. The unlinked contribution corresponds to diagrams {\bf 9a} and {\bf 9b} in Fig.\ \ref{fig:HJI_doubles_unlinked}, 
whose disconnected parts are intersected by a fictitious resolvent line at the top. However, diagram {\bf 9a} is zero by the C condition.

The unlinked diagram {\bf 9b} is canceled by the similar unlinked diagram {\bf 16a} in the right-hand side of the $T_2$-amplitude equation, but this cancellation
 is not exact. This is because they differ from each other in the $T_1$-amplitude entering the disconnected part on the left;
it is $(t_I)^a_i$-amplitude in {\bf 9b} and $(t_J)^a_i$-amplitude in {\bf 16a}.
Their perturbation approximations 
are depicted in Fig.\ \ref{fig:CC_T2}. The first-order approximations of the $(t_I)^a_i$-amplitude are $(f_I)^a_i/(\epsilon_i - \epsilon_a)$, whereas that of the $(t_J)^a_i$-amplitude
is $(f_J)^a_i/(\epsilon_i - \epsilon_a)$. Furthermore, since $|J\rangle = |I^b_j\rangle$, $(f_J)^a_i = (f_I)^a_i + (v)^{ab}_{ib} -  (v)^{aj}_{ij}$, 
the lowest-order approximations to {\bf 9b} and {\bf 16a} can be diagrammatically related to each other 
in the bottom row of Fig.\ \ref{fig:CC_T2}. It can be seen that the unlinked portions of diagrams {\bf 9b} and {\bf 16a} are common ({\bf 58a}) and cancel each other exactly, 
leaving connected remnants such as {\bf 58e} and {\bf 58f}. (These remnants can furthermore be absorbed by diagrams similar to {\bf 23c}, lifting their index restrictions.)
In the main text, we described that these unlinked diagrams {\it substantively} cancel each other out, meaning that the remnants of the inexact cancellation are connected. 

We have thus shown that $\Delta$CCSD theory is diagrammatically size-extensive. 

\section*{supplementary materials}
The numerical data used to generate the figures are provided. 
A proof of the equivalence of VCCS and HF theories is also included.

\acknowledgments
The author sincerely thanks Professors Rodney J. Bartlett, Anna I. Krylov, Mart\'{i}n Mosquera, and Piotr Piecuch for their critical reading of the manuscript, technical advice, and stimulating discussions.
This work has been supported by the U.S. Department of Energy (DoE), Office of Science, Office of Basic Energy Sciences under Grant No.\ DE-SC0006028.. 

\section*{Data Availability}

The data that support the findings of this study are available within the article and its supplementary materials.

\bibliography{library}
\end{document}